# Homeostasis: Design and Implementation of a Self-Stabilizing Compiler

AMAN NOUGRAHIYA, Department of CSE, IIT Madras, India
V. KRISHNA NANDIVADA, Department of CSE, IIT Madras, India

Mainstream compilers perform a multitude of analyses and optimizations on the given input program. Each analysis (such as points-to analysis) may generate a *program-abstraction* (such as points-to graph). Each optimization is typically composed of multiple alternating phases of *inspection* of such program-abstractions and *transformations* of the program. Upon transformation of a program, the program-abstractions generated by various analyses may become inconsistent with the modified program. Consequently, the correctness of the downstream inspection (and consequent transformation) phases cannot be ensured until the relevant program-abstractions are *stabilized*; that is, the program-abstractions are either invalidated or made consistent with the modified program. In general, the existing compiler frameworks do not perform automated stabilization of the program-abstractions and instead leave it to the compiler pass writers to deal with the complex task of identifying the relevant program-abstractions to be stabilized, the points where the stabilization is to be performed, and the exact procedure of stabilization. In this paper, we address these challenges by providing the design and implementation of a novel compiler-design framework called *Homeostasis*.

*Homeostasis* automatically captures all the program changes performed by each transformation phase, and later, triggers the required stabilization using the captured information, if needed. We also provide a formal description of *Homeostasis* and a correctness proof thereof. To assess the feasibility of using *Homeostasis* in compilers of parallel programs, we have implemented our proposed idea in IMOP, a compiler framework for OpenMP C programs. Further, to illustrate the benefits of using *Homeostasis*, we have implemented a set of standard data-flow passes, and a set of involved optimizations that are used to remove redundant barriers in OpenMP C programs. Implementations of none of these optimizations in IMOP required any additional lines of code for stabilization of the program-abstractions. We present an evaluation in the context of these optimizations and analyses, which demonstrates that *Homeostasis* is efficient and easy to use.

## 1 INTRODUCTION

Modern compilers often span millions of lines of code, and perform a multitude of analyses and optimizations on the given input program. Each analysis (such as points-to analysis, call-graph construction, and so on) derives some meaningful information about the program, in the form of a *program-abstraction* (such as points-to graph, call-graph, and so on). An optimization typically involves multiple alternating phases of inspections of such program-abstractions, and transformations of the program. In the *inspection phase*, the optimization invokes the required analyses, if needed, and inspects various program-abstractions to discover opportunities of optimization in the program. In the *transformation phase*, the optimization pass may use the results of the inspection phase to modify the program. Consequently, the transformation phase may render various

Authors' addresses: Aman Nougrahiya, amannoug@cse.iitm.ac.in, Department of CSE, IIT Madras, India; V. Krishna Nandivada, nvk@iitm.ac.in, Department of CSE, IIT Madras, India.

```
1 DeadBranchRemover()
2 begin
3     genPointsToIfNotAvailable();
4     foreach f ∈ IfStmts do
5         c:= f.getPredicate();
6         if isNeverZero(c) then
7             replace(f, f.thenBranch);
8         else if isAlwaysZero(c) then
9             replace(f, f.elseBranch);
```

(a) Optimization: Dead-branch remover using points-to information.

```
int a, b, c, *p, *q;
p = &a; q = &c;
L1: if (q!=&c) {p = &b;}
L2: if (p==&a) S1 else S2
```

(b) Input program ($P_1$)

| Node | Points-to Map |
|---|---|
| L1: (q != &c) | p:{a}, q:{c} |
| L2: (p == &a) | p:{a, b}, q:{c} |
| ... | ... |

(c) Program-abstraction: Points-to map.

```
int a,b,c,*p,*q;
p=&a; q=&c;
L2: if (p==&a) S1
    else S2
```

(d) Generated program ($P_2$).

```
int a,b,c,*p,*q;
p=&a; q=&c;
S1
```

(e) Expected program ($P_3$).

Fig. 1. An example optimizer pass, and a program-abstraction. After removing the if-statement at L1 from the input program $P_1$ to obtain $P_2$, the optimizer will not be able to remove the *else*-branch of L2, thereby not producing the expected optimal code $P_3$: In the absence of stabilization, the points-to map from Fig. 1c does not get updated after removal of the statement at L1. Consequently, with the stale points-to information, the optimizer cannot infer that pointer $p$ may point only to $a$ at the predicate at L2 in $P_2$.

existing program-abstractions inconsistent with the modified program. As a result, the existing program-abstractions must be *stabilized* before their next use; that is, the program-abstractions must be regenerated or updated such that they correspond to the modified program. Otherwise, the correctness of the future inspection phases of the current or downstream optimizations cannot be ensured. This, in turn, can negatively impact the optimality and even the correctness of the overall compilation. We illustrate these problems using an example.

*Example 1.1.* In Fig. 1a, we show the pseudocode of a standard compiler optimization that utilizes points-to information to remove dead branches from a program. For simplicity, we assume that the predicates do not contain any side-effects, and that there are no arbitrary `goto`-statements in the input program. The `isNeverZero` method returns `true` when it can be statically determined that the given expression can never evaluate to zero; similarly, the `isAlwaysZero` method returns `true` only when the expression would always evaluate to zero. Further, assume that both these methods use the points-to information to statically evaluate the given expression. Hence, in the inspection phase, the optimization pass invokes a flow-sensitive points-to analysis pass to obtain the required points-to map (a program-abstraction), if not already available. Further, in this phase, the optimization utilizes the methods `isNeverZero` and `isAlwaysZero` to determine whether an if-statement can be replaced with either of its branches. Accordingly, the transformation phase may be triggered to perform the replacement using the `replace` method (Lines 7 and 9). Note that these transformation and inspection phases may alternate in the presence of multiple if-statements in the program.

This simple and intuitive code of dead-branch remover may not work as expected: Consider a sample input program $P_1$ shown in Fig. 1b, and a part of its points-to map in Fig. 1c. Say the optimization processes the if-statement at label L1 first, and relying on the points-to map at the predicate, replaces the if-statement with its `else`-branch (i.e., an empty body), transforming program $P_1$ to $P_2$ (Fig. 1d). This transformation renders the points-to map stale: at the predicate `(p == &a)`, the pointer p continues to point to {a, b}, instead of {a}, the more precise points-to set. In the absence of stabilization of the points-to map, the optimizer pass in its next iteration cannot determine that the predicate at L2 will never be zero. As a result, the optimizer will not generate the optimal program $P_3$; instead, $P_2$ will be the final output of the optimizer. Note that such issues of sub-optimality (and even correctness) may also be observed in the downstream optimization passes, until the points-to map is stabilized.

| Upon adding a new optimization $O_n$ | Upon adding a new program analysis $\mathcal{A}_n$ |
|---|---|
| **Q$_1$.** *Which* existing program-abstractions need to be stabilized by $O_n$? | **Q$_1^T$.** *Which* existing optimizations need to stabilize the program-abstraction of $\mathcal{A}_n$? |
| **Q$_2$.** *Where* to invoke the stabilization code in $O_n$, for the existing program-abstractions? | **Q$_2^T$.** *Where* to invoke the stabilization code for $\mathcal{A}_n$, in each of the existing optimizations? |
| **Q$_3$.** *How* to stabilize each of the existing program-abstractions, in $O_n$? | **Q$_3^T$.** *How* to stabilize the program-abstraction of $\mathcal{A}_n$, in each of the existing optimizations? |

Fig. 2. Key stabilization-related question to be addressed while adding new compiler passes.

To perform the correct stabilization manually and efficiently, we need to identify all the places in the dead-branch remover where the points-to map may require stabilization. Further, we also need to be concerned about the stabilization of all the other program-abstractions as well: (i) We need to identify the program-abstractions (for example, use-def information [Muchnick 1998]) that may become stale because of the current optimization, and may be required by other downstream optimizations and analyses. (ii) For each of these program-abstractions, we need to identify the places in the dead-branch remover where the program-abstraction needs to be stabilized. In practice, in the presence of multiple optimizations and analyses (hereafter, collectively referred to as *compiler passes*), this problem compounds as we need to be concerned about stabilization of all the program-abstractions in all the optimizations.

In summary, we observe that to avoid the generation of such sub-optimal and/or incorrect programs upon compilation, three key stabilization-related questions shown in Fig. 2 need to be addressed while adding/modifying an optimization or analysis. In conventional compilers, such as LLVM [Lattner and Adve 2004], GCC [Stallman and GCC-Developer-Community 2009], Soot [Vallée-Rai et al. 2010], Rose [Quinlan et al. 2013], JIT compilers (OpenJ9 [IBM 2017], HotSpot [Oracle 1999], V8 [Google 2001]), and so on, the onus of addressing these key questions lies mostly on the compiler pass writers. Some such instances are shown in Fig. 3. This manual process often leads to complex and error-prone codes. Further, as the number and complexity of compiler passes increase (for example, LLVM has more than 347 passes), addressing these questions precisely and efficiently becomes progressively more difficult. This may lead to correctness/efficiency bugs (see bug-fixing commits on GitHub [LLVM-Developer-Community 2019a,b,c, 2020a,b,c,d,e,f,g,h,i, 2021a,b,c]). With the compiler development effort spanning multiple decades involving (sometimes) hundreds of developers, it becomes extremely challenging to manually solve these problems, as no developer of a compiler pass might possess a clear understanding of the semantics of all the hundreds of other passes already present in the compiler.

To mitigate these issues, the focus of this paper is to perform automatic stabilization of all the relevant program-abstractions, when needed, in response to any program modification. We term a compiler that provides this guarantee as a *self-stabilizing* compiler.

There have been various attempts [Blume et al. 1995; Brewster and Abdelrahman 2001; Carle and Pollock 1989; Carroll and Polychronopoulos 2003; Nilsson-Nyman et al. 2009; Reps et al. 1983] towards enabling automated stabilization of *specific* program-abstractions, in response to program transformations in serial programs. However, these works suffer from different drawbacks. For example, Carle and Pollock [1989]; Nilsson-Nyman et al. [2009]; Reps et al. [1983] require that program-abstractions have to be expressed as context-dependent attributes of the language constructs; this is too restrictive. Similarly, besides the non-object-oriented nature of the approach of Carroll and Polychronopoulos [2003], it is unclear how their restrictive techniques can (i) be

```
void getAnalysisUsage(AnalysisUsage &AU) const
    override {
  AU.setPreservesCFG();
  AU.addRequired<AAResultsWrapperPass>();
  AU.addRequired<TargetLibraryInfoWrapperPass>();
  AU.addPreserved<GlobalsAAWrapperPass>();
  AU.addRequired<DominatorTreeWrapperPass>();
  AU.addPreserved<DominatorTreeWrapperPass>();
  AU.addRequired<PostDominatorTreeWrapperPass>();
  ...
```

(a) `DeadStoreElimination.cpp:2068-2077` (LLVM)

```
/* We do not update postdominators,
   so free them unconditionally.*/
free_dominance_info(CDI_POST_DOMINATORS);
/* If we removed paths in the CFG, then
 we need to update dominators as well.
 I haven't investigated the possibility
 of incrementally updating dominators.*/
if (cfg_altered)
  free_dominance_info(CDI_DOMINATORS);
```

(b) `tree-ssa-dce.c:1693-1700` (GCC)

```
_somethingChanged |= inlineCallSites();
// Use/def and value number information must be recalculated for later optimization passes.
if (_somethingChanged || _invalidateUseDefInfo) {
  optimizer()->setUseDefInfo(NULL); _useDefInfo = NULL;}
if (_somethingChanged) {
   optimizer()->setValueNumberInfo(NULL);
   requestOpt(OMR::treeSimplification); requestOpt(OMR::globalValuePropagation);
```

(c) `EscapeAnalysis.java:1357-1371` (OpenJ9)

Fig. 3. Example snippets from LLVM, GCC, and OpenJ9 demonstrating manual stabilization. The comments are written by the pass writers. Snippet (a) shows how pass writers manually specify pass dependencies in LLVM. Snippet (b) shows how some selected program-abstractions are invalidated manually in GCC. Snippet (c) depicts a case where the pass writer has manually written additional code for stabilization after a transformation phase.

applied to compilers that use non-hierarchical IRs, which is a common practice in modern compilers (such as LLVM, Soot, OpenJ9, IMOP [Nougrahiya and Nandivada 2019], and so on), (ii) be used for data-flow analyses that are not based on structural analysis [Muchnick 1998], and (iii) automate the resolution of pass dependencies, or handle manually-defined dependencies. Blume et al. [1995]; Brewster and Abdelrahman [2001] handle only a small set of specific program-abstractions (such as symbol and scope information); this is insufficient. To the best of our knowledge, there are no compiler designs or implementations[1] that address the challenges discussed above for any arbitrary IR, optimization and analysis.

**Proposed solution.** In this paper, we propose a novel and efficient compiler-design framework called *Homeostasis* that addresses the above discussed issues related to stabilization, in the context of object-oriented compilers. *Homeostasis* efficiently captures a summary of all the program changes performed by each transformation phase. Later, if/when an attempt is made to read the possibly stale program-abstraction of a program analysis, *Homeostasis* triggers the required stabilization using the captured information, before returning the correct value of the program-abstraction. With *Homeostasis*, no additional coding effort is required by the optimization writers to guarantee self-stabilization of the existing (or future) program-abstractions. For example, in an implementation of the dead-branch remover (Fig. 1a) in a *Homeostasis*-enabled compiler, *Homeostasis* would ensure self-stabilization of the points-to map, without requiring any additional code in the optimization; this will lead to generation of the expected optimal program $P_3$.

In *Homeostasis*, for ensuring self-stabilization, analysis writers need to spend minimal effort to only conform to the design of *Homeostasis*, while being oblivious to all the (existing/future) optimizations. As part of this conformance, if the analysis writer decides to use an incremental update algorithm for stabilization (instead of full recomputation), the writer has to clearly demarcate the code to incrementally update the program-abstraction. Note that such a code is needed even when the stabilization is performed manually; hence this is not an additional overhead. In essence, while working with a *Homeostasis*-enabled compiler, a compiler pass writer need not manually address any of the key questions listed in Fig. 2, and hence incurs no additional overhead.

[1] Our observation is based on our manual inspection of more than 50 mainstream and experimental compilers [Nougrahiya and Nandivada 2023].

For the ease of exposition, we have discussed *Homeostasis* using IMOP [Nougrahiya and Nandivada 2019] a Java-based compiler in the context of a C-like language. We believe that the underlying principles of the *Homeostasis* are quite generic in nature, and that they can be extended to any object-oriented compilers for any serial/parallel languages.

Note that the goal of this paper is not to propose any new incremental analysis technique (or any new analysis technique or optimization for that matter). The key goal of this paper is to present a scheme which ensures that correct values of various program-abstractions are obtained efficiently by all the compiler passes, without needing any additional manual effort in these passes.

**Contributions:**

- We present *Homeostasis*, a generalized compiler-design framework for *self-stabilization*, which can be added to any object-oriented compiler infrastructure to ensure that all program-abstractions are automatically kept consistent with the modified program during the compilation process. The underlying concepts of *Homeostasis* can be used in any object-oriented compiler for serial/parallel languages to realize self-stabilization.
- To overcome the possible overheads resulting from the naive tracking of the program changes, we present a novel compression algorithm that reduces the memory requirements significantly.
- We give a formal description of *Homeostasis* and a correctness proof thereof, which guarantees that in a *Homeostasis*-enabled compiler, any compiler pass attempting to access the program-abstraction of an analysis will see only the stable value of that program-abstraction.
- We have implemented all the key components of *Homeostasis* in IMOP, a source-to-source Java-based compiler infrastructure for OpenMP C programs, as a successful proof-of-concept. Further, to demonstrate the benefits of using *Homeostasis*, we have implemented (i) HIDFA, a generic, self-stable, inter-thread, flow-sensitive, context-insensitive iterative data-flow analysis (IDFA) and six instantiations thereof, (ii) a set of optimization passes, collectively termed as `BarrElim`, that removes redundant barriers from OpenMP C programs; `BarrElim` uses a set of nine optimization passes, each of which may involve multiple alternating phases of inspection and transformation. As expected, the implementation of the set of `BarrElim` optimization passes required zero additional lines of code for stabilization, and the stabilization of the analyses required minimal effort.
- We present an evaluation over a set of twenty-four benchmarks from four benchmark suites NPB [Van der Wijngaart and Wong 2002], SPEC OMP [Aslot et al. 2001], Sequoia [Seager, M 2008] and IMSuite [Gupta and Nandivada 2015], performed over two different platforms. We show that *Homeostasis* makes it easy to write optimization and analysis passes, and it leads to faster compilation times compared to the possible alternative of typical manual stabilization performed by experts. We also show that our proposed optimizations lead to significant memory savings.

The rest of the manuscript is organized as follows: We give a brief background of the relevant concepts used in the manuscript in Section 2. We present our design of *Homeostasis* in Section 3. We give a formal description of *Homeostasis*, along with a correctness argument in Section 4. We discuss some salient features of *Homeostasis* in Section 5. In Section 6, we discuss an instantiation of *Homeostasis* in a real-world compiler. In Section 7, we briefly describe our implementation and present a detailed evaluation of *Homeostasis*. In Section 8, we discuss some of the relevant prior work, before concluding in Section 9.

## 2 BACKGROUND AND TERMINOLOGY

We now give a brief description of certain relevant concepts and terminologies used in this paper.

**Representation of program IR.** Without loss of generality, in this paper we assume each program's intermediate representation (IR) is represented as a graph. For instance, when the control flow graph (CFG) is the IR, nodes in the graph represent the basic-blocks, and the edges model the flow of control. Similarly, if three-address code (TAC) is the IR, then the TAC instructions are the nodes, and the edges may be of many types, such as control-flow edges, SSA edges [Muchnick 1998], and so on.

**Analysis, Analysis-instances, and Program-abstractions.** In the context of compilers written in an object-oriented language, we assume that each analysis (for example, Andersen's points-to analysis [Andersen 1994]) is implemented as a class. During compilation, one or more of these classes are instantiated to provide analysis-instances. Each such analysis-instance, say $A_i$, can be seen as a function from the set of programs to a set of program-abstractions. A program-abstraction may refer to any data-structure that denotes some meaningful information generated by an analysis-pass, about the input program at compile time. In this manuscript, we use $A_i(P)$ to denote the program-abstraction generated by running the underlying analysis of $A_i$ on a program $P$. For example, Fig. 1c shows a part of the points-to map, the program-abstraction of the corresponding points-to analysis.

**Stable program-abstraction.** At any point during compilation, if the current program IR is $P$, and the stored program-abstraction for an analysis-instance $A_i$ is $v$, then we term $v$ as the *stable* program-abstraction of $A_i$, iff $A_i(P) = v$.

**Elementary transformations.** Based on the grammar of the language under consideration, corresponding to each type of program node with a *body* (such as a block in LLVM IR, or a while-statement construct), we identify a fixed set of primitive transformations which can add/delete/modify the logical components of the program node (such as an instruction of a block, or the predicate of a while-statement); we term such transformations as *elementary transformations*.

## 3 HOMEOSTASIS : DESIGNING SELF-STABILIZING COMPILERS

Considering the stabilization challenges inherent in the design of conventional compilers, we now present a novel solution in the form of a new object-oriented compiler-design framework named *Homeostasis* (in Sections 3.1, 3.2, and 3.3). In order to access/maintain stable program-abstractions in a *Homeostasis*-enabled compiler, while the analysis pass writers have to spend minimal effort, the optimization pass writers need not spend any effort (elaborated in Section 3.4). We formalize the design of *Homeostasis* and present a correctness argument in Section 4.

### 3.1 Overview of *Homeostasis*

*Homeostasis* realizes self-stabilization by tracking the program modifications resulting from different transformations. These program modifications are then used by *Homeostasis* to perform the stabilization of relevant program-abstractions if/when required.

*3.1.1 Design pattern employed by* Homeostasis. To track program modifications resulting from different transformations, *Homeostasis* uses an extension of the popular Observer pattern [Gamma et al. 1995]. The Observer pattern defines a one-to-many dependency between multiple entities, where an update in one entity (termed *subject*), is notified to all the other registered entities (termed *observers*). Fig. 4 shows a class diagram (in Object Modeling Technique, or OMT, notation [Rumbaugh et al. 1991]) which gives an overview of the key design elements of *Homeostasis*: (i) Node class and its subtypes (such as WhileStmt) correspond to the subject(s), (ii) all the program analyses (such as CallGraphGenerator) along with their base class (BasePA) correspond to the observers, and (iii) program modifications (such as WhileStmt::elemTransformReplaceBody) correspond to the various updates to the subject(s). The use of the Observer pattern ensures that these updates (program modifications) are correctly notified to all the observers (program analysis passes).

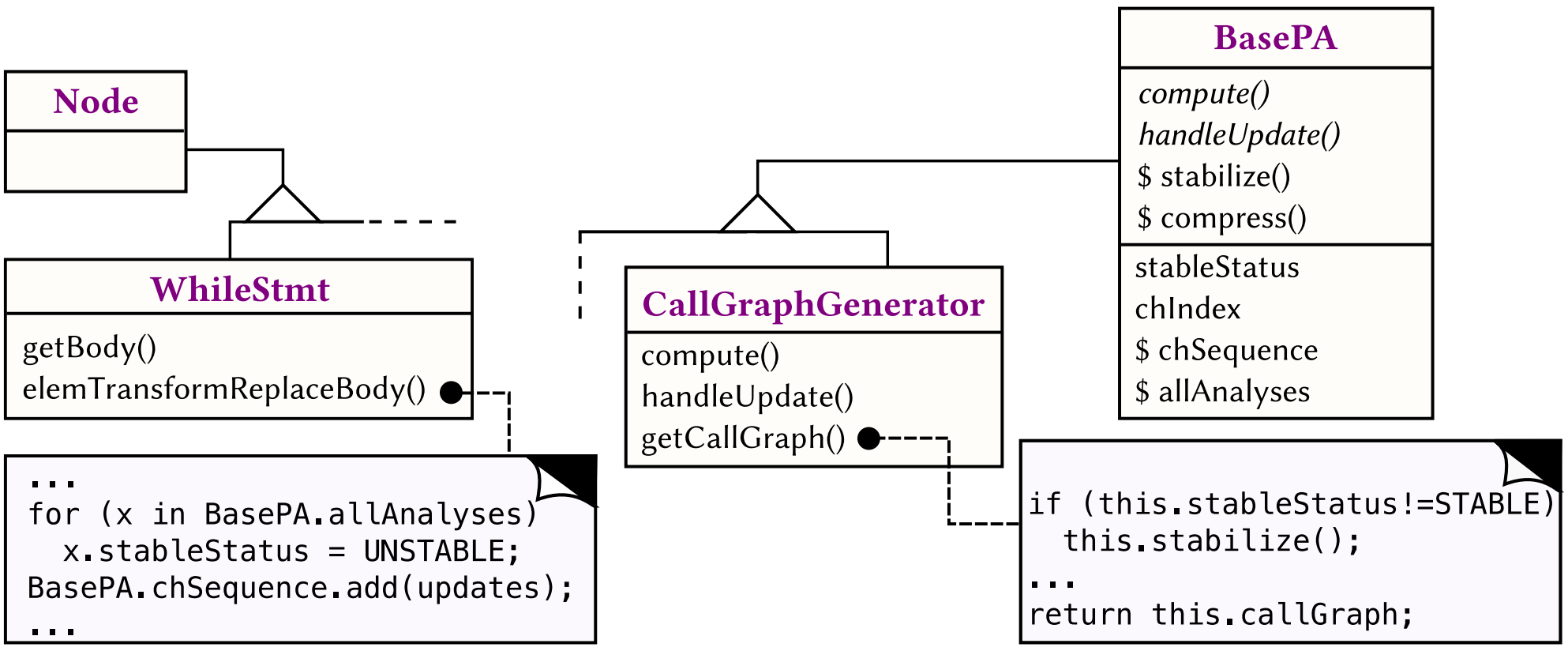

Fig. 4. Class-diagram of *Homeostasis*, depicting key classes and methods, in the OMT notation: class names are shown in **bold**; abstract methods in *italics*; static members are preceded with a $; inheritance is shown using a triangle on the edge, pointing towards the base class; and dashed arrows show code snippets for methods.

*Homeostasis* goes beyond the standard design of the Observer pattern in two key ways. First, instead of triggering updates of the values (i.e., program-abstractions) maintained by various program analyses immediately upon a program modification, *Homeostasis* simply marks the program-abstractions as stale (or UNSTABLE). The actual update of each program-abstraction is performed *lazily*, by deferring the update to a later stage where an attempt is made to read from the program-abstraction, using its *getters* (such as CallGraphGenerator::getCallGraph). Second, instead of storing a copy of the program changes corresponding to such pending updates individually with each program analysis (each of which may need to handle a different list of pending updates), *Homeostasis* maintains a central copy of these program changes in an efficient data-structure (discussed in Section 3.3).

*3.1.2 Nodes of intermediate representation (IR).* In Fig. 4, the Node class represents the super class of all the types of nodes in the program's intermediate representation (IR). Each such class may define a fixed set of *elementary transformation* methods that may be invoked to add/modify/delete any of the logical component(s) of the node. For example, in Fig. 4, elemTransformReplaceBody is an elementary transformation method defined on the node WhileStmt; this transformation replaces a component (the loop body) of the associated while-statement with the given argument. *Homeostasis* requires that (i) all program modifications are performed via a sequence of invocations of elementary transformations, (ii) each elementary transformation marks each program-abstraction in the system as stale by resetting the corresponding stableStatus flag, and (iii) each elementary transformation stores the details of the resulting IR changes in the global sequence of program-changes, chSequence, maintained by BasePA, to be read by the individual program analyses as and when required. Note that the first requirement does not restrict the availability of high-level transformation APIs to the optimization-pass writers; each such API may be adapted by the compiler-writers such that it invokes one or more elementary transformations directly and/or via invocation of other similar high-level APIs. For example, a high-level transformation API to delete a block of statements can be implemented as a sequence of invocations of an elementary transformation that deletes individual statements. Note that updating the high-level APIs is a one-time effort on the part of compiler-writers; no additional effort is needed from the optimization-pass writers.

*3.1.3 Program analysis pass classes.* In *Homeostasis*, each program analysis is implemented as a class that inherits from the abstract class BasePA, as shown in Fig. 4; this class contains various key data structures and methods that are required to enable self-stabilization of the associated program-abstractions. BasePA maintains a global set, allAnalyses, of all instances of the analysis classes (henceforth, referred to as *analysis-instances*). Further, a sequence of all changes to the IR, resulting from various elementary transformations, is maintained in a global list, named chSequence. During stabilization trigger for any analysis-instance, if needed, the exact set of IR changes whose impact needs to be incorporated in its program-abstraction, is obtained using chIndex, an index to some location in the list chSequence. Note that both allAnalyses and chSequence are initialized to empty set/list. *Homeostasis* may periodically employ the compress method to reduce the space requirements of chSequence, depending upon the chIndex value of various analysis-instances in allAnalyses (details in Section 3.2).

Fig. 4 shows CallGraphGenerator as a concrete class of BasePA, which implements the call-graph generation routines and maintains the program-abstraction denoting the call-graph information of the program. In *Homeostasis*, each program-abstraction can only be queried through its corresponding *getter* methods (such as getCallGraph in CallGraphGenerator). Before returning the requested information, each getter method should check if the program-abstraction is stable (by inspecting the value of stableStatus flag) and trigger stabilization if needed (by invoking the method stabilize). Depending upon the mode of stabilization, discussed next, the stabilize method provided by *Homeostasis* internally invokes either the compute or the handleUpdate method, passing them the correct arguments, if any.

*3.1.4 Modes of stabilization.* An important defining characteristic of self-stabilization is the time and manner in which a program-abstraction is stabilized under program modifications. Accordingly, the stabilization mode can vary along the following two dimensions: eager versus lazy, and invalidate versus update.

***Eager versus lazy.*** After an elementary transformation, the stabilization of an abstraction may either get triggered (i) immediately (eager stabilization), or (ii) only in response to the first read request made on the program-abstraction after the transformation (lazy stabilization). Note that in practice, not every program-abstraction is read after each program modification. Hence, to avoid the overheads due to the redundant invocation of the stabilization routines in case of eager stabilization, *Homeostasis* advocates the use of lazy modes of stabilization.

***Invalidate versus update.*** In response to one or more program modifications, the resulting stabilization of a program-abstraction may either (i) involve the complete invalidation of the program-abstraction (leads to regenerating the program-abstraction from scratch), or (ii) be able to incrementally update the program-abstraction based on the modifications. *Homeostasis* attains the fully-automated stabilization for invalidate modes without requiring *any* stabilization-specific code from the analysis-pass writers. In invalidate modes, the stabilize method reinitializes the program-abstraction, and invokes the compute method to regenerate the program-abstraction from scratch; note that the code for this method would exist even in the absence of *Homeostasis*. In contrast, to realize the update modes of stabilization, *Homeostasis* provides a well-defined method, handleUpdate, to be implemented by the analysis-pass writers, which it later uses during automated stabilization triggers, by passing the correct set of program changes to be handled. Though the update modes may seem much more efficient than the invalidate modes, in practice the difference in their performance depends on a number of factors, such as the number of program modifications, the complexity of the associated incremental update, and so on. Further, note that enabling the update

```
1  Function replaceX(newNode) // Method to replace the component X with newNode
2      oldX = getComponentX();
3      // Step A: Save nodes and edges affected by the removal of old node.
4      Set removedNodes = {oldX};
5      Set removedEdges = {..outgoing edges from, and incoming edges to, oldX..};
6      Set addedEdges = {..new edges formed due to the removal of oldX..};
7
8      // Step B: Perform the actual program update by invoking appropriate writer(s).
9      Replace the component X with newNode;
10
11     // Step C: Save nodes and edges affected by the addition of new node.
12     Set addedNodes = {newNode};
13     removedEdges = removedEdges ⋃ {..old edges removed due to the addition of newNode..};
14     addedEdges = addedEdges ⋃ {..outgoing edges from, and incoming edges to, newNode..};
15
16     // Step D: Communicate relevant information to each analysis.
17     for anl ∈ BasePA.allAnalyses do anl.stableStatus = UNSTABLE;
18     BasePA.chSequence.addFourTuple(addedNodes, removedNodes, addedEdges, removedEdges);
```

Fig. 5. Template of an elementary transformation in *Homeostasis* that replaces the component X of a program node with newNode.

modes of stabilization for certain program-abstractions can be a challenging task. Considering such issues, *Homeostasis* supports both invalidate as well as update modes of (lazy) stabilization.

On the basis of these two dimensions, *Homeostasis* supports the following two modes of stabilization for any program-abstraction : (i) Lazy-Invalidate (**LZINV**), and (ii) Lazy-Update (**LZUPD**).

## 3.2 Components of *Homeostasis*

In this section, we expand on the design details of different components of *Homeostasis* (in Sections 3.2.1, 3.2.2, and 3.2.3). In Section 3.2.4, we discuss the details of how effective changes to the IR, resulting from a sequence of elementary transformations, are obtained during stabilization in *Homeostasis*. Finally, in Section 3.2.5, we demonstrate how stabilization is performed in the presence of *Homeostasis*, using the example from Fig. 1a.

*3.2.1 Structure of* Node *and its subclasses.* Stabilization of program-abstractions in response to the modifications performed by an optimization pass, can intuitively be done in two ways: (i) the optimization pass directly modifies the program-abstraction, or (ii) the program analysis pass performs the stabilization internally, possibly based on the exact modifications that have been performed on the program. The first option can be complicated, and goes against the spirit/design principles of object-oriented programming. Hence, we use the latter option in *Homeostasis*.

**Elementary transformations.** *Homeostasis* uses elementary transformations to communicate the program modifications performed by different optimization passes to the program analyses. As mentioned in Section 3.1, in *Homeostasis*, all program modifications happen only via elementary transformations. Each elementary transformation in *Homeostasis* (i) collects the information about addition/deletion of IR nodes and edges, and (ii) makes the collected information visible to every program analysis object. We elaborate this using an example template of a generic elementary transformation method that replaces a component C with a new node, as shown in Fig. 5. The templates for the elementary transformation methods that add (or remove) components can be trivially derived from the above template. The template method in Fig. 5 has four main steps.

| | | |
|---|---|---|
| *Elementary transformation* | : | WhileStmt::setBody(Statement newBody); |
| oldBody | = | this.body |
| *Set* removedEdges | = | {(this.predicate, oldBody), (oldBody, this.predicate), ...} |
| *Set* addedEdges | = | {(this.predicate, newBody), (newBody, this.predicate), ...} |
| *Set* removedNodes | = | {oldBody} |
| *Set* addedNodes | = | {newBody} |

Fig. 6. Values for the various sets used in Fig. 5, for a concrete elementary transformation that replaces the body of a while-statement in the IMOP compiler framework (WhileStatementCFGInfo.java: 81-146).

**Step A:** This step applies to all those elementary transformations that may remove a node from the IR. Besides storing the removed node in `removedNodes`, this step records all the edges that have been removed (in `removedEdges`) or added (in `addedEdges`) as a result of removal of the node oldX.

**Step B:** This step performs the actual modification to the IR. For example, in Fig. 5, the component `X` of a program node is replaced with the provided new node. This step contains the code that would comprise the body of an elementary transformation in the absence of *Homeostasis*.

**Step C:** This step applies to only those elementary transformations that add a node to the IR. Besides storing the added node in `addedNodes`, this step records all the edges that have been removed (in `removedEdges`) or added (in `addedEdges`) as a result of addition of the node newNode.

**Step D:** In this step, *Homeostasis* first marks the status of each program analysis as `UNSTABLE`, so that the corresponding stabilization routine is invoked whenever its program-abstraction is read next (using the getter methods). As discussed in Section 3.1.4, for using the `LZUPD` mode of stabilization a program analysis requires information about all the added/removed nodes and edges. Instead of maintaining a copy of this information with each program analysis, for efficiency, *Homeostasis* maintains it globally in the list chSequence, in the `BasePA` class. This list is 0-indexed. Each element of chSequence is a four-tuple composed up of four sets: nodes added, nodes removed, edges added, and edges removed. At the end of each elementary transformation, the method addFourTuple appends the collected list of IR changes to the end of chSequence.

We now highlight three important points related to these four steps. (1) The exact definitions of Steps A-C depend on (i) the exact type of the node being transformed, (ii) the specific component being added/removed/replaced, and (iii) the semantics of the underlying language. For instance, in case of OpenMP compilers, addition/removal of a node containing flush-directives, would necessitate changes to the inter-task edges (or their equivalents) in the IR; these edges are used to denote inter-task communication via shared variables. The exact set of edges added/removed depend on the underlying semantics of OpenMP constructs. (2) An elementary transformation may also result in addition or removal of edges without any changes in the nodes. In such cases, Steps A and C capture the impacted edges accordingly. Further, if an elementary transformation changes only the type, label, or annotation of an edge, then this can be represented as the replacement of the old edge with a distinct new edge; the former is added to the set removedEdges in Step A, and the latter is added to the set addedEdges in Step C. (3) In Step D, the list chSequence may grow prohibitively large with increasing number of invocations of elementary transformations; this list is maintained efficiently by *Homeostasis* using a *compression optimization*, explained later in Section 3.3.

*Example 3.1 (Concrete elementary transformation).* Fig. 6 shows the values for the various sets used in Fig. 5, for an elementary transformation method that replaces the body component of a while-statement with a given argument (newBody). This example has been taken from our implementation of *Homeostasis* in the IMOP [Nougrahiya and Nandivada 2019] compiler framework for OpenMP C programs. Note that in addition to the elements shown, the sets removedEdges and addedEdges

```
/* BasePA constructor */
public BasePA::BasePA(StabilizationMode mode) {
  BasePA.allAnalyses.add(this);
  this.compute();
  this.stabilizationMode = mode;
  if (this.stabilizationMode == LZINV) {
   this.chIndex = -1;}
  else { // this.stabilizationMode = LZUPD
   this.chIndex = BasePA.chSequence.size() - 1;}
```

(a) Constructor of the base analysis class.

```
/* Template of a program-abstraction getter */
public V Analysis_Y::get() {
  if (this.stableStatus==UNSTABLE) {
   this.stabilize();}
  return this.abstraction;}
```

(b) Template of a getter method of a sample analysis.

```
/* Program-abstraction stabilizer */
private void BasePA::stabilize() {
  this.stableStatus = STABLE;
  if (this.stabilizationMode == LZINV) {
   this.compute();}
  else {// this.stabilizationMode = LZUPD
   this.handleUpdate();
   this.chIndex = BasePA.chSequence.size() - 1;}}
```

(c) Implementation of the stabilizer used by *Homeostasis*.

Fig. 7. Constructor and stabilizer methods of BasePA, and a template getter of a program-abstraction.

also include all the jump edges (due to jump statements, such as continue, break, etc.) that will be removed/added from the IR as a result of this elementary transformation. Further, these sets also contain inter-task edges[2] that are used as a part of the IR to model the parallel semantics of OpenMP C.

*3.2.2 Structure of* BasePA *and its subclasses.* *Homeostasis* maintains a global set `allAnalyses` of program analysis instances, by adding the reference of each program analysis instance (at the time of its construction) to `allAnalyses`; see Fig. 7a. In addition, the constructor (i) invokes the compute method, overridden by each program analysis class, to run the analysis from scratch and populate the corresponding program-abstraction, and (ii) sets the initial value for chIndex, as per the given mode of stabilization. In the case of LZINV mode of stabilization, chIndex is set to $-1$ as this field is not used in LZINV mode. In the case of LZUPD mode of stabilization, the value of chIndex is used to identify those elements in the global list chSequence whose impact needs to be taken into account in order to obtain the stable program-abstraction. These elements are the change elements that lie at indices *beyond* chIndex in chSequence.

The getter method of any program analysis follows the template shown in Fig. 7b, and invokes the stabilize method, if needed, to return the expected value. See a snippet of the method getCallGraph in Fig. 4 for an example.

*3.2.3 Structure of program-abstraction stabilizers.* As discussed in Section 3.1.4, *Homeostasis* allows each program-abstraction to be stabilized in one of the two supported modes of stabilization: LZINV

[2]IMOP uses the concept of inter-task edges between flush operations of a parallel region to statically model the communication that may happen using shared variables between different threads.

```
1  Function getNetChanges(chIndex) // Obtain the net IR changes to be handled for a given chIndex
2      netChanges = <>;
3      foreach ind from (chIndex + 1) to (BasePA.chSequence.size() − 1) do
4          netChanges = mergeChanges(netChanges, BasePA.chSequence.get(ind));
5      return netChanges;
6  Function mergeChanges(ch1, ch2) // Obtain the net IR changes for the given changes
7      mergedChanges = <>;
8      mergedChanges.addNodes = (ch1.addNodes\ch2.remNodes) ∪ (ch2.addNodes\ch1.remNodes);
9      mergedChanges.remNodes = (ch1.remNodes\ch2.addNodes) ∪ (ch2.remNodes\ch1.addNodes);
10     mergedChanges.addEdges = (ch1.addEdges\ch2.remEdges) ∪ (ch2.addEdges\ch1.remEdges);
11     mergedChanges.remEdges = (ch1.remEdges\ch2.addEdges) ∪ (ch2.remEdges\ch1.addEdges);
12     return mergedChanges;
```

Fig. 8. Algorithms to obtain the net IR changes for those elements in chSequence whose impact needs to be taken into account by the handleUpdate method to obtain the stable program-abstraction.

and LZUPD. Naturally, the procedure to stabilize a program-abstraction depends on the corresponding mode of stabilization. For the LZINV mode of stabilization, the stabilize method (Fig. 7c) simply reruns the analysis from scratch by invoking the compute method. Thus in the LZINV mode, the program analysis writer needs to write no additional code to stabilize the program-abstraction.

For the LZUPD mode of stabilization, the stabilize method performs self-stabilization of the abstraction by invoking the method handleUpdate. To utilize the arguably more efficient LZUPD mode, the program analysis writer needs to ensure that the desired incremental update algorithm is implemented in the handleUpdate method. This method should define the impact of the net IR changes, given in terms of the four sets: nodes added, nodes removed, edges added, and edges removed. The stabilize method invokes this handleUpdate method by providing the correct argument in the form of net IR changes (obtained as discussed next), followed by setting the chIndex to the index of the last element of chSequence.

*3.2.4 Obtaining Net IR changes.* In order to obtain the net IR changes to be taken into account by the handleUpdate method during a stabilization trigger, *Homeostasis* provides getNetChanges function, whose algorithm is shown in Fig. 8. The net IR change is obtained by merging the IR changes denoted by all the elements (if any) that lie to the right of the location indexed by chIndex in the global list chSequence. The exact merge operation is specified in the function mergeChanges in Fig. 8. We demonstrate the working of these functions using a concrete example.

*Example 3.2 (Net IR changes).* Fig. 9 illustrates a concrete example of how the net IR changes are obtained by *Homeostasis*, when required by the handleUpdate method during stabilization trigger of some analysis-instance. The value of chIndex for the analysis-instance is $m$, and the size of chSequence list is $(m + 4)$. Assume, the last three elementary transformations are:

(i) Addition of node n1, resulting in addition of edges {e1, e2}.
(ii) Replacement of node n3 with node n2, resulting in addition of edges {e3, e4}, and removal of edges {e5, e6}.
(iii) Replacement of the previously-added node n1 with node n4, resulting in addition of edges {e7, e8}, and removal of the previously-added edges {e1, e2}.

The figure shows the corresponding set of IR changes, whose impact needs to be taken into account during the next stabilization trigger. The net IR changes are obtained by merging all the three IR changes using the mergeChanges function from Fig. 8 (represented using $\oplus$ binary operator).

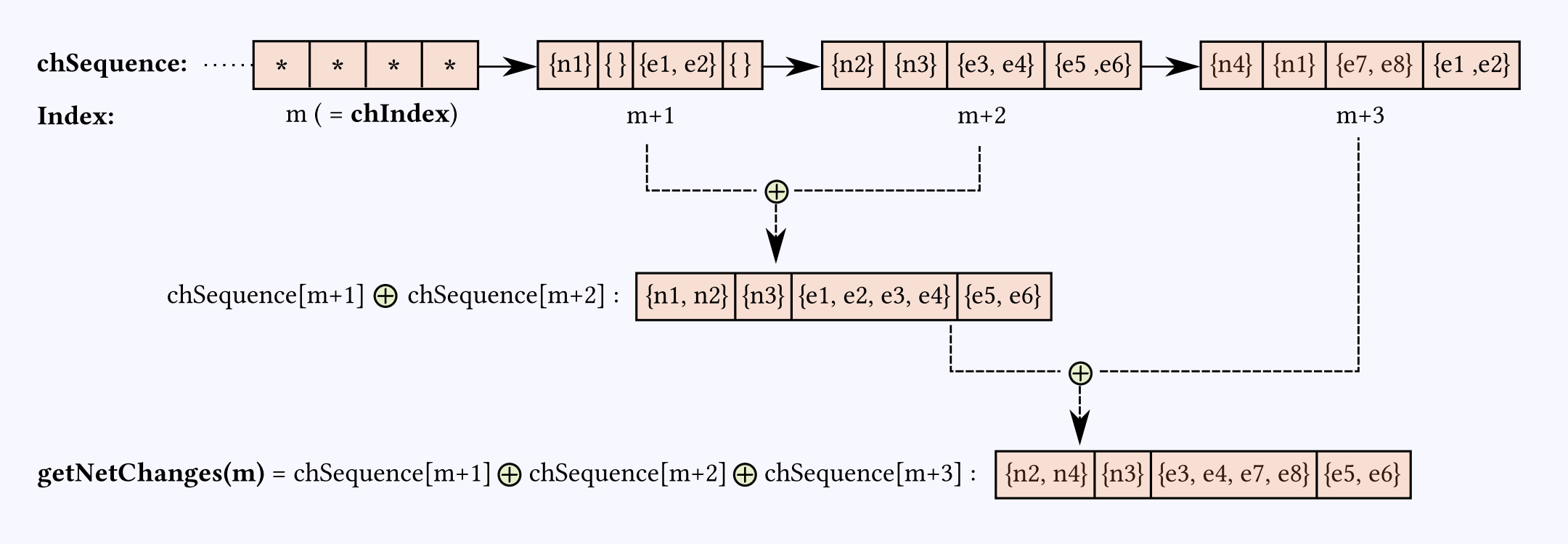

Fig. 9. A concrete example demonstrating how the net IR changes are obtained in the LZUPD mode of stabilization for some analysis-instance with value of chIndex as $m$. Size of the global list chSequence is $(m + 4)$. Each element of chSequence shows a four-tuple, corresponding to the following sets, in order: nodes added, nodes removed, edges added, and edges removed. The $\oplus$ operator denotes the application of mergeChanges function from Fig. 8.

Now, we demonstrate how self-stabilization achieved by *Homeostasis*, as discussed in this section, fixes the issues in compilation of the example shown in Fig. 1a, without requiring any changes to the code written by the optimization-pass writer.

*3.2.5 Example Compilation with* Homeostasis. Let us revisit the example of the dead-branch remover optimization pass given in Fig. 1a, implemented in a *Homeostasis*-enabled compiler. Here, the `replace` method would invoke a sequence of elementary transformations, directly or indirectly through other high-level transformation APIs. These elementary transformations will capture the information about the resulting program modifications (say, from $P_1$ to $P_2$), and notify it to `BasePA`, thereby making the information available to all program analysis passes, including the points-to analysis pass. Later, when an attempt is made to read the points-to map using a getter of the analysis pass, say from within the methods `isNeverZero` and `isAlwaysZero`, the getter would automatically trigger stabilization of the points-to map internally such that the map correctly corresponds to $P_2$. Hence, the compiler will be able to precisely deduce that the pointer `p` at `L2` will now point only to variable `a`; this would trigger the removal of the `false` branch of the `if-statement` at `L2`, thereby generating the expected program, $P_3$.

### 3.3 Compression Optimization: Efficient Tracking of Program Changes

The number of elementary transformations that a program undergoes during compilation can be substantial – compilation of large programs may easily involve several hundreds of elementary transformations. Hence, storing the individual IR-changes resulting from each elementary transformation separately in the global list chSequence can be prohibitively expensive in terms of space requirements. In order to address this challenge, *Homeostasis* employs a compression optimization that reduces the size of chSequence, relying on the following two key observations on the value of chIndex of analysis-instances that employ the LZUPD mode of stabilization.

**Observation 1.** *If the IR-changes corresponding to an elementary transformation have been taken into account by all the analysis-instances in the compiler, then the IR-changes need not be saved in* chSequence. In other words, if no analysis-instance has chIndex less than 0, then the first element (at index 0) in chSequence can be safely removed (while adjusting the chIndex of all the analysis-instances by decrementing 1 from their values).

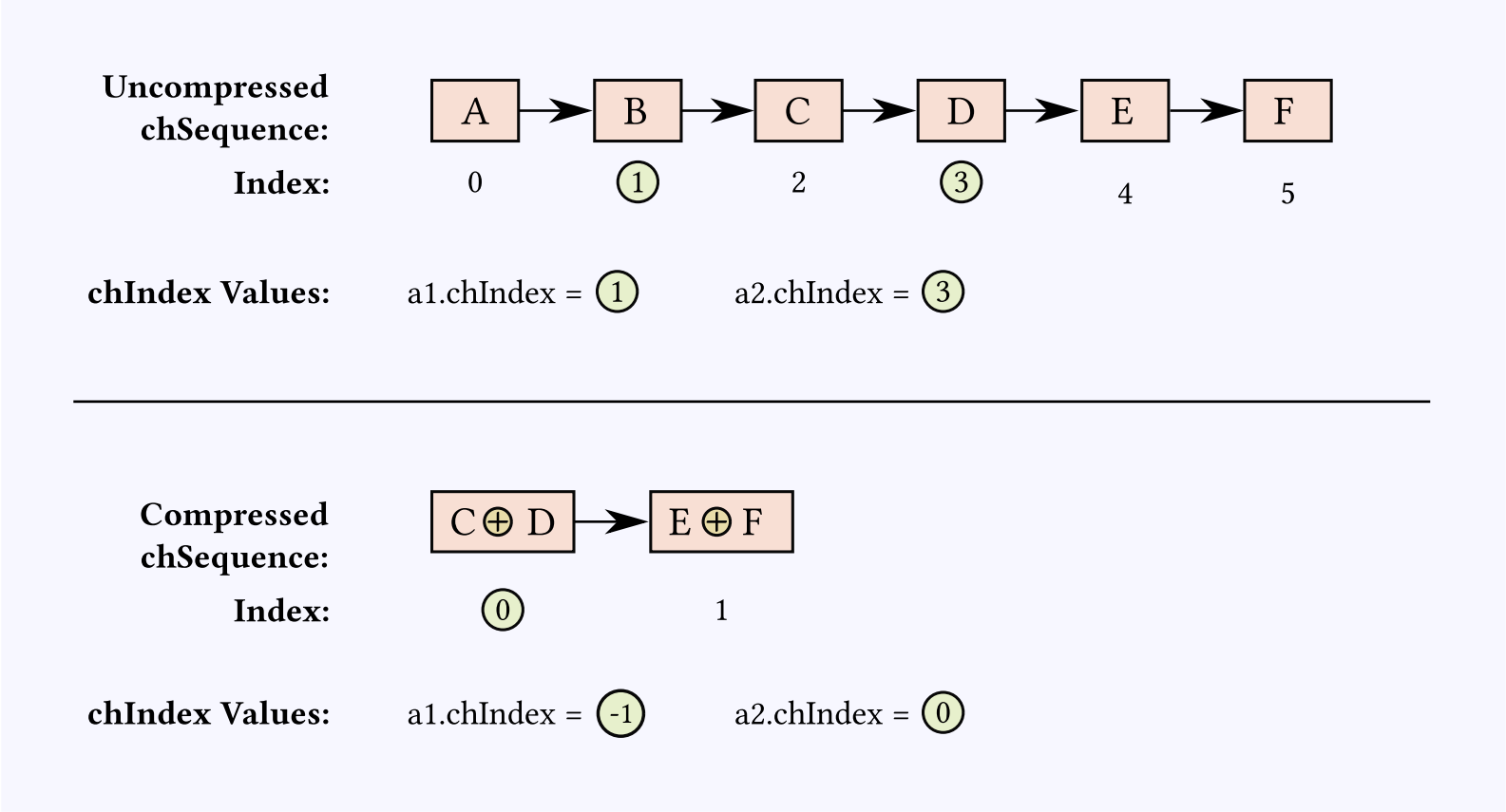

Fig. 10. Concrete example of compression optimization, demonstrating application of Observations 1 and 2. Those indices in chSequence, which are indexed to by chIndex of any analysis-instance in the compiler, are shown encircled.

*Correctness argument.* Recall that chSequence is a 0-indexed list. Further, as per the definition of getNetChanges from Fig. 8, the net IR changes to be used in the next stabilization trigger of an analysis-instance are obtained by merging the IR changes denoted by all the elements that lie beyond the index *chIndex* (of the analysis-instance) in chSequence. Hence, if there is no analysis-instance with chIndex less than 0, then the first element of chSequence (at index 0) will not be used to obtain the net IR changes during the stabilization trigger of any analysis-instance.

**Observation 2.** *If no analysis-instance has a* chIndex *that indexes into the location of an IR-change element in the* chSequence *list, then it is safe to merge the IR-change element to its successor, if any, in the list (while also adjusting the* chIndex *of various analysis-instances appropriately).*
*Correctness argument.* Consider an index $i$ in chSequence, such that no analysis-instance has its chIndex as $i$. In this case, those analysis-instances whose chIndex is greater than $i$, will not consider the element at index $i$ while obtaining their net IR changes during future stabilization triggers (see Fig. 8). Whereas, those analysis-instances whose chIndex is less than $i$, will anyway merge the elements at index $i$ and $(i + 1)$ (if the element exists) to obtain the net IR changes for their next stabilization trigger. Hence, merging the elements at $i$ and $(i + 1)$ in chSequence, will not affect the net IR changes for any analysis-instance during any of their future stabilization triggers.

The compression optimization employed by *Homeostasis* compresses the global chSequence list by repeated application of these two observations, until fixed-point. Note that *Homeostasis* does not dictate how often the compression optimization is invoked during the compilation. The exact details of the compression optimization, including details on how various corner cases are handled, are discussed formally in Section 4.2. In this section, we demonstrate the application of these two observations using a concrete example.

*Example 3.3 (Compression Optimization).* In Fig. 10, we demonstrate the working of compression optimization on a concrete example. Consider only two analysis-instances in the compiler, namely $a1$ and $a2$. Their chIndex values are 1 and 3, respectively. Upon repeated applications of Observation 1, we find that IR-changes A and B both can be safely removed from chSequence. Accordingly, both the chIndex values are decremented by 2. Further, the IR-changes C and D (and later E and F), can be merged, by repeated applications of Observation 2. Due to the merging of C and D, the chIndex of $a2$ is further decremented by 1, to obtain 0.

| Type of Developer | Additional Tasks | Efforts | Frequency |
|---|---|---|---|
| Compiler-base writers | - Define elementary transformations<br>- Modify high-level transformation APIs<br>- Create the super-class BasePA | Moderate | Once per compiler |
| Analysis-pass writers | - Follow the structure defined by BasePA | Minimal | Once per analysis |
| Optimization-pass writers | *None* | Nil | – |

Fig. 11. Summary of the impact of enabling *Homeostasis*, on different types of compiler writers.

**Impact of compression optimization on space complexity of tracking IR-changes.** In the absence of compression, the list *chSequence* would contain as many elements as the number of elementary transformations employed (say $k$). The size of each element can be $O(s)$, in terms of the size $s$ of the input program (calculated as some function of the number of nodes and edges in the IR). Hence, the total space complexity of an uncompressed *chSequence* can be $O(k \times s)$. In contrast, from Observations 1 and 2, we note that the number of elements in the fixed-point compressed state of chSequence cannot be greater than $\mathtt{min}(k, p)$, where $p$ is the number of analysis-instances in the compiler. Since $p$ is usually significantly smaller than $k$, the worst-case space complexity of chSequence with compression optimization, drops to $O(p \times s)$.

**Time complexity of the compression algorithm.** When the compression algorithm is invoked, the fixed-point state obtained upon repeated applications of Observations 1 and 2 may compress chSequence to a single-element, in the worst-case. Hence, if the total number of elementary transformations performed is $k$, then the compression algorithm may perform at most $k$ merge operations (using function `mergeChanges`, Fig. 8). If the program size is given by $s$, then the worst-case time complexity of each merge operation will be $O(s)$. Hence, the worst-case time complexity for the compression algorithm is $O(k \times s)$.

### 3.4 Working in a Homeostasis-enabled compiler

We now describe the additional effort that is needed to enable and use *Homeostasis*, and the impact of using *Homeostasis* on the key questions from Fig. 2. For this discussion, let us assume a hypothetical compiler $\mathbb{C}$ and its *Homeostasis*-enabled counterpart $\mathbb{C}_{HS}$. We present this discussion with respect to three key kinds of compiler developers: (i) *Compiler-base writers* obtain $\mathbb{C}_{HS}$ from $\mathbb{C}$ by implementing those core components of *Homeostasis* which are common across all the analysis/optimization passes. (ii) *Analysis-pass writers* write analysis passes in $\mathbb{C}_{HS}$. (iii) *Optimization-pass writers* write optimization passes in $\mathbb{C}_{HS}$. Fig. 11 summarizes the impact of enabling and using *Homeostasis*, on these key developers.

**Impact on compiler-base writer.** To obtain $\mathbb{C}_{HS}$ from $\mathbb{C}$, the compiler-base writer will have to additionally perform the following tasks: (i) identify the elementary transformations and develop them in accordance with the design discussed in Section 3.2.1, (ii) ensure that all high-level transformation APIs in the compiler internally rely on the identified set of elementary transformations to perform any program modifications, and (iii) create a base class for all analysis passes, equivalent to `BasePA` from Sections 3.2.2 and 3.2.3. Note that these tasks are one-time efforts and involve modifying only some of the IR-related classes.

**Impact on analysis-pass writer.** Given any analysis class AC in $\mathbb{C}$, the analysis writer would have to additionally ensure the following, for adapting AC to $\mathbb{C}_{HS}$: (i) AC is a subtype of `BasePA`, (ii) the code of AC needed to generate the corresponding program-abstraction is written in the `compute` method, (iii) the corresponding program-abstraction for AC can be accessed only through a getter as defined in Section 3.2.2, and (iv) optionally, if the analysis writer wants to support the `LZUPD` mode of stabilization, the method handleUpdate has to be defined as per Section 3.2.3.

In terms of additional coding effort, ensuring point (i) is trivial; so is ensuring points (ii) and (iv), since the relevant code would be common to both $\mathbb{C}$ and $\mathbb{C}_{HS}$, and would only possibly differ in their method names. We argue that point (iii) would be automatically ensured in an object-oriented compiler, and hence requires no additional effort. Consequently, we claim that the amount of additional coding effort required to write an analysis in $\mathbb{C}_{HS}$ is minimal. In terms of the key questions shown in Fig. 2, in a *Homeostasis*-enabled compiler, an analysis-pass writer does not need to address questions $\mathbf{Q}^{\mathrm{T}}_{1-3}$, while adding any new program analysis in the `LZINV` mode of stabilization; for the `LZUPD` mode of stabilization, the analysis writer needs to address *only* $\mathbf{Q}^{\mathrm{T}}_{3}$, by providing the definition of the method `handleUpdate`. Note that this is not an additional overhead when using *Homeostasis* – such a code is needed even in the case of manual stabilization in incremental update modes.

**Impact on optimization-pass writer.** An optimization-pass writer need not write any additional code to adapt an optimization in $\mathbb{C}$ to $\mathbb{C}_{HS}$. In terms of the key questions shown in Fig. 2, in a *Homeostasis*-enabled compiler, an optimization pass writer need not address the key questions $\mathbf{Q}_{1-3}$, while adding any new optimization in a *Homeostasis*-enabled compiler. For instance, in Fig. 1, a *Homeostasis*-enabled compiler can generate the expected program $P_3$ without requiring any changes to the code of the optimization (DeadBranchRemover) shown in Fig. 1a. Note that the compiler-base writer would have updated the code for high-level transformation APIs (including the `replace` method), such that all program-changes are expressed directly or indirectly using elementary transformations; a one-time effort.

***Overall Summary:*** Once a *Homeostasis*-enabled compiler has been designed (requires one-time moderate effort on the part of the compiler-base writers), designing and implementing program analyses and optimizations require minimal/zero additional coding effort to maintain stable program-abstractions.

## 4 FORMAL DESCRIPTION AND CORRECTNESS OF *HOMEOSTASIS*

We now present a formal description of various components of *Homeostasis*. In Section 4.1, we formally define an abstract *Homeostasis*-enabled compiler, and argue the *correctness guarantee* provided by *Homeostasis*: in a *Homeostasis*-enabled compiler, any compiler pass accessing the program-abstraction of an analysis will only read the stable value of that program-abstraction (see Theorem 4.12). Later, in Section 4.2, we first formally describe how *Homeostasis* maintains the compressed list of program-changes, and then we prove that the compression is semantics-preserving.

### 4.1 An Abstract *Homeostasis*-enabled Compiler

Consider a *Homeostasis*-enabled compiler for an IR language $\mathbb{L} = \{P_1, P_2, \ldots\}$, where each $P_i$ is a valid (syntactically and semantically correct) IR program. Without loss of generality, we assume each program IR to be in the form of a graph. Let the fixed set of elementary transformations in the compiler be represented as $\Gamma = \{\tau_1, \tau_2, \ldots, \tau_m\}$, where each transformation $\tau_i$ is a map $\mathbb{L} \longrightarrow \mathbb{L}$. Note that each elementary transformation updates an IR program by adding/removing a set of nodes and/or edges in the graph to obtain the modified IR program. Fig. 12 summarizes the sets and maps used in the formalism.

*Definition 4.1 (Program-change).* Given two programs, $P_i$ and $P_j \in \mathbb{L}$, the *program-change* between $P_i$ and $P_j$ (denoted by $\Delta(P_i, P_j)$) is defined as a 4-tuple of the form $\langle \delta_{na}, \delta_{nr}, \delta_{ea}, \delta_{er} \rangle$, where to obtain $P_j$ from $P_i$, $\delta_{na}$ is the minimal set of program nodes to be added to $P_i$, $\delta_{nr}$ is the minimal set of nodes to be removed from $P_i$, $\delta_{ea}$ is the minimal set of edges to be added to $P_i$, and $\delta_{er}$ is the minimal set of edges to be removed from $P_i$. Specifically, for $P_i = (N_i, E_i)$ and $P_j = (N_j, E_j)$, the following holds: $\delta_{na} = N_j \setminus N_i$, $\delta_{nr} = N_i \setminus N_j$, $\delta_{ea} = E_j \setminus E_i$, and $\delta_{er} = E_i \setminus E_j$. We denote the set of all possible program-changes in $\mathbb{L}$ by $C$. Further, we term a sequence of program-changes, of any length, as a *change-sequence*. Without loss of generality, for the ease of exposition, we assume that all such sequences are 0-indexed. □

| Set | Description | Typical Element Symbol |
|---|---|---|
| $\mathcal{N}$ | set of all possible IR nodes | $n$ |
| $\mathcal{E}$ | set of all possible IR edges | $e$ |
| $\mathbb{L}$ | set of all valid *IR programs* $\subseteq \mathbb{P}(\mathcal{N}) \times \mathbb{P}(\mathcal{E})$ | $P_i$ |
| $\Gamma$ | set of all the *elementary transformations*, where each element is of the form $\mathbb{L} \to \mathbb{L}$ | $\tau_i$ |
| $C$ | set of all *program-changes* $\subseteq \mathbb{P}(\mathcal{N}) \times \mathbb{P}(\mathcal{N}) \times \mathbb{P}(\mathcal{E}) \times \mathbb{P}(\mathcal{E})$; if program $P_i = (N_i, E_i)$ and program $P_j = (N_j, E_j)$, then program-change $\Delta(P_i, P_j) = c = (N_j \setminus N_i, N_i \setminus N_j, E_j \setminus E_i, E_i \setminus E_j)$ | $c$ |
| $C^*$ | set of all *change-sequences* of elements in $C$, of any size | $\bar{c}$ |
| $\mathcal{A}$ | set of all *analysis-instances* in the compiler = $\{A_1, A_2, \ldots\}$, where each element is of the form $\mathbb{L} \to R_i$ | $A_i$ |
| $R_i$ | set of all *program-abstraction* values for analysis-instance $A_i$ | $v$ |
| $\mathcal{R}$ | set of all the program-abstraction values = $R_1 \cup R_2 \cup \ldots$ | – |
| StbSt | set of *stabilization-status* values = {Stable, Unstable} | $st$ |
| StbMd | set of lazy *modes of stabilization* = {LZINV, LZUPD} | $m$ |
| $D_i$ | set of all *analysis-states* for the analysis-instance $A_i$ = $R_i \times$ StbSt $\times$ StbMd $\times \mathbb{N}_{-1}$ | $d$ |
| $\mathcal{D}$ | set of all the analysis-states = $D_1 \cup D_2 \cup \ldots$ | – |
| $\mathbb{S}$ | set of *compiler states* = $\mathbb{L} \times \mathbb{P}(\mathcal{D}) \times C^*$ | $S$ |

Fig. 12. Sets and maps used in the formalism. $\mathbb{N}_{-1}$ is the set of natural numbers $\cup \{0, -1\}$. $\mathbb{P}(X)$ denotes the power set of $X$.

*Definition 4.2 (Merge operator).* If $\Delta(P_a, P_b) = c_p$ and $\Delta(P_b, P_c) = c_q$, then we use $c_p \oplus c_q$ to provide the *net* program-change between $P_a$ and $P_c$. The set $C$ is closed under the binary operator $\oplus$ ($C \times C \to C$). This operator corresponds to the function mergeChanges from Fig. 8.

In a compiler written in OO-style, we assume that each analysis is implemented as a class, extending BasePA. During the invocations of the compiler, one or more analysis classes get instantiated (one or more times). Let $\mathcal{A} = \{A_1, A_2, \ldots\}$ be the set of all analysis-instances (such as an instance of points-to analysis, an instance of alias analysis, and so on) in the compiler, where each analysis-instance is of the form $A_i : \mathbb{L} \to R_i$, where $R_i$ denotes the set of meaningful program-abstraction values for $A_i$. We term $A_i(P)$ as the *stable program-abstraction* of the analysis-instance $A_i$ for program $P$.

*Definition 4.3 (Analysis State).* While compiling a program $P$, at any point during the compilation process, for an analysis-instance $A_i$, we define its *analysis state* as a 4-tuple of the form $d_i = \langle v, st, m, r \rangle$, where $v \in R_i$ is the value currently maintained by the compiler for the program-abstraction corresponding to $A_i$, $st \in$ StbSt = {Stable, Unstable}, denotes the current stabilization status of $A_i$, $m \in$ StbMd = {LZINV, LZUPD}, determines the selected lazy mode of stabilization for $A_i$, and $r$ is either $-1$ (for LZINV mode of stabilization) or an integer that refers to an index in the global change-sequence (for LZUPD mode of stabilization); $r$ is termed as the *change-index* of $A_i$. For the LZUPD mode of stabilization, intuitively, the non-negative **change-index** of an analysis-instance $A_i$ denotes that position in the global change-sequence list until and including which the impact of all elementary transformations have been taken into account, to arrive at the program-abstraction value $v$ for analysis-instance $A_i$.

Let the set $\mathcal{R} = R_1 \cup R_2 \cup \ldots$ be the set of all possible values for all program-abstractions. Further, let $\mathcal{D}$ be the set of all analysis states, for all the analyses-instances. □

*Definition 4.4 (Compiler state).* Consider a compilation process, where $k$ elementary transformations have been performed on the input program $P$. We define a *compiler state* as a 3-tuple of the form $S_i = \langle P_i, D_i, \overline{c_i} \rangle$, where $P_i$ represents the current program (obtained upon applying the $k$ transformations on $P$), $D_i = \{d_1, d_2, \ldots\}$ represents the current set of *analysis states* for different analysis-instances (each a 4-tuple as defined above), and $\overline{c_i} = \langle c_0, c_1, c_2, \ldots c_{k-1} \rangle$, denotes the sequence of $k$ program-changes obtained upon application of each of the $k$ elementary transformations (in order) on the input program $P$. We denote the set of all compiler states by $\mathcal{S}$. Note that for an input program $P$, the *initial* compiler state when compilation starts is $\langle P, \emptyset, \emptyset \rangle$. □

Now, we state two axioms that are derived from the assumption that the analysis writer has provided the correct implementations of `compute` and `handleUpdate` methods; the latter is required only for the LZUPD mode of stabilization. Axiom 4.5 formally states that the exhaustive algorithm of an analysis-instance has been correctly implemented by the analysis writer in a method named `compute`. Axiom 4.6 formally states that in order to realize the incremental-update mode of stabilization, the analysis writer has provided a correct definition of the `handleUpdate` method, specifying how an abstraction would get updated in response to the given program-changes (between the current program and the one to which the stale program-abstraction corresponds). For the ease of presenting our formal argument, we assume that `compute` and `handleUpdate` methods also return the set of abstractions that were stabilized during the evaluation of these methods. This set is used to only facilitate the formal treatment of our proposed design, and is not required otherwise to perform the actual stabilization.

Axiom 4.5. *In a* Homeostasis-*enabled compiler, consider an analysis-instance $A_k$ with* LZINV *stabilization mode. The analysis writer of $A_k$ provides a correct implementation of the method* `compute`*, such that when* `compute` *is invoked at any compilation point,* `compute` *returns a two-tuple $(v, \mathcal{L})$, where $v = A_k(P_i)$, and $\mathcal{L} \subseteq \overline{\mathcal{A}}$, is the set of abstractions which were stabilized during the evaluation of the* `compute` *method.*

Axiom 4.6. *Consider an analysis $A_k$ with* LZUPD *mode of stabilization. Consider any compiler state $S_i = \langle P_i, D_i, \overline{c_i} \rangle$, where the corresponding analysis state for $A_k$ in $D_i$ is denoted by $d_k = \langle v_k, \mathrm{st}_k, \mathsf{LZUPD}, r_k \rangle$. Further, let, $\overline{c_i} = \langle c_0, c_1, \ldots, c_n \rangle$. In a* Homeostasis-*enabled compiler, the analysis writer of $A_k$ provides the correct implementation of method* `handleUpdate`*, such that if there exists a program $P_j \in \mathbb{L}$, such that $v_k = A_k(P_j)$ and $\Delta(P_j, P_i) = c_{r_k+1} \oplus c_{r_k+2} \ldots \oplus c_n$, then, when* `handleUpdate` *is invoked on the compiler state $S_i$, it returns a pair $(v, \mathcal{L})$, where (i) $v = A_k(P_i)$, and (ii) $\mathcal{L} \subseteq \overline{\mathcal{A}}$, the set of abstractions which were stabilized during the evaluation of the* `handleUpdate` *method.*

In Fig. 13, we provide the definition of *stabilization-function*, $\sigma : \mathcal{S} \times \mathcal{D} \longrightarrow \mathcal{D} \times \mathbb{P}(\mathcal{A})$, which, given a compiler state and a (possibly stale) analysis state for some analysis-instance (in that compiler state), utilizes the definitions of `compute` and `handleUpdate` methods, to provide a two-tuple: (i) the stabilized analysis state of the analysis-instance, and (ii) the set of other analysis-instances on which the stabilization-function was invoked internally during the evaluation of `compute` or

$$\frac{S = \langle P, D, \bar{c} \rangle \quad d_k \in D \quad d_k = \langle *, \textsc{Stable}, *, * \rangle}{\sigma(S, d_k) = \langle d_k, \emptyset \rangle} \text{ [Stb]}$$

$$\frac{\begin{array}{c} S = \langle P, D, \bar{c} \rangle \quad d_k \in D \quad d_k = \langle v, \textsc{Unstable}, \textsf{LZINV}, -1 \rangle \\ \textbf{compute}(A_k, P) = (v', \mathcal{L}) \quad d'_k = \langle v', \textsc{Stable}, \textsf{LZINV}, -1 \rangle \end{array}}{\sigma(S, d_k) = \langle d'_k, \mathcal{L} \rangle} \text{ [Stb-Inv]}$$

$$\frac{\begin{array}{c} S = \langle P, D, \bar{c} \rangle \quad \bar{c} = \langle c_0, c_1, \ldots, c_n \rangle \quad d_k \in D \quad d_k = \langle v, \textsc{Unstable}, \textsf{LZUPD}, r \rangle \quad r < n \\ \textbf{handleUpdate}(A_k, P, v, c_{r+1} \oplus c_{r+2} \oplus \ldots \oplus c_n) = (v', \mathcal{L}) \quad d'_k = \langle v', \textsc{Stable}, \textsf{LZUPD}, n \rangle \end{array}}{\sigma(S, d_k) = \langle d'_k, \mathcal{L} \rangle} \text{ [Stb-Upd]}$$

Fig. 13. Evaluation of the stabilization-function, $\sigma : \mathcal{S} \times \mathcal{D} \to \mathcal{D} \times \mathbb{P}(\mathcal{A})$, as derived from the template of stabilier from Fig. 7c.

`handleUpdate` method. Given such a two-tuple $\Theta$, we use $\Theta$.fst and $\Theta$.snd to refer to the first and second elements, respectively, of the tuple.

If the given analysis state is already Stable, the stabilization-function returns the analysis state as it is, along with an empty set of analysis-instances (rule [Stb]). Otherwise, when the stabilization mode of the analysis state is LZINV, the method `compute` is invoked in order to obtain the correct abstraction value ($v'$) by re-running the corresponding analysis from scratch on the current program (rule [Stb-Inv]). In contrast, as per rule [Stb-Upd], when the stabilization mode is LZUPD, the method `handleUpdate` is invoked to obtain the correct abstraction value ($v'$), given the following arguments: (i) a reference to the analysis-instance ($A_k$), (ii) the current program, (iii) the unstable/stale abstraction value ($v$), and (iv) the net program-changes obtained by using the merge operator on the appropriate elements from the change-sequence. Note that the change-index for the returned analysis state is set to $n$ (index of the last element in the global change-sequence).

*Definition 4.7 (Compiler action).* Recall that *Homeostasis* allows changes to the IR program only through elementary transformations. Similarly, the program-abstraction of an analysis can be accessed only through its appropriate getters. Accordingly, we define three kinds of *compiler actions* that can be issued by an optimization or analysis pass, and may alter the current compiler state, $S = \langle P, D, \bar{c} \rangle$.

(i) Given an analysis class AC, the compiler action **`construct`**$(\textsf{AC}, m)$ creates an instance of AC, invokes the `compute` method to populate the program-abstraction, and sets the stabilization mode of the constructed analysis-instance to $m \in$ StbMd.
(ii) The compiler action **`get`**$(A_k)$ returns the value $A_k(P)$.
(iii) The compiler action **`transform`**$(\mathcal{T}_i)$ transforms the current program $P$ to $\mathcal{T}_i(P)$.

We represent the set of all possible compiler actions with $\mathcal{W}$. □

Note that in object-oriented compilers, the `construct` compiler action for an analysis will always be invoked before the first invocation of the corresponding `get` compiler action.

*Definition 4.8 (`apply` function).* We define a compilation step using the function **`apply`**: $\mathcal{S} \times \mathcal{W} \Longrightarrow \mathcal{S} \times (\mathcal{R} \cup \{\epsilon\})$, which applies a given compiler action $w$ on the compiler state $s$, to generate (i) a possibly-modified compiler state $s'$, and (ii) a program-abstraction value or $\epsilon$.

For a *Homeostasis*-enabled compiler, the definition for function `apply` is shown in Fig. 14a, Fig. 14b, and Fig. 14c, for the design templates of various components shown in Fig. 7a (constructor), Fig. 7b-7c (getters and stabilizers), and Fig. 5 (elementary transformations), respectively.

$$\frac{\begin{array}{c} S = \langle P, D, \bar{c} \rangle \quad A_i = \text{new AC}() \\ \texttt{compute}(A_i, P) = (v, \mathcal{L}) \quad \mathcal{L} = \{A_1, A_2, \ldots, A_n\} \quad A_i \notin \mathcal{L} \\ D_a = D - \{d_1, d_2, \ldots, d_n\} \quad D_b = \{\sigma(S, d_1).\text{fst}, \sigma(S, d_2).\text{fst}, \ldots, \sigma(S, d_n).\text{fst}\} \\ d_i = \langle v, \textsc{Stable}, m, ((m = \texttt{LZUPD})?(|\bar{c}| - 1) : -1) \rangle \end{array}}{\texttt{apply}(S, \texttt{construct}(\text{AC}, m)) \Rightarrow (\langle P, D_a \cup D_b \cup \{d_i\}, \bar{c} \rangle, \epsilon)} \text{[C-Apply]}$$

(a) Evaluation of apply on the "construct" compiler action, as derived from Fig. 7a. The helper method newInstanceOf gives a fresh uninitialized analysis-instance for the given analysis.

$$\frac{\begin{array}{c} S = \langle P, D, \bar{c} \rangle \quad d \in D \\ \sigma(S, d_k) = (d'_k, \mathcal{L}) \quad d'_k = \{v', \textsc{Stable}, *, *\} \quad \mathcal{L} = \{A_1, A_2, \ldots, A_n\} \quad A_k \notin \mathcal{L} \\ D_a = D - \{d_1, d_2, \ldots, d_n\} - \{d_x\} \quad D_b = \{\sigma(S, d_1).\text{fst}, \sigma(S, d_2).\text{fst}, \ldots, \sigma(S, d_n).\text{fst}\} \end{array}}{\texttt{apply}(S, \texttt{get}(A_k)) \Rightarrow (\langle P, D_a \cup D_b \cup \{d'_k\}, \bar{c} \rangle, v')} \text{[G-Apply]}$$

(b) Evaluation of apply on the "get" compiler action, as derived from the template of getters from Fig. 7b.

$$\frac{\begin{array}{c} S = \langle P, D, \bar{c} \rangle \quad P' = \tau_l(P) \quad \overline{c'} = \bar{c} \cdot \langle \Delta(P, P') \rangle \\ D = \{\langle v_1, s_1, m_1, r_1 \rangle, \langle v_2, s_2, m_2, r_2 \rangle, \ldots, \langle v_x, s_x, m_x, r_x \rangle\} \\ D' = \{\langle v_1, \textsc{Unstable}, m_1, r_1 \rangle, \langle v_2, \textsc{Unstable}, m_2, r_2 \rangle, \ldots, \langle v_x, \textsc{Unstable}, m_x, r_x \rangle\} \end{array}}{\texttt{apply}(S, \texttt{transform}(\tau_l)) \Rightarrow (\langle P', D', \overline{c'} \rangle, \epsilon)} \text{[T-Apply]}$$

(c) Evaluation of apply on the "transform" compiler action, as derived from the template of elementary transformations given in Fig. 5.

Fig. 14. Definition for the apply function, derived from the design of *Homeostasis* explained in Section 3.2.

- When a `construct` compiler action is issued for an analysis-instance $A$, an initial analysis state for $A$ is generated and added to the set of analysis states in the current compiler state (rule [C-Apply]). Note that in the case of LZUPD mode, the change-index of $A$ is set to the index of the last element in the global change-sequence; it is set to $-1$ if the list is empty.
- When a `get` compiler action is issued for $A_k$, the design of *Homeostasis* ensures that the analysis state of $A_k$ is changed as per the mapping rules shown in Fig. 14b and 13. Using Axioms 4.5 and 4.6, rules [Stb-Inv] and [Stb-Upd] set the stabilization status of $A_k$ to Stable, as can be inferred from the design of an abstraction getter, and stabilizer.
- When a `transform` compiler action with elementary transformation $\tau_l$ is issued on a compiler state $S = \langle P, D, \bar{c} \rangle$, then as per rule [T-Apply] in Fig. 5, the design of *Homeostasis* changes the compiler state to a tuple where, (i) the current program $P$ is modified to $\tau_l(P)$, (ii) the stabilization status of all analyses is set to Unstable, and (iii) $\Delta(P, \tau_l(P))$ is added to the sequence of program-changes.

*Definition 4.9 (Valid compiler state).* We term a compiler state of the form $S_i = \langle P_i, D_i = \{d_1, d_2, \ldots\}, \overline{c_i} = \langle c_0, c_1, \ldots, c_n \rangle \rangle$ as a *valid compiler state*, if and only if the following condition, termed as **stability-condition**, holds for the state $d_j = \langle v_j, st_j, m_j, r_j \rangle \in D_i$ of each analysis-instance $A_j$:

(i) if $st_j$ = STABLE, then $r_j = n$ and $v_j = A_j(P_i)$, else
(ii) if $st_j$ = UNSTABLE, then $n \geq 0$, $r_j < n$, and $\exists P_o \in \mathbb{L}$, such that $v_j = A_j(P_o)$ and $\Delta(P_o, P_i) = c_{r_j+1} \oplus c_{r_j+2} \ldots \oplus c_n$.

Intuitively, a compiler state is considered *valid* when each program-abstraction is either stable, or has access to the complete information that can be used to make it stable. □

The following Lemma gives the correctness guarantee for the value returned by a `get` compiler action.

LEMMA 4.10. *If a compiler state $S_i = \langle P_i, D_i, \overline{c_i} \rangle$ is valid, then for each analysis-instance $A_k$ with corresponding analysis state $d_k \in D_i$, the value obtained upon invocation of* `get` *compiler action for $A_k$ (i.e.,* `get`$(A_k)$*) on $S_i$, will always be same as $A_k(P_i)$ (i.e., the stable program-abstraction value). Formally, for each analysis $A_k$ and valid compiler state $S_i$, $\exists S_j \in \mathbb{S}$ such that* `apply`$(S_i,$ `get`$(A_k)) \Longrightarrow (S_j, A_k(P_i))$.

PROOF. Let $\overline{c_i} = \langle c_0, c_1, \ldots, c_U \rangle$, and let $D_i = \{d_1, d_2, \ldots, d_k, \ldots\}$, where $d_k = \langle v_k, st_k, m_k, r_k \rangle$ be the analysis state of $A_k$ in $S_i$. Assume, `apply`$(S_i,$ `get`$(A_k)) \Longrightarrow (S_j, v'_k)$. Hence, it is sufficient to prove that $v'_k = A_k(P_i)$. As the compiler state $S_i$ is valid, the analysis state $d_k$ satisfies the stability-condition from Definition 4.9. Two cases arise based on the value of $st_k$:

**CASE A:** $st_k$ = STABLE. From the stability-condition of $d_k$, we obtain $v_k = A_k(P_i)$. Hence, from rule [STB], we derive $v'_k = v_k = A_k(P_i)$. Hence proved.

**CASE B:** $st_k$ = UNSTABLE. Two cases arise based on the value of $m_k$:

`Case B.1`: $m_k$ = `LZINV`. From rule [STB-INV], we obtain $(v'_k, \mathcal{L})$ = `compute`$(A_k, P_i)$. From Axiom 4.5, $v'_k$ = `compute`$(A_k, P_i)$.fst $= A_k(P_i)$. Hence proved.

`Case B.2`: $m_k$ = `LZUPD`. From rule [STB-UPD], we have $(v'_k, \mathcal{L})$ = `handleUpdate`$(A_k, P_i, v_k, c_{r_k+1} \oplus c_{r_k+2} \ldots \oplus c_U)$. From the stability-condition of $d_k$, we derive the following: $U \geq 0$, $r_k < U$, and that there exists a program $P_j$ such that $v_k = A_k(P_j)$ and $\Delta(P_j, P_i) = c_{r_k+1} \oplus c_{r_k+2} \ldots \oplus c_U$. Using Axiom 4.6, we obtain $v'_k = A_k(P_i)$. Hence proved.

□

The following Lemma gives the correctness guarantee that the invocation of any compiler action (`construct`, `get`, or `transform`) on a valid compiler state will result in a valid compiler state.

LEMMA 4.11. *For every sequence $W = \langle w_1, w_2, \ldots \rangle$ of compiler actions, the compiler state, starting with the initial compiler state $\langle P, \phi, \phi \rangle$, remains valid after application of each action from the sequence.*

PROOF. *(Proof by natural induction on the number of elements in $W$.)*

**Base Case:** $|W| = 0$. For an input program $P$, the initial compiler state is always $\langle P, \emptyset, \emptyset \rangle$, which is trivially valid.

**Inductive hypothesis:** The compiler state, say $S_i = \langle P_i, D_i, \overline{c_i} \rangle$, obtained after application of all compiler actions from the string $W_m$, of length $m$, is valid. Let $D_i = \{d_1, d_2, \ldots, d_k = \langle v_k, st_k, m_k, r_i \rangle, \ldots\}$, and $\overline{c_i} = \langle c_0, c_1, \ldots, c_U \rangle\rangle$.

**Inductive step:** Consider a string of compiler actions, $W_{m+1}$, of length $m$+1, obtained by appending a compiler action $w_{m+1}$ to $W_m$. We now need to prove that if `apply`$(S, w_{m+1}) = (S', v)$, then $S'$ is a valid compiler state. Three cases occur for possible form of compiler action $w_{m+1}$:

**CASE A:** $w_{m+1}$ is `construct`$(\text{AC}, m_s)$ for some analysis class AC. Say the created analysis-instance is $A_s$ and the initial analysis state for $A_s$ is $d_s = \langle v_s, st_s, m_s, r_s \rangle$. As per rule [C-APPLY], $S' = \langle P_i, D_i \cup \{d_s = \langle A_s(P_i), \text{STABLE}, m_s, U \rangle\}, \overline{c_i} \rangle$. $S'$ is trivially valid as $d_s$ satisfies the stability-condition from Definition 4.9. Hence proved.

**Case B:** $w_{m+1}$ is `get`$(A_x)$ for some $d_x \in D_i$. As $S$ is a valid compiler state, $\forall d_j \in D_i$ satisfies the stability-condition. Let `get`$(A_x)$ change a subset of analysis states $D_y \subseteq D_i$. Let $d_y (\in D_y)$ be given by $\langle v_y, st_y, m_y, r_y \rangle$. Assume, for each $d_y$, the changed analysis state $d'_y$ is given by $\langle v'_x, st'_x, m_x, r'_x \rangle$. To prove that $S'$ is a valid compiler state, it is sufficient to prove that each such $d'_y$ satisfies the stability-condition. Let us consider any such $d'_y$. Two sub-cases arise based on the values of $st_y$:

`Case B.1`: $st_y$ = Stable. As $d_y$ satisfies the stability-condition, we derive the following from rule [Stb]: $st'_y = st_y$ = Stable, $r'_y = r_y = U$, and $v'_y = v_y = A_x(P_i)$. Hence, $d'_y$ satisfies the stability-condition. Hence proved.

`Case B.2`: $st_y$ = Unstable. From rules [Stb-Inv] and [Stb-Upd], we derive $st'_y$ = Stable. Similarly, we derive $r'_y = -1$ or $r'_y = U$, from rules [Stb-Inv] and [Stb-Upd], respectively. Depending upon value of $m_y$, two sub-cases arise:

*Case B.2.i:* $m_y$ = LZINV. From rule [Stb-Inv], $v'_y$ = `compute`$(A_y, P_i)$.fst. From Axiom 4.5, $v'_y = A_y(P_i)$. Hence, $d'_y$ satisfies the stability-condition, and $S'$ is a valid compiler state.

*Case B.2.ii:* $m_y$ = LZUPD. From rule [Stb-Upd], $v'_y$ = `handleUpdate`$(A_y, P_i, v_y, c_{r_y+1} \oplus c_{r_y+2} \ldots \oplus c_U)$.fst. From Axiom 4.6, $v'_y = A_y(P_i)$, if there exists a program $P_k \in \mathbb{L}$, such that (i) $v_y = A_y(P_k)$, and (ii) the net program-change (i.e., last argument of `handleUpdate`) is same as $\Delta(P_k, P_i)$. Since $d_y$ satisfies the stability-condition, such a program $P_k$ exists. (Note that the $\Delta(P_k, P_i)$ is the same as the last argument of `handleUpdate`.) Hence, $v'_y = A_y(P_i)$, and $S'$ is a valid compiler state.

**Case C:** $w_{m+1}$ is `transform`$(\mathcal{T}_t)$ for some $\mathcal{T}_t \in \Gamma$. Let $S' = \langle P'_i, D'_i, \overline{c'_i} \rangle$. From rule [T-Apply], we have: $P'_i = \mathcal{T}_t(P_i)$, and $\overline{c'_i} = \langle c_0, c_1, \ldots, c_U, \Delta(P_i, \mathcal{T}_t(P_i)) \rangle \rangle$. Further, for each $d \in D_i$, there exists a modified analysis state $d' \in D'_i$. To prove that $S'$ is a valid compiler state, we need to prove that each analysis state $d' \in D'_i$ satisfies the stability-condition. Consider any analysis-state $d_x \in D$, and its corresponding modified analysis-state $d'_x \in D'_i$. Let $d_x = \langle v_x, st_x, m_x, r_x \rangle$, and let $d'_x = \langle v'_x, st'_x, m'_x, r'_x \rangle$. Now, two sub-cases arise based on the value of $st_x$:

`Case C.1`: $st_x$ = Stable. Since $d_x$ satisfies the stability-condition, we have: $r_x = U$ (if $m_x$ = LZUPD) or $r_x = -1$ (if $m_x$ = LZINV), and $v_x = A_x(P_i)$. From rule T-Apply, we have: $v'_x = v_x$, $m'_x = m_x$, $r'_x = r_x$, and $st'_x$ = Unstable. From Definition 4.9, $d'_x$ satisfies the stability-condition, and $S'$ is a valid program state.

`Case C.2`: $st_x$ = Unstable. Since $d_x$ satisfies the stability-condition, we know that: $U \geq 0, r_x < U$, and that a program, say $P_j$, exists such that $v_x = A_x(P_j)$ and $\Delta(P_j, P_i) = c_{r_x+1} \oplus c_{r_x+2} \ldots \oplus c_U$. From rule T-Apply, we have: $v'_x = v_x$, $m_x = m'_x$, $r'_x = r_x$, and $st'_x$ = Unstable. Hence, $r'_x < U + 1$. Further from the definition of $\oplus$, $\Delta(P_j, P'_i) = \Delta(P_j, P_i) \oplus \Delta(P_i, P'_i)$. Hence, $\Delta(P_j, P'_i) = \Delta(P_j, P_i) \oplus c_{U+1} = c_{r_x+1} \oplus c_{r_x+2} \ldots \oplus c_U \oplus c_{U+1}$. By Definition 4.9, $S'$ is a valid compiler state. □

Now, we first state the correctness theorem for *Homeostasis*. Intuitively, Theorem 4.12 states that at any point of time during compilation, the analysis results given by *Homeostasis* match those obtained by rerunning the complete analyses, provided the methods `compute` and `handleUpdate` provided by the analysis writer are correct (i.e., Axioms 4.5 and 4.6 hold).

Theorem 4.12 (Correctness of *Homeostasis*). *At any compilation point of a* Homeostasis-*enabled compiler with compiler state* $S_i = \langle P_i, D_i, \overline{c_i} \rangle$, *for each analysis* $A_k$ *with corresponding analysis state* $d_k \in D_i$, *the value obtained upon invocation of* `get` *compiler action for* $A_k$ *(i.e.,* `get`$(A_k)$*) on* $S_i$, *will always be same as* $A_k(P_i)$. *Formally, for each analysis* $A_k$ *and compiler state* $S_i$, $\exists\, S_j \in \mathbb{S}$, *such that* `apply`$(S_i,$ `get`$(A_k)) \Longrightarrow (S_j, A_k(P_i))$.

Proof. *Follows directly from Lemma 4.10 and Lemma 4.11.* □

$$\frac{|\overline{c}| = 0}{\texttt{compress}((D,\overline{c})) = (D,\overline{c})}\ \text{[CS-BASE]} \qquad \frac{|\overline{c}| = k \quad k \geq 1 \quad \overset{k-2}{\underset{i=-1}{\forall}}, \exists d \in D, d = \langle *,*,*,i\rangle}{\texttt{compress}((D,\overline{c})) = (D,\overline{c})}\ \text{[CS-COMPRESSED]}$$

$$\frac{\begin{array}{c} D = D_{\mathsf{INV}} \bigcup D_{\mathsf{UPD}} \quad \forall d \in D_{\mathsf{INV}}, d = \langle *,*,\mathbf{LZINV},-1\rangle \quad \forall d \in D_{\mathsf{UPD}}, d = \langle *,*,\mathbf{LZUPD},*\rangle \\ \overline{c} = c_1 \cdot \overline{c'} \quad \nexists d \in D_{\mathsf{UPD}}, d = \langle *,*,\mathbf{LZUPD},-1\rangle \\ D_{\mathsf{UPD}} = \{\langle v_1,s_1,\mathbf{LZUPD},r_1\rangle, \langle v_2,s_2,\mathbf{LZUPD},r_2\rangle, \ldots, \langle v_k,s_k,\mathbf{LZUPD},r_k\rangle\} \\ D'_{\mathsf{UPD}} = \{\langle v_1,s_1,\mathbf{LZUPD},r_1{-}1\rangle, \langle v_2,s_2,\mathbf{LZUPD},r_2{-}1\rangle, \ldots, \langle v_k,s_k,\mathbf{LZUPD},r_k{-}1\rangle\} \end{array}}{\texttt{compress}((D,\overline{c})) = (D_{\mathsf{INV}} \bigcup D'_{\mathsf{UPD}}, \overline{c'})}\ \text{[CS-DELFIRST]}$$

$$\frac{\begin{array}{c} D = D_{\mathsf{INV}} \bigcup D_{\mathsf{UPD}} \quad \forall d \in D_{\mathsf{INV}}, d = \langle *,*,\mathbf{LZINV},-1\rangle \quad \forall d \in D_{\mathsf{UPD}}, d = \langle *,*,\mathbf{LZUPD},*\rangle \\ \overline{c} = \overline{c_a} \cdot c_i \cdot \overline{c_b} \quad |\overline{c_a}| = i \quad i \geq 1 \quad |\overline{c_b}| \geq 1 \\ \overset{i-1}{\underset{j=-1}{\forall}}, \exists d \in D_{\mathsf{UPD}}, d = \langle *,*,\mathbf{LZUPD},j\rangle \quad \nexists d \in D_{\mathsf{UPD}}, d = \langle *,*,\mathbf{LZUPD},i\rangle \\ \overline{c_b} = \langle c_{i+1}, c_{i+2}, \ldots, c_n\rangle \quad \overline{c'_b} = \langle c_i \oplus c_{i+1}, c_{i+2}, \ldots, c_n\rangle \quad \overline{c'} = \overline{c_a} \cdot \overline{c'_b} \\ D_{\mathsf{UPD}} = \{\langle v_1,s_1,\mathbf{LZUPD},r_1\rangle, \langle v_2,s_2,\mathbf{LZUPD},r_2\rangle, \ldots, \langle v_k,s_k,\mathbf{LZUPD},r_k\rangle\} \\ D'_{\mathsf{UPD}} = \{\langle v_1,s_1,\mathbf{LZUPD},((r_1{>}i)?(r_1{-}1){:}r_1)\rangle, \langle v_2,s_2,\mathbf{LZUPD},((r_2{>}i)?(r_2{-}1){:}r_2)\rangle, \\ \ldots, \langle v_k,s_k,\mathbf{LZUPD},((r_k{>}i)?(r_k{-}1){:}r_k)\rangle\} \end{array}}{\texttt{compress}((D,\overline{c})) = (D_{\mathsf{INV}} \bigcup D'_{\mathsf{UPD}}, \overline{c'})}\ \text{[CS-MERGEMID]}$$

Fig. 15. Definition for the function **compress**: $\mathbb{P}(\mathcal{D}) \times C^* \rightarrow \mathbb{P}(\mathcal{D}) \times C^*$.

## 4.2 Compression of Change-Sequence in *Homeostasis*

*Homeostasis* keeps the global change-sequence compact by internally performing the compression operation on it, as discussed in Section 3.3. In this section, we formalize this functionality, and argue that the correctness of *Homeostasis* continues to hold when we use the proposed idea of change-sequence compression.

Fig. 15 presents the formal definition of the compress function. This function takes a two-tuple of the form $(D, \overline{c})$, where $D$ is a set of analysis-states, and $\overline{c}$ is a change-sequence. The compress function returns another two-tuple, say $(D', \overline{c'})$, such that (i) the change-sequence $\overline{c'}$ is obtained by optionally compressing a single element of the input change-sequence $\overline{c}$, and (ii) the set $D'$ contains all the analysis-states from $D$, with required modifications in case if the input change-sequence was compressed.

Recall that the key intuition behind compression is as follows: If no analysis-state in the compiler has a change-index equal to some index in the change-sequence, then the element at that index in the change-sequence can be safely merged (using the ⊕ operator) to its successor (if any) in the change-sequence. Further, if no analysis-state in the compiler has a change-index less than zero, then the first element (at index 0) of change-sequence can be safely removed. Following are the cases that may arise when performing compression of the input change-sequence one element at a time:

- When the change-sequence $\overline{c}$ is empty, it is already in its compressed form. Hence, as per rule [CS-BASE], $D$ and $\overline{c}$ are returned without any changes.
- Similarly, if corresponding to each element in the change-sequence (except, optionally, the last) if there exists at least one analysis-state that has its change-index same as the index of that element, and if there exists at least one analysis-state that has its change-index as −1, then the change-sequence is already in its compressed form ([CS-Compressed]).

• Otherwise, if no analysis-state has a change-index of −1, then we purge the first element from the change-sequence, and update all the analysis-states by decrementing their change-index by one (rule [CS-DelFirst]).

• Otherwise, the element at the least index number that is not present as the change-index of any analysis-state is merged to its successor element (if any) in the change-sequence. Further, the change-index of all those analysis-states whose change-index is greater than the compressed index number, is decremented by one (rule [CS-MergeMid]).

Note that no action is performed during compression when only the last element in the change-index is not indexed to by any of the analysis-states; this case falls under the rule [CS-Compressed], once the fixed-point application of the `compress` is complete. Further note that for efficiency purposes, in practice, in a change-sequence, all those $k$ consecutive elements that are not indexed by any change-index are compressed together and change-indices of the affected analysis-instances are decremented by $k$, in one go.

In *Homeostasis*, the `compress` function can be called during any compilation point, such as, immediately after a stabilization trigger, after an elementary transformation, and so on. Regardless of the number and places of invocations for this function, the following lemma holds.

Lemma 4.13. *Consider a valid compiler state, say $S = \langle P, D, \overline{c}\rangle$, and another state obtained after a single application of the* `compress` *function, say $S' = \langle P, D', \overline{c'}\rangle$, such that* `compress((`$D, \overline{c}$`))` $= (D', \overline{c'})$. *Assume, $\overline{c} = \{c_0, c_1, \ldots, c_{|\overline{c}|-1}\}$, and $\overline{c'} = \{c'_0, c'_1, \ldots, c'_{|\overline{c'}|-1}\}$. For each analysis-state in* LZUPD *mode, say $d = \langle v, s, \texttt{LZUPD}, r\rangle \in D$, and the corresponding modified analysis-state, say $d' = \langle v, s, \texttt{LZUPD}, r'\rangle \in D'$, the following equality holds.*

$$c_{r+1} \oplus c_{r+2} \ldots \oplus c_{|\overline{c}|-1} = c'_{r'+1} \oplus c'_{r'+2} \ldots \oplus c'_{|\overline{c'}|-1}. \tag{1}$$

Proof. Let $k$ be the smallest integer in the set $\{-1, 0, 1, \ldots, |\overline{c}| - 1\}$ such that there does not exist any analysis-state in $D$ whose change-index is same as $k$. That is, $\nexists d = \langle *, *, *, r\rangle \in D$, such that $r = k$. Following three cases arise:

**Case I:** *No such $k$ exists.* This case refers to the fixed-point state for the `compress` function, as described by Rules [CS-Base] and [CS-Compressed]. As per these rules, note that $\overline{c'} = \overline{c}$, and $D' = D$. Hence, Equation 1 trivially holds.

**Case II:** $k = -1$. Following two sub-cases arise based on the size of $\overline{c}$.

`Case II.a.:` $|\overline{c}| = 0$. In this case, Rule [CS-Base] is applicable. Since $\overline{c'} = \overline{c}$, and $D' = D$, Equation 1 trivially holds.

`Case II.b.:` $|\overline{c}| \geq 1$. In this case, Rule [CS-DelFirst] is applicable. As per this rule, $|\overline{c'}| = |\overline{c}| - 1$,

$$r' = r - 1 \qquad \text{and} \qquad \forall_{i=0}^{|\overline{c'}|-1}, c'_i = c_{i+1}$$

Under these equalities we infer that, (i) the number of terms on the LHS and RHS of Equation 1 is same, and (ii) each term on the LHS is same as the corresponding term on the RHS. Hence, Equation 1 holds.

**Case III:** $k \geq 0$. When $k = |\overline{c}| - 1$, Rule [CS-Compressed] is applicable. As before, Equation 1 will hold in that case. Otherwise, when $0 \leq k \neq |\overline{c}| - 1$, Rule [CS-MergeMid] is applicable. In this case, $|\overline{c'}| = |\overline{c}| - 1$. Further, as per this rule, the following holds:

$$\forall_{i=0}^{k-1}, c'_i = c_i, \qquad c'_k = c_k \oplus c_{k+1} \qquad \text{and} \qquad \forall_{i=k+1}^{|\overline{c'}|-1}, c'_i = c_{i+1} \tag{2}$$

Following two sub-cases arise based on the different values of $r$:

`Case III.a.` $r < k$. As per Rule [CS-MergeMid], $r' = r$. In this case, the number of elements on the LHS of Equation 1 is one more than the number of elements of its RHS. Clearly, $c_k$ and $c_{k+1}$ are present in the LHS, whereas in the RHS, $c_k$ and $c_{k+1}$ have been replaced with another element $c'_k$. Using Equation 2, we derive that all terms on the left of $c_k$ are same as the corresponding terms on the left of $c'_k$. Similarly, all terms on the right of $c_{k+1}$ are same as the corresponding terms on the right of $c'_k$. Additionally, since $c'_k = c_k \oplus c_{k+1}$, we infer that Equation 1 holds.

`Case III.b.` $r > k$. As per Rule [CS-MergeMid], $r' = r - 1$. In this case, the number of elements on the LHS of Equation 1 is same as the number of elements of its RHS. Clearly, $c_k$ does not appear in the LHS. As per Equation 2, each term on the LHS is same as the corresponding term on the RHS. Hence Equation 1 holds.

□

Corollary 4.14. *Consider a valid compiler state, say $S = \langle P, D, \overline{c} \rangle$. If the fixed-point application of the* `compress` *function on the input $(D, \overline{c})$ is $(D', \overline{c'})$, that is,* `compress`$((D, \overline{c}))^\star = (D', \overline{c'})$, *then the compiler state $S' = \langle P, D', \overline{c'} \rangle$ is valid.*

Proof. The proof follows directly from Definition 4.9 and Lemma 4.13. □

Note that as a direct implication of Corollary 4.14, Lemma 4.10, and Lemma 4.11, the correctness guarantee of *Homeostasis*, stated in Theorem 4.12, will hold regardless of the number of compilation points where the `compress` function is invoked by *Homeostasis*.

## 5 DISCUSSION

We now present some salient characteristics of *Homeostasis*, followed by an account of how we estimate the amount of stabilization-effort required in the absence of *Homeostasis*. Then, we briefly discuss a generic incremental iterative data-flow analysis pass, which we have implemented and instantiated in *Homeostasis* to aid our evaluation.

### 5.1 Salient Characteristics of *Homeostasis*

**Merging program changes across transformations.** To stabilize a program-abstraction in update (UPD) mode, the `handleUpdate` method of the corresponding analysis requires *net* changes performed on the program (in terms of added and removed nodes/edges) since the last stabilization of that program-abstraction. A simple union of all such changes performed across multiple transformations, would not suffice. For instance, if a program node $n$ (or edge $e$), is added by a transformation, and then deleted by another transformation before the stabilization gets triggered, the node $n$ (or edge $e$) should not be considered as an added/removed node (or edge). In *Homeostasis*, `BasePA` internally merges the changes saved across all transformations to provide the relevant information when requested. In Section 5.5, we give an example of how the provided information is used in the `handleUpdate` method of a specific analysis.

**Resolving pass-dependencies.** *Homeostasis* automatically resolves the pass-dependencies of all compiler passes on program analysis passes by internally stabilizing the program-abstractions in the topological order of their dependencies. This is achieved through the design of getters and stabilizers in *Homeostasis* (see Fig. 7). For example, the getter method of an `UNSTABLE` analysis, say $A$, does not return until its stabilizer has been run. If analysis $A$ is dependent upon some other analysis $B$ (that is, if the computation of the program-abstraction of $A$ requires the program-abstraction of $B$), then the `compute` or `handleUpdate` methods of $A$ invoked from the stabilizer of $A$ would, in turn, invoke the getter of $B$. This will ensure that the stabilizer of $B$ is run before the completion of $A$, regardless of the order in which the stabilizations of both analyses were triggered.

### 5.2 Using *Homeostasis* in Compilers with Multi-Level IRs

A majority of the real-world compilers employ multiple levels of IR, and invoke a host of IR-level-specific analyses and optimizations on the program at each IR level. For such compilers, we can easily see that *Homeostasis* can be used for stabilizing the program-abstractions of the analyses within each such IR level, in response to the optimizations performed at that level. During compilation when the program is lowered from a higher-level IR (say, HIR) to a lower-level IR (say, LIR), (i) the program-abstractions of HIR are no longer relevant for the compiler passes of LIR (and hence need not be stabilized), and (ii) all the necessary program-abstractions for LIR are computed on the current state of the program, when needed. Thus, program-abstractions need not be stabilized *across* different levels of IR. Consequently, compilers supporting multi-level IRs can realize self-stabilization by using *Homeostasis* at each level of the IR independently.

### 5.3 Using *Homeostasis* in Non-Object-Oriented Compilers

While this paper presents the design of *Homeostasis* in the context of object-oriented compilers, we believe that the proposed design can be adapted to allow self-stabilization in compilers that are written even in procedural languages. For example, in languages like C, the information about each analysis pass can be represented as a structure containing six members: (i) a program-abstraction of type "`void*`", (ii) an `enum` field `stabilizationMode`, (iii) an integer field `chIndex`, (iv) a function pointer named `compute` pointing to the function that populates the corresponding program-abstraction, and (v) a function pointer named `handleUpdate` pointing to the function that incrementally stabilizes the corresponding program-abstraction, (vi) a function pointer named `get` pointing to a method that returns a stable program-abstraction (by invoking the corresponding `compute` or `handleUpdate` functions, as needed). The first five members correspond to the instance fields and methods of the class `BasePA`, and the last member corresponds to the `get` method shown in Fig. 7. All instances of this six-member structure are maintained in a global list. We have to identify the set of elementary transformations and modify them to conform to the code shown in Fig. 5.

Since procedural languages do not support the Observer pattern naturally, we would need that the compiler-writer follows a disciplined approach in modifying the IR, and accessing the program-abstractions. To use *Homeostasis* in compilers written in such a language, (i) the compiler writer must perform all IR modifications (directly/indirectly) only via the fixed set of elementary transformations, (ii) perform the initialization of the analysis instance via a generic method (akin to the constructor of `BasePA`, Fig. 7a), and (iii) access the program-abstraction of an analysis, only via the invocation of its `get` function. We believe that a similar exercise can be carried out for compilers written in functional programming languages. We leave it as a future work to actually instantiate *Homeostasis* in such non-OO compilers.

### 5.4 Estimating Manual Stabilization Effort in the Absence of *Homeostasis*

To contrast the savings in effort due to the self-stabilization scheme of *Homeostasis*, we now present a scheme to estimate the effort required for manual stabilization in the absence of *Homeostasis*.

**Critical program-abstractions, and change-points.** Note that in order to perform manual stabilization in the context of an optimization $O$, the pass writer needs to inspect the following two parameters in relation to questions $\mathbf{Q}_1$ and $\mathbf{Q}_2$ (see Fig. 2) : (i) *critical program-abstractions:* those program-abstractions which may be rendered stale by $O$, and may be read later by an existing or

```
public void unrollLoop (WhileStatement loop) {
  DumpSnapshot.dumpPointsTo(loop, "original"); // L1
  Statement newBody = ... /* code to obtain the unrolled body */
  loop.setBody(newBody); // L2
  CompoundStatementNormalizer.removeExtraScopes(loop); // L3
  DumpSnapshot.dumpPointsTo(loop, "afterUnroll");} // L4
```

Fig. 16. Concrete optimization pass: Loop unroller from IMOP (`WhileStatementInfo.java:60-83`).

future optimizing pass (and hence may need to be stabilized); we use $S_a$ to denote the set of critical program-abstractions, and (ii) *change-points:* those program points in the code of the optimization $O$ which may directly or indirectly trigger an elementary transformation (and hence may necessitate stabilization); we use $S_p$ to denote the set of change-points. We demonstrate these parameters using an example.

*Example 5.1.* Fig. 16 shows a snippet of a concrete optimization pass that performs loop unrolling in the IMOP compiler framework. We have added the calls to the `dumpPointsTo` method to simply print the points-to information corresponding to the specified node, symbolising the clients that may need points-to information, before and after unrolling. Method `setBody` is an elementary transformation method that replaces the current body of the receiver while-statement with the provided body. Method `removeExtraScopes` is used to remove unnecessary wrapping of statements within blocks (for example, {{S}} ⟶ {S}). For this snippet, $S_p$ = {L2, L3}, and $S_a$ is the set of those program-abstractions that may be impacted by the statements in $S_p$; for example, CFG, call-graph, use-def chains, and so on. Note that there may be other change-points within the implementation of the common utility `removeExtraScopes`; we do not consider them here as they are not present in the code written as a part of this pass.

**Relevant change-Points (RP).** An inexperienced programmer may perform manual stabilization naïvely by re-computing each of the program-abstractions (or a likely superset of $S_a$, for correctness), after each change-point in $S_p$ – a highly inefficient scheme, both in terms of its effect on performance, and the number of program points where the stabilization code needs to be invoked. An experienced programmer, on the other hand, is more likely to insert stabilization code only at a subset of change-points present in the code of the optimization pass. This reduction stems from two key insights: (i) Stabilization is required only after the last change-point in a series of transformations, after which one or more critical program-abstractions may be read. (ii) Further, a program-abstraction need not be stabilized at such a last change-point, if the net program-changes at that change-point is guaranteed to not alter the existing values of the program-abstraction. For example, consider the addition/deletion of a node for an assignment statement (of the form `x=y`). If the variable `x` is a pointer-type variable, then such an addition/deletion does not change the points-to map of the variable `x`, if both `x` and `y` point to the same set of 'may points-to' locations, even in the absence of this statement. Similarly, if the variable `x` is of integer type, then the addition/deletion of such a node will not impact the points-to information of any variable (if the expert can assert that the value of the variable `x` is not used as an offset for any pointer dereference).

The second key insight leads to cases where even if the program-abstraction is marked as `UNSTABLE`, in reality no stabilization is required. Considering the challenges in obtaining an expert-based study in a large compiler framework, and the difficulty in automatically identifying the cases related to the second key insight, for the purpose of our evaluation, we approximate the set of expert-marked change-points by the set of *relevant change-points* that is based on the first key insight. A relevant change-point is a program point in the optimization pass that corresponds to the last change-point in a series of transformations, after which one or more critical program-abstractions may be read; hence, one or more program-abstractions may require manual stabilization at a relevant change-point. For example, in Fig. 16, the set of relevant change-points, from among the change-points {L2, L3}, is a singleton set {L3} – L3 is the last change-point in the series of transformations (L2;L3) after which a program-abstraction, the points-to map, will be read (at L4).

| | Full recomputation of the analysis from scratch | Incremental update of the program-abstraction using the stored changes to the program |
|---|---|---|
| Stabilization triggers eagerly at each elementary transformation | EGINV | EGUPD |
| Stabilization triggers at *relevant change-points* in the optimization pass | RPINV | RPUPD |
| Stabilization triggers lazily upon an attempt to read from the stale program-abstractions | LZINV | LZUPD |

Fig. 17. Summary of various modes of stabilization used in the evaluation of our implementation of *Homeostasis* in IMOP.

**Modes of manual stabilization**. Besides the lazy modes of stabilization preferred by *Homeostasis*, a few other modes of stabilization are easily conceivable. For example, the stabilization of all program-abstractions can be triggered during each elementary transformation of the program eagerly (see Section 3.1.4). Like the two modes of lazy stabilization, this eager scheme leads to two modes of stabilization: eager-invalidate (EGINV) and (EGUPD). An efficient alternative to the eager modes of stabilization is to invoke stabilization at only the set of relevant change-points. Like before, two modes of stabilization are possible – RPINV and RPUPD, corresponding to the invalidate and incremental update options, respectively.

These stabilization modes are very similar to the custom codes manually written by the compiler pass writers in conventional compilers: (i) UPD modes in the presence of incremental update, whereas INV modes otherwise, and (ii) EG-modes during naïve (and inefficient) but easy manual stabilization, whereas RP-modes during relatively efficient stabilizations performed by an experienced compiler writer. For the purpose of evaluation (see Section 7), along with the two lazy-modes of self-stabilization advocated by *Homeostasis*, we have implemented these four modes of stabilization (EGINV, EGUPD, RPINV, and RPUPD) to approximate different modes of manual stabilization. Fig. 17 summarizes all the implemented modes of stabilization.

### 5.5 Implementing Iterative Data-Flow Analyses in *Homeostasis*

To conveniently implement multiple analyses that can use LZUPD mode of stabilization, we implemented a generic pass of an analysis that can be used to implement any incremental, inter-thread, flow-sensitive, iterative data-flow analysis (IDFA) for OpenMP parallel programs. We term this generic analysis as HIDFA and implemented it as a class that extends BasePA.

Fig. 18 shows the steps used by the handleUpdate method of the HIDFA class in the context of forward data-flow analyses; the steps in the context of backward analysis are similarly derived (not shown). To realize self-stabilization, it populates a set (seeds) of nodes, starting which the flow maps may need an update as a result of program transformations. These seed nodes are passed

```
1 Function handleUpdate()
2     Set seeds = .. empty set ..;
3     IRChanges netIRChanges = getNetChanges(this.chIndex);
4     foreach e in netIRChanges.removedEdges do Add the destination node of the removed edge e to seeds;
5     foreach n in netIRChanges.removedNodes do Add the (old) successors of the removed node n to seeds;
6     foreach n in netIRChanges.addedNodes do Add the added node n itself, as well as its successors to seeds;
7     foreach e in netIRChanges.addedEdges do Add the destination node of the added edge e to seeds;
8     incrementalIDFA(seeds);
```

Fig. 18. Definition of the `handleUpdate` function for forward iterative data-flow analysis.

to the `incrementalIDFA` method for performing the stabilization (Line 8). The implementation of `incrementalIDFA` is inspired from the work by Marlowe and Ryder [1989]; Ryder et al. [1988], and is a simple extension of the two-phase incremental update algorithms for iterative versions of data-flow analysis given by Pollock and Soffa [1989], adapted to handle the parallel semantics of the OpenMP parallel programs.

For completeness, we now give brief details of our implementation of `incrementalIDFA`. This method starts with the reconstruction of the SCC Decomposition Graph (SDG) of the updated control-flow graph of the program. Each SCC-node in the SDG that contains any of the seed nodes is *marked* for processing. All the marked SCC-nodes are processed one-by-one, as per their topological-sort order in the SDG. Within each SCC-node, the processing continues using a worklist-based approach in two-phases: (i) *exaggerate* phase: in this phase, the flow-maps of various program-nodes in the SCC-node are initialized to an under-approximated value (such as $\top$), starting which the precise flow-map can be eventually obtained, and (ii) *adjust* phase: in this phase, the program-nodes are processed until fixed-point, using the data-flow equations similar to the exhaustive iterative data-flow algorithm. At the end of the second phase, the flow-maps of all program-nodes within the SCC-node correspond to the maximum fixed-point (MFP) solution. During the processing of an SCC-node, if it is found that the program-nodes of any successor SCC-nodes need to be added to the worklist, then such successor SCC-nodes are marked for processing. This method continues till there are no more marked SCC-nodes left for processing. We refer the reader to the work by Pollock and Soffa [1989] for more details on the underlying two-phase incremental update algorithm.

## 6 INSTANTIATION OF *HOMEOSTASIS*

In order to assess the feasibility and generality of *Homeostasis*, we have retrofitted *Homeostasis* to a research compiler, and have implemented/adapted a number of analysis passes and optimization passes in the compiler, to conform to the design of *Homeostasis*. In this section, we discuss the relevant instantiation details.

### 6.1 Retrofitting *Homeostasis* to a Real-World Compiler

The design of *Homeostasis* is applicable to all the object-oriented compilers. Retrofitting *Homeostasis* in mainstream real-world compilers, like LLVM, OpenJ9, IMOP, and so on, will require rewriting all the existing program transformations using the identified fixed set of elementary transformations, and modifying all the existing analysis passes to conform to the structure prescribed by *Homeostasis* (see Fig. 11). We have implemented our prototype in IMOP. We believe that a similar exercise can be undertaken for other object-oriented compiler frameworks as well. To further strengthen this belief, in Appendix B we briefly discuss the guidelines on how the LLVM compiler framework can be extended with *Homeostasis*. The one-time effort required to retrofit *Homeostasis* to IMOP is discussed next.

**Identifying elementary transformations.** The IMOP IR comprises two kinds of nodes – *leaf* nodes and *non-leaf* nodes. All OpenMP or C constructs with “body”, such as while-loop, if-statement, and OpenMP parallel-region, are represented as *non-leaf nodes*. Similarly, nodes such as simple statements of C, and OpenMP directives such as barriers and flushes, are considered as *leaf* nodes. Each non-leaf node defines a nested flow of control, using edges among its components, which can themselves be non-leaf nodes. We identified the pre-existing fundamental transformations in IMOP that are used to add, replace, or delete various components of a non-leaf node, as the set of elementary transformations. In total, we identified and modified 47 elementary transformations spread across 17 non-leaf nodes to make them conform to the design shown in Fig. 5.

We observe that the modification of these elementary transformations did not necessitate any changes at their call-sites. From Fig. 5, note that Step B, which corresponds to the pre-existing code of an elementary transformation (in the absence of *Homeostasis*) does not undergo any changes while retrofitting the compiler with *Homeostasis* – the IR transformations are performed as before. Steps A, C, and D, interact only with stabilization-specific data structures to capture the program-changes resulting from Step B; they do not alter the meaning of the elementary transformation. Consequently, the compiler-pass writers can continue to use the elementary transformation as before, without having to consider their impact on self-stabilization.

**Modifying higher-level transformation APIs.** In order to allow for easier expression of transformation passes, IMOP provides four basic CFG transformations: (i) insertion of a successor node, (ii) insertion of a predecessor node, (iii) insertion of a node on an edge, and (iv) removal of a node. To retrofit *Homeostasis* to IMOP, we have modified the implementation of all these four transformations such that they express all IR changes directly or indirectly in terms of the identified set of elementary transformations. IMOP provides many higher-level transformation APIs such as duplicating a block of statements, renaming of free variables, and so on. All these transformation APIs employ elementary transformations, the above-mentioned CFG transformations, or other higher-level transformations to express any changes to the IR. Since all the transformations APIs deployed in IMOP directly or indirectly express all the program changes via elementary transformations, the implementation of none of the existing optimization passes in IMOP required any specific modifications to adapt them to use *Homeostasis*.

**Creating/identifying the super-class `BasePA`.** IMOP already contains a class named `BasePA`, which serves as the super-class of all the analysis passes. We modified `BasePA` as per Fig. 7 to conform to the design of *Homeostasis*. This mainly involved (i) modifying the constructor as shown in Fig. 7a, (ii) adding the `stabilize` method from Fig. 7c, (iii) adding an abstract method `compute` to be implemented by each concrete subclass of `BasePA`, and (iv) adding a dummy method for `handleUpdate` to be overridden by each concrete subclass of `BasePA` that seeks to employ `LZUPD` stabilization-mode. Further, to allow efficient maintenance of the centralized list of program-changes, as detailed in Fig. 15, we added the appropriate data structures and methods.

## 6.2 Adapting Analysis Passes to Use *Homeostasis*

In order to assess the applicability of *Homeostasis* to various kinds of program-abstractions, we have (i) extended existing analysis passes of IMOP to conform to the design of *Homeostasis*, and (ii) implemented HIDFA and its various instantiations in IMOP.

Fig. 19 provides a brief description of the program-abstractions for some of the key analysis passes in IMOP that now use *Homeostasis*. In particular, we have implemented a set of six standard flow-sensitive context-insensitive inter-thread iterative data-flow analyses as subclasses of the HIDFA class: (i) points-to analysis, (ii) reaching-definitions analysis, (iii) liveness analysis (a backward IDFA), (iv) copy propagation analysis, (v) lockset analysis, and (vi) dominance analysis. None of these IDFA passes required any stabilization-specific code in order to support `LZUPD` stabilization-mode. For their super-class HIDFA, as well as for other analyses listed in the figure, we had to

| | Program-abstraction (Relevant analysis-pass class) | Description |
|---|---|---|
| 1. | Points-to maps (PointsToAnalysis*) | points-to information |
| 2. | Reaching-definitions (ReachingDefinitionAnalysis*) | reaching definitions |
| 3. | Liveness maps (LivenessAnalysis*) | liveness information of variables |
| 4. | Copy maps (CopyPropagationAnalysis*) | information about variables that are copies of each other |
| 5. | Must-lock sets (LockSetAnalysis*) | information about locks that must have been taken at a node |
| 6. | Dominator sets (DominanceAnalysis*) | dominator information |
| 7. | Control-flow graphs (CFGInfo) | control-flow graph |
| 8. | Inter-task edges (InterTaskEdge) | information representing inter-task communication |
| 9. | Strongly-connected components (SCC) | SCC Decomposition Graph (SDG) and enclosing SCC-nodes |
| 10. | Phase-flow graphs (NodePhaseInfo) | may-happen-in-parallel (MHP) analysis |
| 11. | Call graphs (CallStatementInfo) | call graph |
| 12. | Symbol-/type-tables (CompoundStatementInfo) | information about the symbols and user-defined types in a scope |
| 13. | Label-lookup sets (StatementInfo) | information about labels of a statement |
| 14. | Single-valued expressions (SVEChecker) | information about expressions that evaluate to same value by all the threads |

Fig. 19. List of program-abstractions corresponding to key analysis passes in IMOP that were made conforming to *Homeostasis*. The bracketed text in the first column, gives the name of the IMOP class modified. Classes annotated with * are the analyses that are implemented as subclasses of HIDFA in IMOP.

(1) modify their corresponding getter methods, as per the design from Fig. 7, (2) ensure that the code to generate the program-abstraction from scratch is invoked from the overridden method `compute`, and (3) wherever applicable, ensure that the code to perform an incremental update of the program-abstraction is invoked from the overridden method `handleUpdate`.

### 6.3 Optimization Passes in the Context of *Homeostasis*

Recall that in a *Homeostasis*-enabled compiler framework, the optimization-pass writers can efficiently design and implement new optimizations without having to write *any* stabilization-specific code (see Fig. 11). Fig. 20 lists some key optimization passes (including existing, as well as new ones implemented by us) in IMOP; none of them required any stabilization-specific code. Two common characteristics of these passes are: (i) Each of them expresses the required program transformations using either the elementary transformations and/or other transformation APIs provided by IMOP. (ii) Each of them uses/impacts one or more *Homeostasis*-enabled program-abstractions described in Fig. 19. Now, we briefly describe a subset of these optimization passes (that we implemented for evaluating *Homeostasis*), which we collectively term as `BarrElim`.

**`BarrElim`: a set of optimization passes to aid barrier elimination.** In order to develop a real-world client for evaluating *Homeostasis*, we have implemented a set of three optimization passes that aim to reduce redundant barriers in OpenMP C programs: redundant-barrier-remover, parallel-construct-expander, and selective-function-inliner. The parallel-construct-expander in turn uses five pre-existing optimizations: (i) privatisation of shared variables, (ii) parallel-region scope expander, (iii) merging of consecutive parallel-regions, (iv) parallel-region and enclosing serial-loop interchange, and (v) parallel-region unswitching. Further, the selective-function-inliner uses two pre-existing cleanup-optimizations: extra-scope-remover, and unused-elements-remover. Collectively, we term the three implemented optimization passes as BarrElim.

| | Optimization Pass (Relevant class/method in IMOP) | Description |
|---|---|---|
| 1. | Redundant barrier remover† (RedundantSynchronizationRemoval) | removes those barriers that do not preserve any inter-thread dependences across them |
| 2. | OmpPar expander† (ParallelConstructExpander.expandParRegion) | expands the scope of parallel-regions upwards and downwards |
| 3. | OmpPar merger† (ParallelConstructExpander.mergeParRegions) | merges two consecutive parallel-regions, if safe |
| 4. | OmpPar-loop interchange† (ParallelConstructExpander.interchangeUp) | interchanges a parallel-region with its enclosing loop, if safe |
| 5. | OmpPar unswitching† (ParallelConstructExpander.interchangeUp) | interchanges a parallel-region with its enclosing if-statement, if safe |
| 6. | Variable privatization† (ParallelConstructExpander.expandUpward) | privatizes OpenMP shared variables, if safe |
| 7. | Function inliner† (FunctionInliner) | selectively inlines those monomorphic calls that contain at least one barrier |
| 8. | Scope remover† (CompoundStatementNormalizer) | removes redundant encapsulations of compound-statements within the given node |
| 9. | Unused-elements remover† (NodeInfo::removeUnusedElements) | removes unused functions, types, typedefs, and symbol declarations |
| 10. | Expression simplification (ExpressionSimplifier) | simplifies complex C expressions as per IMOP's normalization rules |
| 11. | OmpConstruct simplification (ExpressionSimplifier) | simplifies combined OpenMP constructs as per IMOP's normalization rules |
| 12. | Loop unroller (IterationStatementInfo::unrollLoop) | unrolls a serial loop |
| 13. | Inter-phase code percolator (BarrierDirectiveInfo::percolateCodeUpwards) | moves instructions across an OpenMP barrier, if safe |
| 14. | Copy eliminator (CopyElimination) | removes redundant variables, relying on copy-propagation analysis |
| 15. | Loop barrier extractor (LoopInstRescheduler::peelFirstBarrier) | peels out a barrier from within a serial loop, if safe |

Fig. 20. List of key optimization passes inspected in IMOP; none of these passes required any stabilization-specific code in the context of *Homeostasis*. The bracketed text in the first column, gives the name of the IMOP class modified. Passes annotated with a † indicate that they belong to, or are used by, the BarrElim set of optimization passes.

The optimization passes in BarrElim extend prior work by Gupta and Schonberg [1996]; Tseng [1995] on parallel-region expansion and barrier removal with function inlining to realize an efficient barrier removal algorithm. These optimization passes perform the following steps, as shown in the block diagram in Fig. 21 : (1) Remove redundant barriers (within a parallel-region), whose removal do not violate any data dependence among the statements across them. The remaining two passes help improve the opportunities for barrier removal within each function. (2) Expand and merge the parallel-regions, while possibly expanding their scope to the call-sites of their enclosing functions, wherever possible. This helps in bringing more barriers (including the implicit ones) within the resulting parallel-region, thereby creating new opportunities for barrier removal. (3) Inline those monomorphic calls whose target function (i) is not recursive, and (ii) contains at least one barrier. These three optimization passes are repeated until fixed-point (no change).

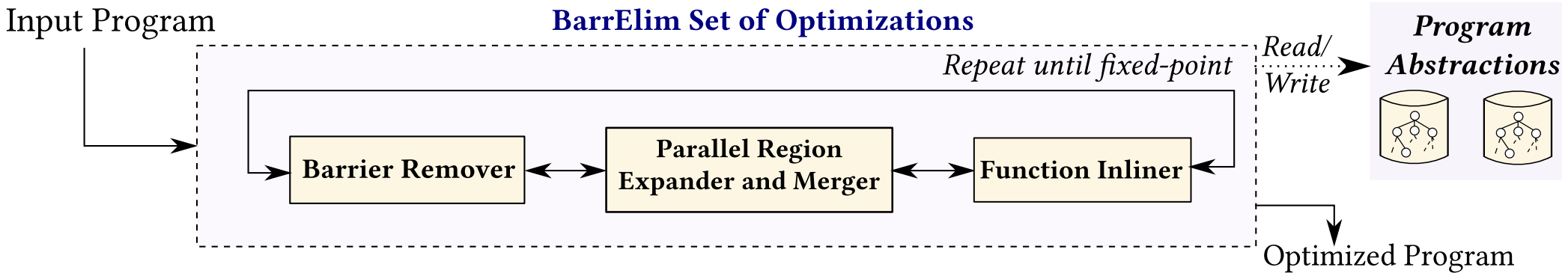

Fig. 21. Block diagram of `BarrElim`, the selected set of client optimization passes, for removal of barriers from OpenMP programs.

Like many similar optimizations [Aloor and Nandivada 2015; Barik et al. 2013; Gupta et al. 2017; Nandivada et al. 2013], the optimization passes in `BarrElim` involve multiple alternating phases of inspections and transformations, which in turn lead to a number of interleaved accesses (reads and writes) to various program-abstractions, such as phase information, points-to information, super-graph (involving CFGs, call-graphs, and inter-task edges), AST, and so on – this interaction is depicted in Fig. 21 using the dotted edge.

We have found that `BarrElim` is an impactful set of optimization passes for OpenMP codes. For example, on the NPB benchmarks [Van der Wijngaart and Wong 2002], the `BarrElim` optimized code yielded up to 5% improvement in execution time; the input benchmarks, as well as the `BarrElim` optimized codes were compiled using the `-O3` switch of `gcc`. Considering that the obtained gains are on top of the many optimizations enabled by the `-O3` switch, it can be seen that the gains are significant. Consequently, from among the various optimization passes present in IMOP, we select the `BarrElim` set of optimization passes (involving nine optimization passes and numerous program-abstractions) to evaluate the ease-of-use and performance of *Homeostasis* (see Section 7).

## 7 IMPLEMENTATION AND EVALUATION

We have implemented all the key components of *Homeostasis* (see Fig. 4, and Section 3.2) in the IMOP compiler framework [Nougrahiya and Nandivada 2019]. IMOP is a new open-source compiler framework for conveniently writing program analyses, source-to-source optimizations, and profiling tools for OpenMP C programs; it spans more than 170$k$ lines of code (LOC) in Java. IMOP has been used successfully in various published works (such as [Viswakaran Sreelatha and Balachandran 2016], [Viswakaran Sreelatha et al. 2018], [Viswakaran Sreelatha and Nasre 2018] and [Krishnakumar et al. 2019]), and is under active development. Our implementation of *Homeostasis* in IMOP involved coding of ~4500 LOC for adding self-stabilization to the elementary-transformation methods, and ~2500 LOC for supporting self-stabilization for existing program-abstractions of IMOP.

We have also implemented HIDFA, our generic inter-thread IDFA pass with incremental update in *Homeostasis* (~3300 LOC; discussed in Section 5.5). We instantiated HIDFA to realize the implementation of following six standard flow-sensitive context-insensitive inter-thread iterative data-flow analyses: (i) points-to analysis (~870 LOC), (ii) reaching-definitions analysis (~190 LOC), (iii) liveness analysis (~140 LOC), (iv) copy-propagation analysis (~200 LOC), (v) lock-set analysis (~240 LOC), and (vi) dominance analysis (~116 LOC). Further, in order to assess the usability of *Homeostasis*, we have implemented a set of optimization passes, collectively termed as `BarrElim`, for barrier removal (~2500 LOC; see Section 6.3). As expected, in the set of optimization passes in `BarrElim` no stabilization-specific code was needed. This underscores the ease-of-use facilitated by the design of *Homeostasis*. We have made our implementation of *Homeostasis* in IMOP open-source and publicly available [Nougrahiya and Nandivada 2021].

| 1 | | 2 | 3 | 4 | 5 | 6 |
|---|---|---|---|---|---|---|
| **Benchmark** | | **#LOC** | **#Nodes** | **#Edges** | **#PC** | **#Barr** |
| 1. | BT (NPB) | 2615 | 4748 | 5016 | 9 | 47 |
| 2. | CG (NPB) | 642 | 1403 | 1485 | 14 | 31 |
| 3. | EP (NPB) | 352 | 775 | 813 | 2 | 4 |
| 4. | FT (NPB) | 899 | 2033 | 2151 | 7 | 14 |
| 5. | IS (NPB) | 333 | 711 | 762 | 2 | 4 |
| 6. | LU (NPB) | 2355 | 4687 | 4974 | 8 | 35 |
| 7. | MG (NPB) | 1278 | 2784 | 2918 | 10 | 19 |
| 8. | SP (NPB) | 2543 | 5364 | 5744 | 7 | 72 |
| 9. | quake (SPEC) | 1489 | 3333 | 3491 | 11 | 22 |
| 10. | art-m (SPEC) | 1691 | 1710 | 1791 | 4 | 4 |
| 11. | amgmk (Sequoia) | 895 | 1867 | 1949 | 2 | 5 |
| 12. | clomp (Sequoia) | 1605 | 4162 | 4289 | 28 | 73 |
| 13. | stream (Sequoia) | 331 | 735 | 762 | 6 | 12 |
| 14. | bellman-ford (IMSuite) | 143 | 546 | 564 | 3 | 6 |
| 15. | bfs-dijkstra (IMSuite) | 104 | 295 | 305 | 2 | 4 |
| 16. | byzantine (IMSuite) | 135 | 444 | 467 | 3 | 6 |
| 17. | dominating-set (IMSuite) | 321 | 1223 | 1277 | 12 | 24 |
| 18. | kcommittee (IMSuite) | 220 | 889 | 925 | 6 | 12 |
| 19. | leader-elect-dp (IMSuite) | 122 | 442 | 460 | 2 | 4 |
| 20. | leader-elect-hs (IMSuite) | 184 | 528 | 553 | 3 | 6 |
| 21. | leader-elect-lcr (IMSuite) | 89 | 301 | 314 | 3 | 6 |
| 22. | mis (IMSuite) | 161 | 549 | 574 | 4 | 8 |
| 23. | mst (IMSuite) | 299 | 1226 | 1279 | 9 | 18 |
| 24. | vertex-coloring (IMSuite) | 172 | 609 | 636 | 3 | 6 |

Fig. 22. Benchmark characteristics. Abbreviations: *#LOC*=number of lines of code, *#Node*=number of executable nodes in the CFG, *#Edge*=number of edges in the CFG, *#PC*=number of static parallel constructs, and *#Barr*=number of static barriers (implicit + explicit).

### 7.1 Experimental Setup

We present our evaluation on a set of twenty-four real-world benchmark programs taken from four popular benchmark suites (listed in Fig. 22) : (i) all eight benchmark programs of NPB-OMP 3.0 suite [Van der Wijngaart and Wong 2002], (ii) `quake`, and `art-m`, the two (out of 3) OpenMP-C benchmark programs from SPEC OMP 2001 [Aslot et al. 2001] that can be handled by IMOP, (iii) all three OpenMP C benchmark programs – `amgmk`, `clomp`, and `stream` – from Sequoia benchmark suite [Seager, M 2008], and (iv) all the eleven benchmark programs from IMSuite benchmark suite [Gupta and Nandivada 2015]. Note that these comprise some of the largest standard open-source benchmark programs for OpenMP C. Since IMOP accepts only OpenMP C programs, we did not consider any other benchmarks from SPEC OMP 2001, or from Sequoia, as they contain a mix of C/C++/MPI code. For `IS` from NPB, the set of client optimization passes, `BarrElim`, could not find any opportunity for optimization; as no program transformation was performed by the pass, no stabilization of data-flow analyses was triggered in any mode of self-stabilization – hence, we omit `IS` from our discussion. In Fig. 22, we list a few static characteristics of the benchmarks, such as the size of each selected benchmark, the number of parallel constructs, and the number of static barriers.

| Component Pass of BarrElim | LOC | #CP |
|---|---|---|
| Parallel-construct expansion | 1675 | 66 |
| Function inlining | 463 | 15 |
| Redundant-barrier deletion | 313 | 2 |
| Driver | 10 | 1 |
| **Total :** | 2461 | **84** |

(a) Maximum number of change-points obtained within the component optimization passes of `BarrElim` upon running them on the benchmarks under study. Abbr. : *LOC*=number of lines of code; *#CP*=number of change-points.

| Program-Abs. | Mode | #RP |
|---|---|---|
| Points-to graphs | UPD | 17 |
| Control-flow graphs | UPD | 70 |
| Call graphs | UPD | 60 |
| Phase information | INV | 61 |
| Inter-task edges | INV | 31 |
| Symbol-/type-tables | UPD | 37 |
| Label-lookup tables | UPD | 60 |

(b) Program-abstractions read by the `BarrElim` passes. Abbr : *Mode*=stabilization-mode; *#RP*=number of *relevant* change-points in the `BarrElim` passes after which the abstraction was read from.

Fig. 23. Stabilization-related information for the set of optimization passes constituting `BarrElim`, obtained upon profiling/inspecting the code.

In order to study the impact of different hardware on the performance improvements obtained with *Homeostasis*, we used two multicore platforms to perform our empirical evaluation: (i) **Nanda**, a 64-thread (dual socket, 16 cores per socket, 2 threads per core) 2.3 GHz Intel Xeon Gold 5218 system with 64 GB of memory, running Ubuntu 22.04.1 LTS and Java HotSpot 64-Bit Server VM 14.0; and (ii) **K2**, a 64-thread (quad-socket, 8 cores per socket, 2 threads per core) 2.3 GHz AMD Abu Dhabi system with 512 GB of memory, running CentOS 6.4 and Java HotSpot 64-Bit Server VM 9.0. The multi-core aspect of these systems is relevant for the part of the evaluation that executes the generated OpenMP code. The OpenMP codes were compiled using the GCC 11.3.0 compiler. Taking inspiration from the insightful paper of Georges et al. [2007], we report the compilation- and execution-time numbers by taking a geometric mean over 30 runs. We present the evaluation in four directions: (i) Ease-of-use of *Homeostasis* (in Section 7.2), (ii) Performance evaluation of the proposed lazy modes of stabilization (in Section 7.3), (iii) Impact of using the compression optimization on the data-structures employed by *Homeostasis* to keep track of the program-changes (in Section 7.4), and (iv) Correctness of the lazy modes of stabilization (in Section 7.5).

### 7.2 Ease of Use: Self-stabilization vs. Manual Stabilization

We now present an empirical study for assessing the impact of *Homeostasis* on writing different compiler passes, by comparing the coding effort required to realize self-stabilization in a *Homeostasis*-enabled compiler, against those required to perform manual stabilization (in the absence of *Homeostasis*). We perform this evaluation in the context of the various component optimization passes of `BarrElim`.

Based on the discussion in Section 5.4, we designed a simple scheme to estimate the additional coding effort that may be required to perform manual stabilization. We profiled the IMOP compiler by instrumenting the implementations of the various optimization passes of `BarrElim`, as well as of various program analyses and elementary transformations. By running this profiled compiler on each benchmark program, we obtained (i) the set of change-points for the `BarrElim` passes, and (ii) the set of program-abstractions that may be impacted by the `BarrElim` passes (see Section 6.3). In this section, we use this data to estimate the manual coding effort that may be required to answer the key questions from Fig. 2.

**Which program-abstractions to stabilize?** As discussed in Section 5.4, since it is difficult to obtain the exact set of *critical program-abstractions* (to be stabilized manually), we manually analyzed the code of various optimization passes of `BarrElim` and estimated that there are seven program-abstractions (listed in Fig. 23b) that may be impacted by the `BarrElim` passes. Thus, in the case of manual-stabilization, on writing each of the `BarrElim` passes, the pass writer needs to identify these seven program-abstractions, from the plethora of available program-abstractions – a daunting task. Note that the `BarrElim` optimization passes may also impact other program-abstractions that may be added to the compiler in future; hence, for each newly added analysis-pass, the analysis-pass writer would have to manually assess whether the corresponding program-abstraction may be rendered stale due to the transformations performed by any of the `BarrElim` optimization passes (and, in fact, by any of the other optimization passes in the compiler).

In contrast, in a *Homeostasis*-enabled compiler, all these tasks are automated – the optimization-pass (or even analysis-pass) writer needs to put no effort into identifying the program-abstractions (existing or new) that may be impacted by an optimization pass.

**Where to invoke stabilization?** In Fig. 23a, we enumerate the number of change-points discovered in the major components of `BarrElim`. In the absence of *Homeostasis*, the pass writer would have to correctly identify these 84 change-points (i.e., on average, *almost 1 for every 28 lines of code!*) across various optimization passes of `BarrElim`, and insert code for ensuring stabilization of the affected program-abstractions. At each change-point, the pass writer may need to handle stabilization of the impacted program-abstractions, irrespective of the chosen mode of stabilization. To identify a more aggressive baseline, we note that not all change-points (referred to in Figure 23a) may warrant stabilization, and a pass writer may need to invoke the stabilization code only at the relevant change-points (see Section 5.4), for only those set of abstractions that may be modified at that point.

Fig. 23b, column 4, lists the number of *relevant* change-points for each program-abstraction impacted by the optimization passes of `BarrElim`; this data too was obtained by profiling the IMOP compiler (profiling details discussed above), while running it on all the benchmarks under study. The figure shows that there are a significant number of places in the components of `BarrElim` where this stabilization code needs to be manually invoked, in the absence of *Homeostasis*. For example, CFG stabilization needs to be performed at 70 places, and call-graphs at 60 places – which can lead to cumbersome and error-prone code. Further, upon addition of any new program analysis to the compiler (or any modification to the existing analysis), the pass writer would have to revisit all the change-points of pre-existing optimizations (for example, 84 change-points across all optimization passes of `BarrElim`) to check if the change-point may have necessitated stabilization of the newly added/modified program-abstraction.

In contrast, in a *Homeostasis*-enabled compiler, all the above tasks are automated – the pass writer needs to spend no effort in identifying the places of stabilization, as she needs to add no additional code as part of the optimization in order to stabilize the program-abstractions.

**How to stabilize?** We use the same stabilization code for both manual stabilization and *Homeostasis*, and hence there is no difference between them, with respect to this key question.

*Summary.* In contrast to the traditional compilers, it is much easier to ensure stabilization of program-abstractions when writing optimizations or analyses in *Homeostasis*.

### 7.3 Performance Evaluation

We conduct the performance evaluation of the proposed lazy modes of stabilization, by studying the parameters related to stabilization time, total compilation time, and memory consumption, in the context of the prior discussed set of optimization passes, `BarrElim`. We do so by presenting a comparison to the RP-modes of stabilization, which correspond to manual stabilizations performed by experienced compiler-pass writers (see Section 5.4). In Fig. 24, we present various parameters related to compilation of the benchmark programs under study, when compiled with the `RPINV` mode of stabilization; these numbers serve as the baseline for the evaluations discussed in this section.

| 1 | 2 | 3 | 4 | 5 | 6 | 7 | 8 | 9 | 10 |
|---|---|---|---|---|---|---|---|---|---|
| **Benchmark** | **#Trig** | **#T-Func** | **STB-Time** (s) | | **Total Time** (s) | | **Memory** (MB) | | **Ch-Size** |
| | | | Nanda | K2 | Nanda | K2 | Nanda | K2 | |
| 1. BT (NPB) | 17 | 214327 | 3.62 | 5.93 | 8.29 | 14.25 | 2862.24 | 8530.41 | 15170 |
| 2. CG (NPB) | 89 | 242019 | 2.14 | 3.46 | 3.65 | 6.11 | 1675.41 | 3850.55 | 115070 |
| 3. EP (NPB) | 2 | 1983 | 0.04 | 0.07 | 0.56 | 1.17 | 421.32 | 1784.43 | 0 |
| 4. FT (NPB) | 61 | 443820 | 4.42 | 7.16 | 7.81 | 13.03 | 2771.97 | 10060.87 | 49871 |
| 5. IS (NPB) | 0 | 806 | 0 | 0 | 0.33 | 0.82 | 330.51 | 1430.9 | 0 |
| 6. LU (NPB) | 33 | 507444 | 8.43 | 14.63 | 11.92 | 21.13 | 2856.21 | 8894.1 | 41361 |
| 7. MG (NPB) | 230 | 2372204 | 24.95 | 58.17 | 34.47 | 74.12 | 2788.71 | 13969.44 | 571691 |
| 8. SP (NPB) | 35 | 672958 | 8.31 | 14.36 | 11.66 | 20.43 | 2828.68 | 16217.09 | 38102 |
| 9. quake (SPEC) | 35 | 437217 | 6.74 | 10.44 | 8.92 | 14.29 | 3220.78 | 13348.05 | 16727 |
| 10. art-m (SPEC) | 5 | 88347 | 1.03 | 2.01 | 2.33 | 4.56 | 610.15 | 3030.15 | 134 |
| 11. amgmk (Sequoia) | 56 | 239940 | 3.2 | 6.54 | 6.15 | 11.36 | 2715.4 | 7760.35 | 52486 |
| 12. clomp (Sequoia) | 63 | 397078 | 6.76 | 11.56 | 13.72 | 23.19 | 2779.42 | 14921.1 | 146511 |
| 13. stream (Sequoia) | 45 | 51020 | 0.55 | 1.03 | 1.4 | 2.79 | 548.6 | 2154.01 | 36426 |
| 14. bellman-ford (IMS) | 23 | 15038 | 0.3 | 0.49 | 0.86 | 1.62 | 488.84 | 2008.73 | 6051 |
| 15. bfs-dijkstra (IMS) | 17 | 4695 | 0.13 | 0.29 | 0.73 | 1.49 | 480.23 | 1965.33 | 3237 |
| 16. byzantine (IMS) | 19 | 7979 | 0.19 | 0.32 | 0.59 | 1.25 | 405.9 | 1768.1 | 4563 |
| 17. dominating-set (IMS) | 75 | 186360 | 1.88 | 3.09 | 3.19 | 5.32 | 1255.84 | 4298.49 | 64687 |
| 18. kcommittee (IMS) | 43 | 44272 | 0.56 | 1.01 | 1.57 | 2.93 | 535.33 | 2147.7 | 30767 |
| 19. leader-elect-dp (IMS) | 5 | 2325 | 0.07 | 0.12 | 0.39 | 0.84 | 340.87 | 1407.41 | 282 |
| 20. leader-elect-hs (IMS) | 11 | 9978 | 0.18 | 0.35 | 0.84 | 1.67 | 500.83 | 1977.11 | 2624 |
| 21. leader-elect-lcr (IMS) | 6 | 1844 | 0.05 | 0.1 | 0.52 | 1.11 | 407.03 | 1698.05 | 402 |
| 22. mis (IMS) | 7 | 6033 | 0.13 | 0.24 | 0.51 | 1.08 | 380.34 | 1650.04 | 540 |
| 23. mst (IMS) | 53 | 217334 | 2 | 3.21 | 3.28 | 5.57 | 899.43 | 3184.65 | 48387 |
| 24. vertex-coloring (IMS) | 13 | 13370 | 0.26 | 0.46 | 0.91 | 1.82 | 504.2 | 2017.47 | 1954 |

Fig. 24. Evaluation numbers for performing the `BarrElim` optimization passes on various benchmark programs, in the context of RPINV mode of stabilization. These numbers serve as the baseline for Section 7.3. Abbreviations: *#Trig*=number of stabilization triggers, *#T-Func*=number of transfer-function applications, and *Ch-Size*=total size of the change-sequence across all stabilization triggers (as a sum of the number of nodes and twice the number of edges in each trigger). *STB-Time* and *Total Time* refer to the IDFA-stabilization time, and overall compilation time, respectively. *Memory* refers to the maximum resident set size of the compilation process when running the `BarrElim` optimization passes. Note that *Ch-Size* is calculated through a special run of the RPINV mode, where we force retention of the change-sequence, even though the mode does not require it.

We have also done an elaborate evaluation compared to the weaker baseline of the eager EG-modes of stabilization, which correspond to naïve (that is, easy but inefficient) manual stabilizations; the details can be found in Appendix A.

**(A) Stabilization time.** In Fig. 24, columns 4 and 5 show the stabilization time for IDFA flow-maps in the context of RPINV mode of stabilization, while performing the set of optimization passes of `BarrElim` on the benchmark programs under study, for Nanda and K2, respectively. The zero (0) entries for the STB-times of IS are due to the fact that no optimization opportunities were found by the `BarrElim` passes for IS.

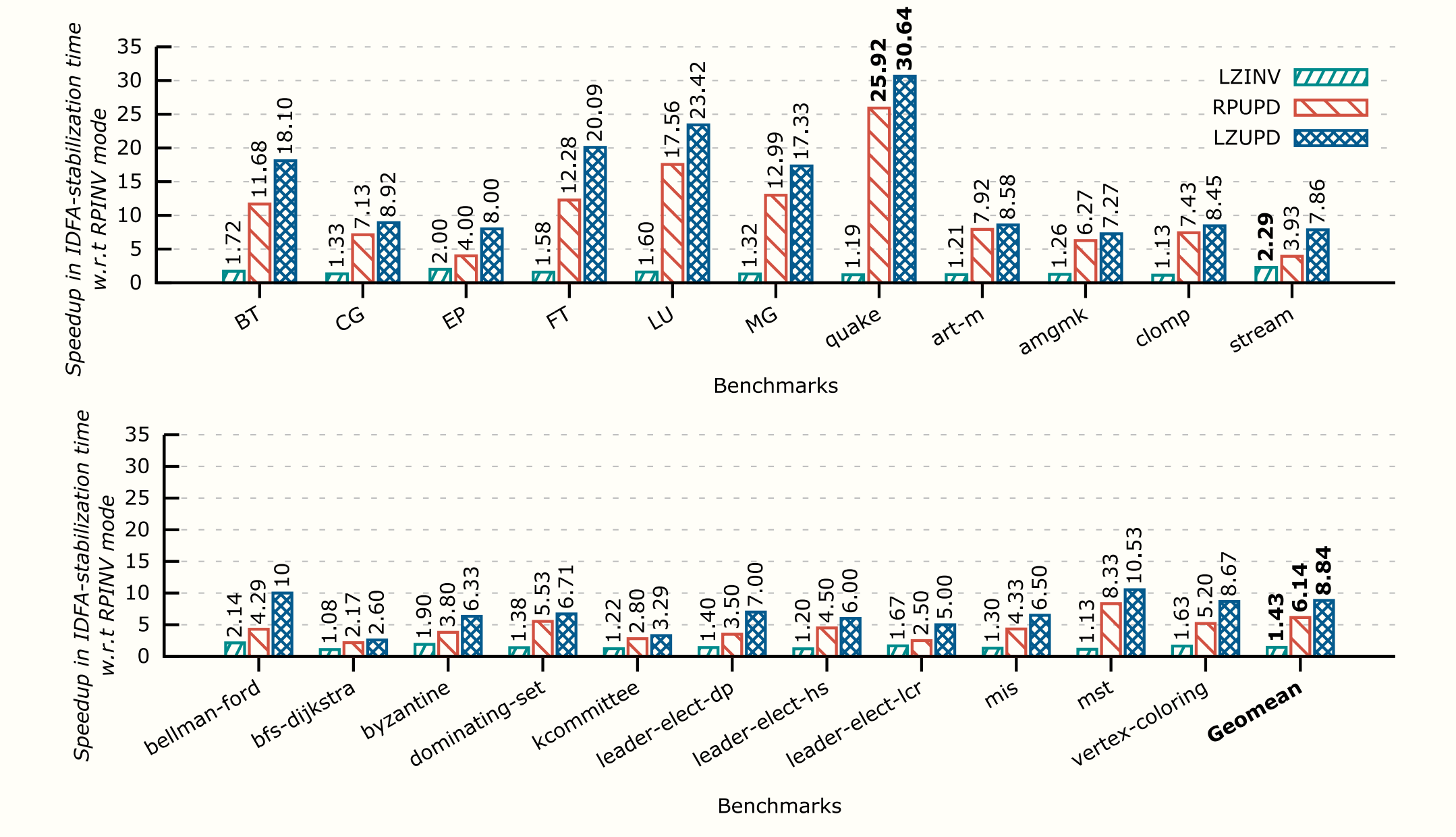

(a) Nanda

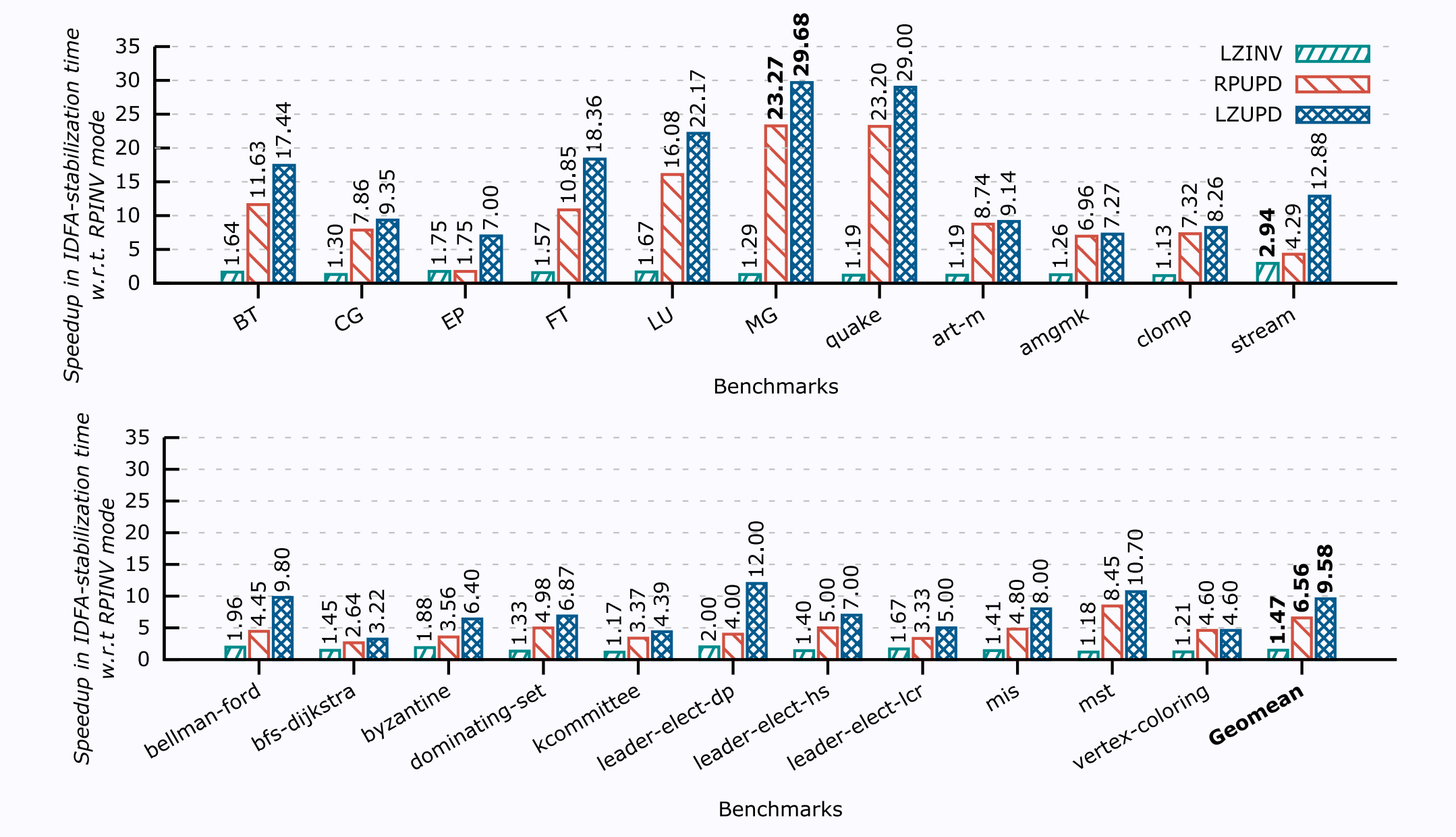

(b) K2

Fig. 25. Speedup in IDFA stabilization-time under various modes of stabilization with respect to the RPINV mode, when applying the client optimization passes of BarrElim. Higher is better.

In Fig. 25, we illustrate the impact of using LZINV, RPUPD, and LZUPD modes of stabilization, by showing their relative speedups with respect to RPINV, in terms of speedups in the IDFA stabilization-time. We exclude SP from this discussion on stabilization time, since there were no stabilization triggers for SP under lazy modes of stabilizations (that is, the stabilization time for lazy modes of stabilization was zero). As expected, the LZUPD mode incurs the least cost for stabilization among all the cases across both the platforms; consequently, it results in the maximum speedup with respect to RPINV – with speedups varying between 2.6× and 30.64× (geomean 8.84×) for Nanda; between 3.22× and 29.68× (geomean 9.58×) for K2.

We have noted that the gains in IDFA stabilization time using a particular mode of stabilization depend on multiple stabilization-mode-specific factors, such as (i) number of triggers of stabilization, (ii) number of times transfer-functions are applied on various program nodes during IDFA stabilization, (iii) cost incurred to process each program node per stabilization, and so on. For our baseline mode of stabilization, RPINV, we show the first two factors in columns 2 and 3 of Fig. 24, respectively. We do not report the numbers for the third factor as it varies significantly based on the different types of statements in the benchmark, and hence is difficult to summarize per benchmark. In Fig. 26, we show the number of stabilization triggers when using the lazy modes of stabilization normalized with respect to the RP-modes. Similarly, in Fig. 27, for various modes of stabilization we show the number of transfer-function applications across all stabilization triggers normalized with respect to those numbers in the case of RPINV mode. We observe that for each benchmark, the speedup obtained across different modes of stabilization closely correlates to these numbers. Across different benchmarks, the relative impact of these factors may vary, depending upon, for instance, the fraction of time spent in applying transfer-function when processing a program node during IDFA stabilization. We now illustrate our observations by comparing the performance of different modes of stabilization.

**LZUPD vs. RPINV.** The LZUPD mode consistently outperforms the RPINV mode across all benchmarks, for both the platforms, as shown in Fig. 25. This can be attributed to the significant reductions in the number of transfer-function applications, as well as in the number of stabilization triggers, when using the LZUPD mode as compared to the RPINV mode, as shown in Fig. 27 and Fig. 26, respectively. The maximum speedup in IDFA stabilization-time for LZUPD was observed in quake for Nanda (30.64×), and in MG for K2 (29.68×), consequent upon the fact that for quake and MG, compared to RPINV, LZUPD re-processes a significantly small fraction of the nodes (1.18% and 0.25%, respectively), over a reduced number of stabilization triggers (82.86% and 75.22%, respectively). In contrast, for bfs-dijkstra, LZUPD attains the least (though still quite significant) speedups of 2.6× in Nanda, and 3.22× in K2, as it results in reprocessing one of the highest fractions of nodes (9.03%), along with a higher count of stabilization triggers (76.47%), as compared to the RPINV mode, across all the benchmarks.

**LZUPD vs. RPUPD.** It is clear from Fig. 25 that though the RPUPD mode consistently performs better than RPINV, and compared to LZUPD it performs either worse or similarly, across all the benchmarks, for both the platforms. This is because, as shown in Fig. 26 and Fig. 27, LZUPD results in significantly fewer stabilization triggers (geomean 36.52% lower) than the RPUPD mode, while also reprocessing considerably fewer nodes (geomean 8.41% lower).

**LZUPD vs. LZINV.** As shown in Fig. 25, we see that in the context of IDFA stabilization-time, LZUPD performs better than LZINV for all the benchmarks (geomean 6.18× better for Nanda; geomean 6.51× better for K2). This observation can be attributed to the fact that the cost of invalidating and recomputing a program-abstraction (such as the results of any instantiation of HIDFA) is, in general, higher than the cost of incrementally updating the program-abstraction. This is evident from Fig. 27, which shows that the LZUPD mode reprocesses only a small fraction of the program nodes reprocessed in the case of LZINV mode (geomean 91.19% fewer).

**LZINV vs. RPINV.** Across all the benchmarks, and both the platforms, we notice a performance improvement with LZINV mode, as compared to the RPINV mode, with geomean speedup as 1.43×, for Nanda, and 1.47×, for K2. The maximum speedup was observed in the case of stream (2.29× in

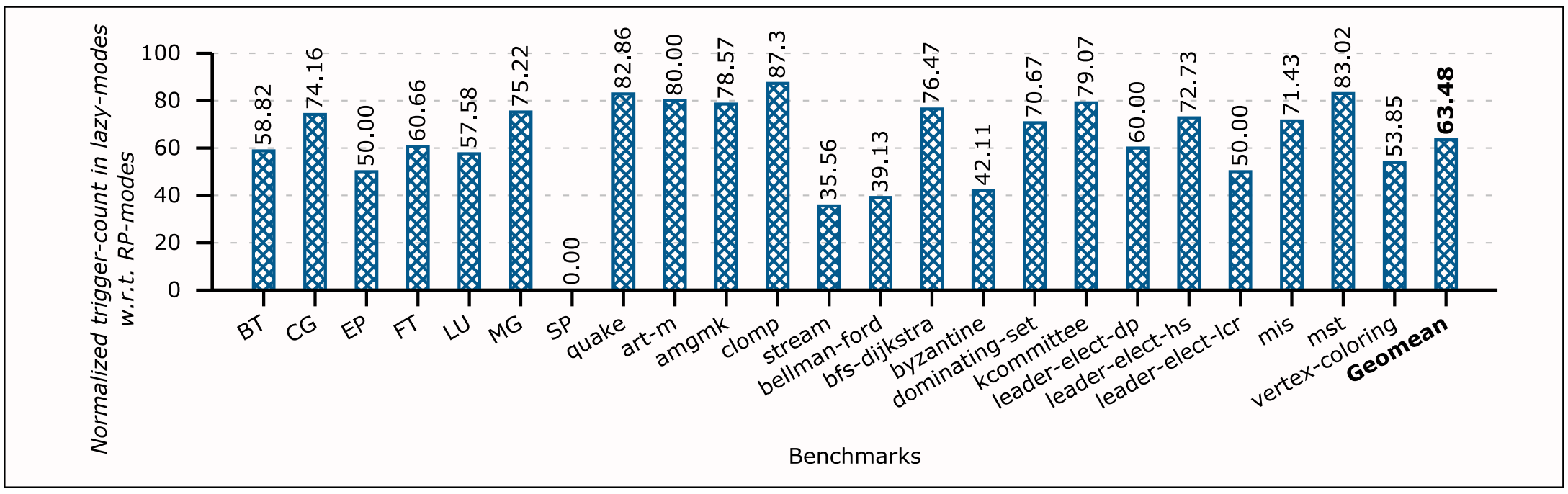

Fig. 26. Number of times IDFA stabilization was triggered under lazy modes of stabilization when performing the BarrElim optimization passes, normalized with respect to the number of triggers in case of RP-modes. Lower is better.

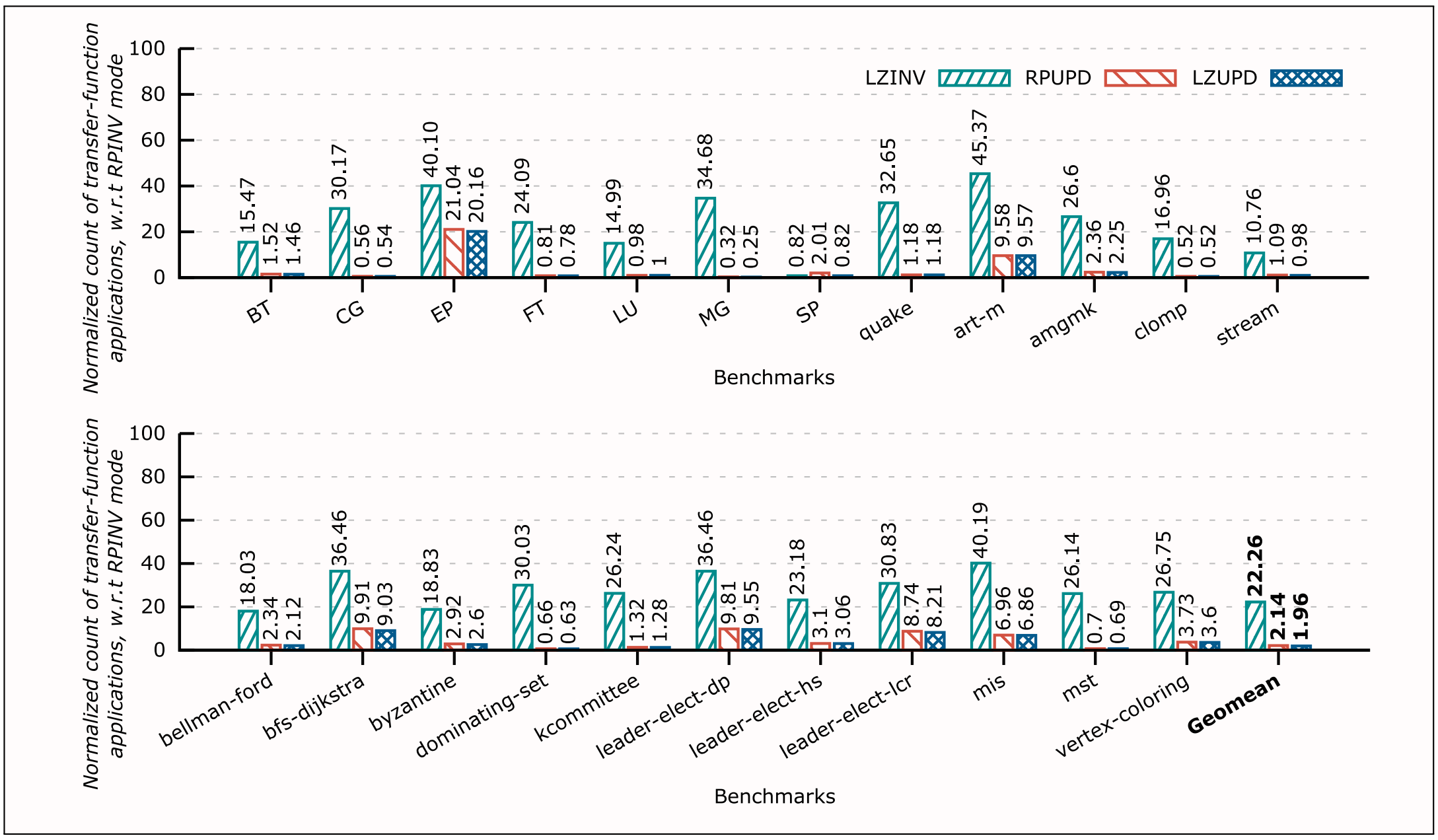

Fig. 27. Total number of applications of transfer-functions across all stabilization triggers when performing the BarrElim optimization passes, for various modes of stabilization normalized with respect to the number of applications in case of the RPINV mode (set to 100). Lower is better.

Nanda, and 2.94× for K2), consequent upon the fact that among the benchmarks shown in Fig. 25, stream corresponds to the largest reduction in the number of stabilization triggers (see Fig. 26) with the lazy modes of stabilization. In contrast, clomp and mst correspond to some of the lowest reductions in number of stabilization triggers, thereby resulting in some of the lowest speedups with the LZINV mode.

**LZINV vs. RPUPD.** From Fig. 25, note that the RPUPD mode consistently outperforms the LZINV mode in the context of IDFA-stabilization time, across all the benchmarks, for both the platforms (except for EP in K2, where the performance is the same). This trend can be explained as follows. From Fig. 26, it is clear that the LZINV mode consistently results in fewer stabilization triggers than the RPUPD mode. However, it is also expected that the incremental update of IDFA flow-maps in UPD-modes will result in fewer applications of the transfer-functions per trigger, as compared to the same in the case of INV-modes which involve invalidation and full recomputation of the IDFA flow-maps. This is evident in Fig. 26, where we notice that RPUPD mode consistently applies significantly fewer transfer-functions (geomean 90.38% fewer) as compared to the LZINV mode. As a result, RPUPD mode outperforms the LZINV mode. For the case of EP on K2, where we observe similar performance for both these modes, we believe that the performance gains expected due to 47.53% reduction in the number of transfer-function applications in case of RPUPD as compared to the LZINV mode (which is also the least of such reductions across all the benchmarks), are negated by the other overheads resulting from double the number of stabilization triggers in case of RPUPD as compared to that in LZINV.

*Summary.* Overall, we found that the LZUPD mode leads to the maximum benefits in stabilization time, across all the four modes of stabilization. Further, we also observed that if the UPD-mode is unavailable for a program-abstraction due to the absence of its incremental-update algorithm, the fully-automated stabilization provided by *Homeostasis* in the form of LZINV mode can still provide good performance improvements over the manual stabilization attained using the RPINV mode. These observations underline the performance benefits in stabilization time when using *Homeostasis*. This may in turn improve the overall compilation speed, as discussed next.

**(B) Total compilation time.** In Fig. 24, columns 6 and 7 show the overall compilation time incurred in the context of RPINV mode of stabilization, for Nanda and K2, respectively. These numbers correspond to an end-to-end compilation that starts with parsing of the input program file, followed by the application of various default simplification/normalization passes employed by IMOP, and ends with the application of the BarrElim set of optimization passes until fixed-point, before the final (optimized) program is written to the disk.

In Fig. 28, we demonstrate the impact of using the lazy modes of stabilization, as compared to the RPINV mode of manual stabilization, in terms of speedup obtained in the total compilation time. Both LZINV and LZUPD perform consistently better than the baseline across all the benchmarks, with maximum and geomean speedups of: 5.58× and 1.25× for Nanda with LZINV; 5.06× and 1.23× for K2 with LZINV; 8.95× and 2.36× for Nanda with LZUPD; and 11.32× and 2.21× for K2 with LZUPD mode of stabilization. We have also observed that LZUPD is never slower than RPUPD mode of stabilization – we skip the plot for RPUPD and the detailed analysis for brevity.

We observe that the speedups obtained in total compilation time using different modes of stabilization, with respect to a baseline mode, depends on the following key factors: (i) fraction of the total compilation time spent in performing the stabilization using the baseline mode (see Fig. 24: columns 4 and 6 for Nanda; columns 5 and 7 for K2); (ii) overall speedup obtained in the stabilization process (shown in Fig. 25); and (iii) changes in memory footprint and available heap size for the VM across the compilation process, both of which may non-deterministically affect the GC overheads. We now illustrate these observations by comparing the relative performances of various modes of stabilization.

**LZUPD vs. RPINV.** In Fig. 28, we observe that LZUPD consistently outperforms RPINV, as expected. The greatest speedups in total compilation time are obtained for MG (8.95× in Nanda, and 11.32× in K2), owing to the facts that MG has (i) one of the highest fractions of total compilation time spent in stabilization (72.38% in Nanda; 78.48% in K2; see Fig. 24), and (ii) one of the highest speedups in stabilization-time (17.33× in Nanda; 29.68× in K2), across all the benchmarks. In contrast, since EP corresponds to the least fraction of total compilation time spent in stabilization (7.14% in Nanda; 5.98% in K2), across all the benchmarks, as well as (comparatively) only moderate speedup in stabilization-time (8× in Nanda; 7× in K2), it obtains the least speedup in total compilation time (1.10× in Nanda; 1.09× in K2).

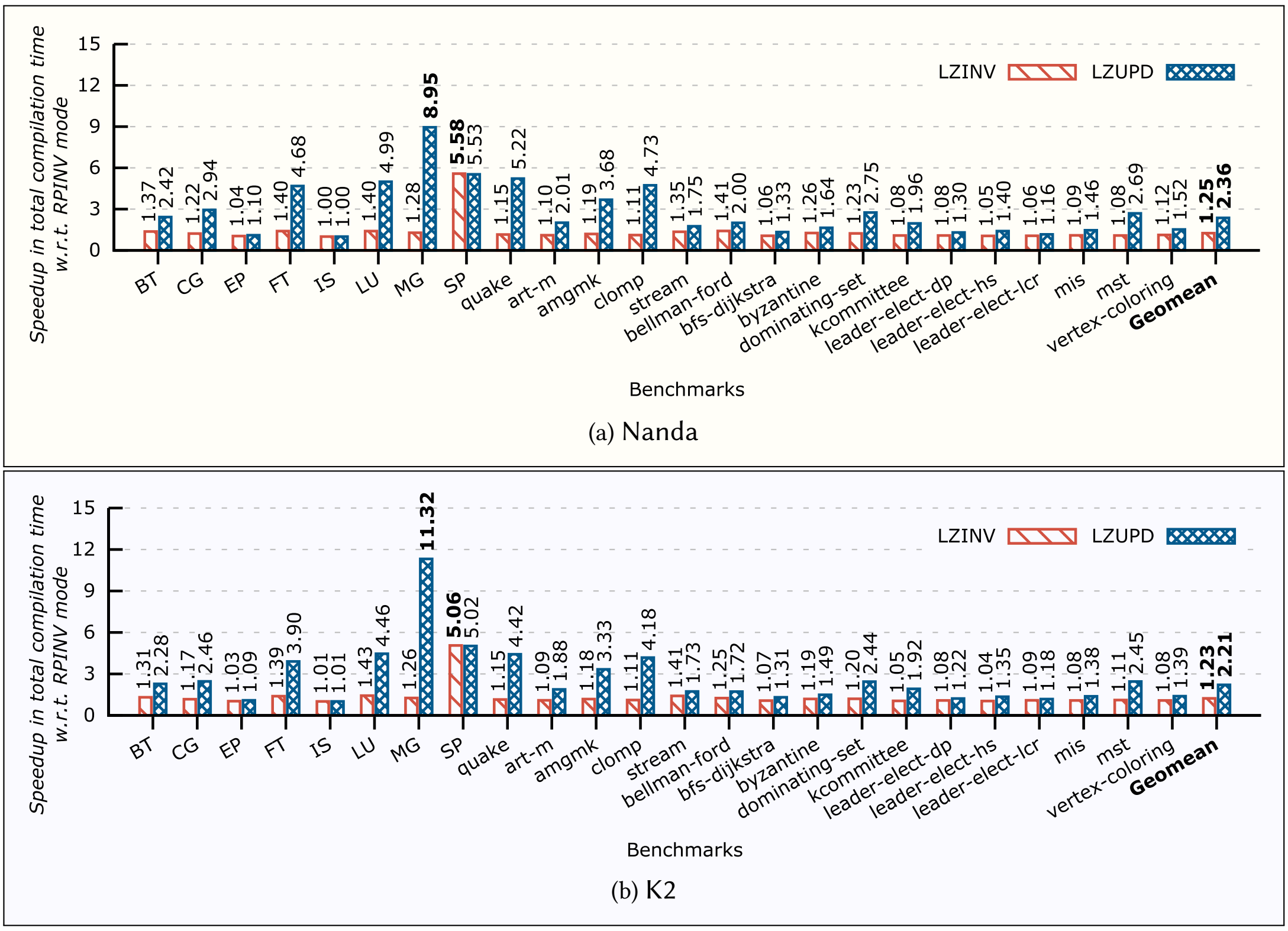

(a) Nanda

(b) K2

Fig. 28. Speedup in total compilation time under both the lazy modes of stabilization facilitated by *Homeostasis*, with respect to the RPINV mode, when applying the BarrElim set of optimizations. Higher is better.

**LZINV vs. RPINV.** Across all the benchmarks, and for both the platforms, LZINV results in better total-compilation time than RPINV (geomean 2.36× in Nanda; 2.21× in K2). In Fig. 26, note that the number of stabilization-triggers is reduced to zero in the case of SP, when using the lazy modes of stabilization; this results in SP attaining the highest gains in total compilation time with LZINV (5.58× in Nanda; 5.06× in K2), among all the benchmarks. In contrast, for similar reasons as in the case of LZUPD, EP corresponds to the least speedup in total compilation time when using LZINV, for both the platforms (1.04× in Nanda; 1.03× in K2).

**LZUPD vs. LZINV.** In Fig. 28, LZUPD consistently performs better than LZINV (except for SP, where the performance difference is negligible), across both the platforms – maximum and geomean values are 6.98× and 1.88× in Nanda, respectively; 8.98× and 1.79× in K2, respectively. We attribute this observation to the fact that, as shown in Fig. 27, the number of applications of the transfer-functions in the case of LZUPD is smaller than that with LZINV for all the benchmarks (except for SP, where the difference is zero); note that the number of stabilization triggers remain the same between both the modes. The maximum speedups in both the platforms are obtained for MG, consequent upon the fact that MG corresponds to the largest drop in the number of stabilization-triggers (138.72×, in Fig. 27) from LZINV to LZUPD, across all the benchmarks.

An interesting observation from our study was that across both the lazy modes of stabilization, the actual benefits obtained in the total compilation time, with respect to the RPINV mode of stabilization, were far higher than the speedups expected in the total compilation time (based on the fraction of the total compilation time spent in the stabilization process, and the speedup within the stabilization process). For instance, in the case of MG in K2, when using the RPINV mode of stabilization (i) around 78% of the total compilation time was spent in performing the IDFA-stabilization (as shown in Fig. 24), and (ii) this fraction of the time spent in IDFA-stabilization process benefitted from the LZUPD mode with an approximate speedup of 30× (shown in Fig. 25) over the RPINV mode. This should result in a speedup of 4.14× (= $1/(1 - 0.78) + (0.78/30)$) in the total compilation time. However, as shown in Fig. 28, the actual speedup observed in MG for K2 was 11.32×! We believe that this behaviour can be attributed to the latent benefits obtained in other parts of the compilation (that is, outside the stabilization process itself), owing to a significant reduction in the memory usage across most of the benchmarks, as discussed in part (C) of this section.

*Summary.* In conclusion, the lazy modes of stabilization used in *Homeostasis* may result in significant gains in the overall compilation time, as compared to the manual stabilization (RPINV mode). Further, the UPD-mode consistently outperforms the INV-mode, between both the lazy modes of stabilization facilitated by *Homeostasis*. These gains in lazy modes are on top of the benefits obtained in terms of ease-of-use (see Section 7.2) due to the nearly-automated nature of *Homeostasis*, requiring minimum (in case of LZUPD) or zero (in case of LZINV) additional effort from the compiler developers.

**(C) Memory consumption.** In Fig. 24 (columns 8 and 9) we show the maximum memory footprint (in MB), in terms of the maximum resident set size, while running the BarrElim optimization passes in the context of RPINV mode, for Nanda and K2, respectively. The values shown are calculated with the help of /usr/bin/time GNU utility (version : 1.7). While this tool is quite effective in drawing a broad picture of the peak memory requirements of a process, there are two main limitations to using this tool. First, the resident set size is not measured continuously, but rather at some indeterminate intervals. This means that the tool might not calculate the actual peak memory usage of the process. Second, the maximum resident set size does not include the size of pages that have been swapped out, which may not accurately reflect the true memory requirements of the process. In addition, the non-deterministic nature of the garbage collector (GC) implemented by the Java Virtual Machine (JVM) can also affect the accuracy of these measurements. For example, a process with higher memory demands may result in the triggering of a full GC, in which the JVM is paused to clean up the garbage from all memory regions of the JVM heap. This can result in a much smaller heap size compared to a process with lower memory demands that does not trigger a full GC. Considering these sources of imprecision in such a measurement, we believe that very small improvements or deteriorations (< 5%) should be ignored.

In Fig. 29, we illustrate the percentage savings in the memory footprint by LZINV, RPUPD, and LZUPD modes, as compared to the RPINV mode, for both Nanda and K2. All these three modes of stabilization perform better (or are more or less comparable), in terms of memory requirements, than the RPINV mode, except for the case of mst with LZINV in Nanda. The geomean improvements in memory consumption for LZINV, RPUPD, and LZUPD modes, as compared to the RPINV mode, are 10.67%, 34.29%, and 36.35%, respectively, for Nanda; 10.69%, 34.25%, and 35.97%, respectively, for K2.

One key factor that determines the impact of various modes of stabilization on the percentage reduction in memory footprint with respect to RPINV is the relative count of the number of times various transfer-functions were applied (see Fig. 27). As discussed above, both *lazy* and *update* options minimize the number of times different transfer functions are applied during stabilization of the data-flow analyses. Consequently, LZUPD requires the least amount of memory (within the margin of error), with maximum percentage savings over the RPINV mode – 81.78% for quake in Nanda, and 85.8% for SP in K2. In SP, the *lazy* modes (LZINV and LZUPD) do not trigger any stabilization of the data-flow analyses (i.e., zero transfer functions are applied during stabilization), attributing to high savings in memory consumption – 78.41% and 78.45%, respectively, for Nanda;

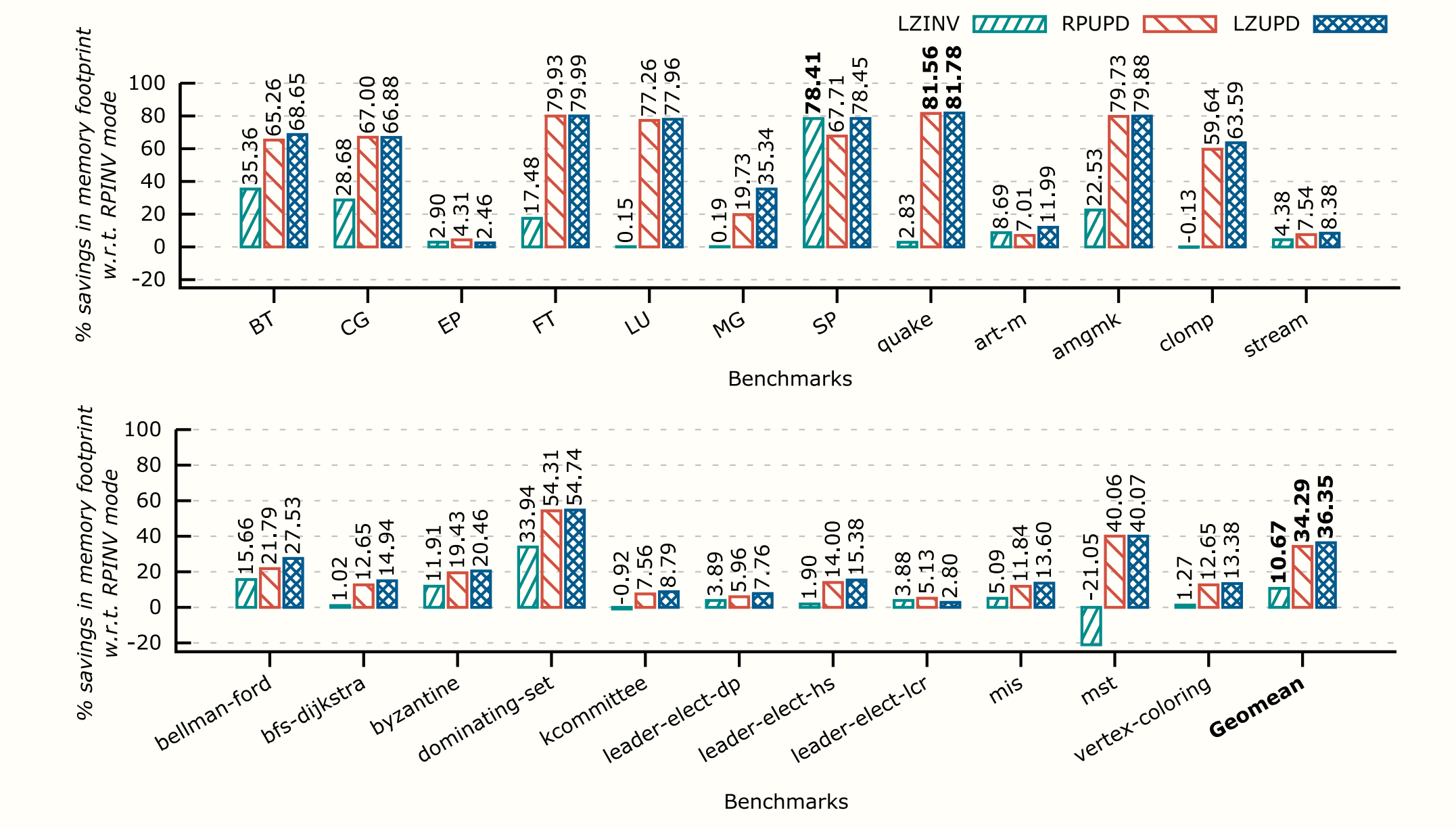

(a) Nanda

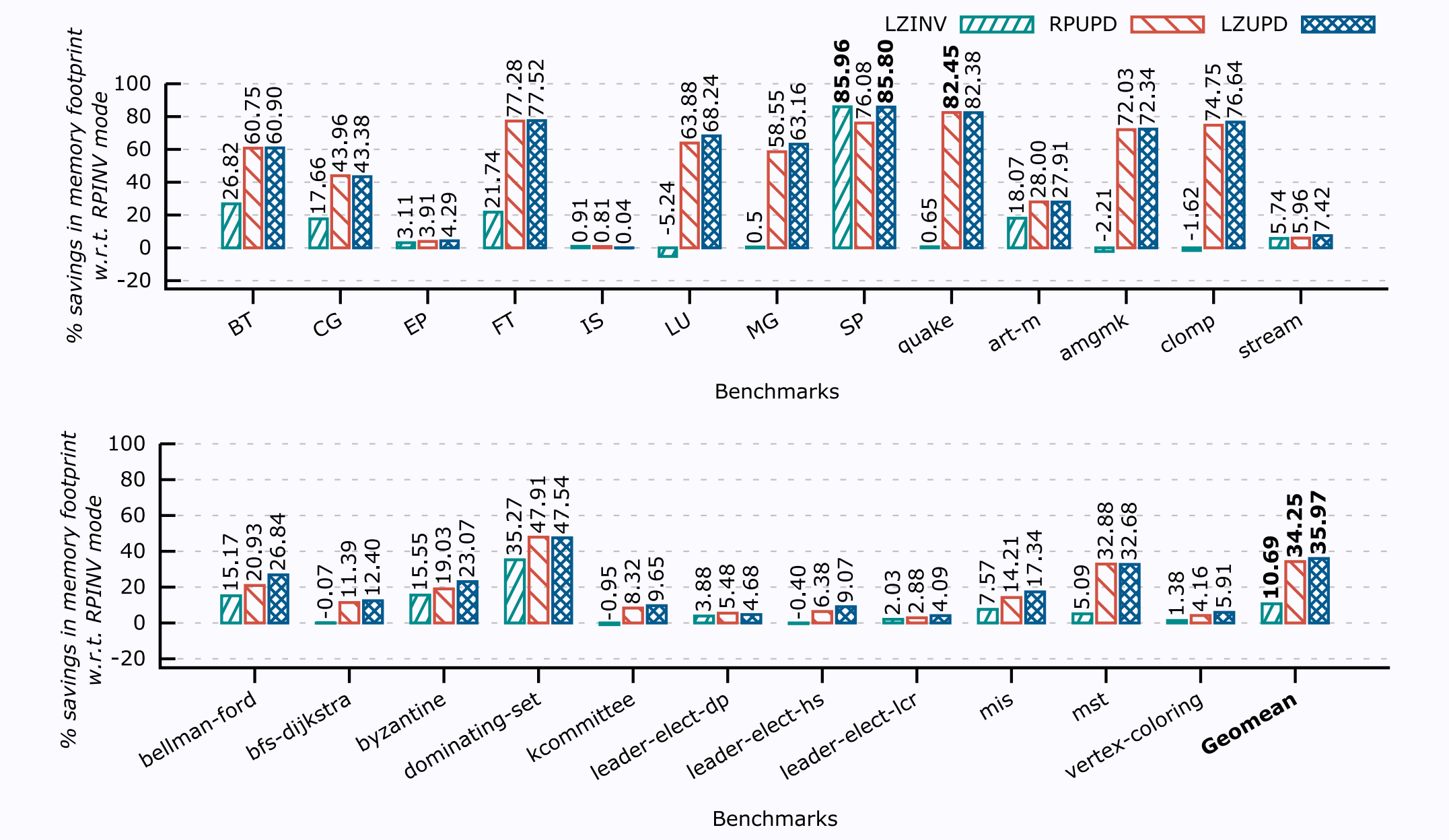

(b) K2

Fig. 29. Percentage savings in memory footprint (in terms of maximum resident set size) under various modes of self-stabilization, with respect to the RPINV mode, while running the client analysis. Higher is better.

85.96% and 85.80%, respectively, for K2. In contrast, since for both the UPD-modes (RPUPD and LZUPD), EP corresponds to the least reduction in the number of transfer-function applications as compared to the `RPINV` mode (see Fig. 27), it witnesses some of the least savings in its memory footprint with the UPD-modes. For the case of mst with `LZINV` mode, we observe an anomalous behaviour, where the memory footprint reduces by 5.09% in K2, but increases by 21.05% in Nanda. Since all the steps involved in compilation remain the same across both the platforms, we believe that such behaviour is due to the differences in the underlying architectures and operating systems, as well as in the JVMs and the GCs.

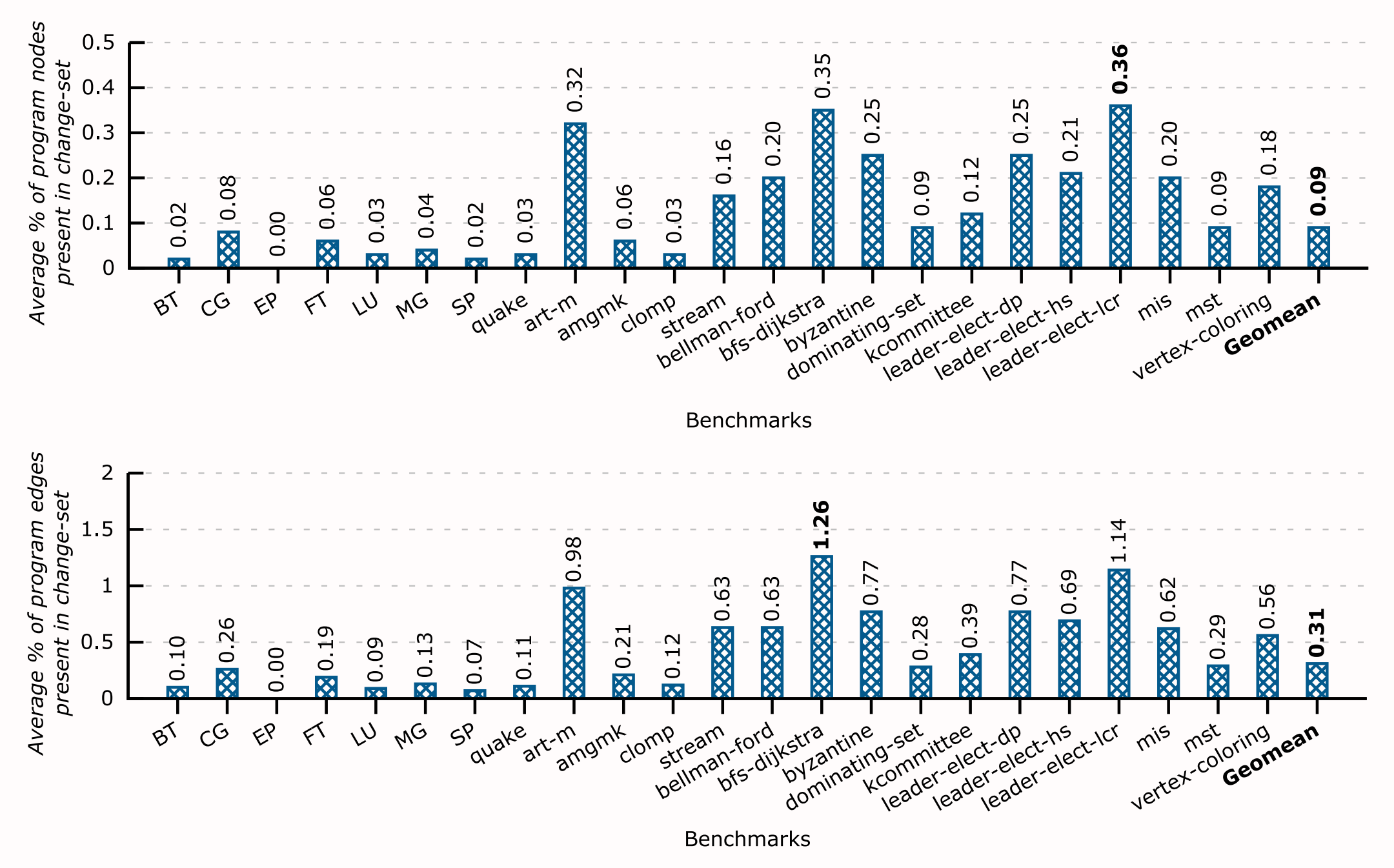

Fig. 30. Size of the compressed change-set in terms of the program nodes and edges, averaged across various stabilization triggers, when applying the set of client optimization passes from `BarrElim`. Lower is better.

*Summary.* Overall, we see that the proposed *lazy* modes of stabilization lead to significant memory savings compared to the naive `RPINV` scheme. This in turn can improve the memory traffic and overall gains in performance.

### 7.4 Impact of Compression Optimization

In this section, we empirically study the memory requirements of the change-set, which represents the program changes that need to be maintained in order to allow UPD-modes of stabilization at a later stage during compilation. Further, we also assess the efficacy of compression optimization employed by *Homeostasis* in order to keep the change-set compact.

**Memory requirements of maintaining change-set.** In Fig. 22, columns 3 and 4 show the size of each benchmark program in terms of the number of program nodes and program edges, respectively. In order to obtain the memory requirements of maintaining the change-set throughout the compilation process, we calculate its size in terms of two parameters at each stabilization-trigger: (i) fraction of the program nodes that are present in the change-set, and (ii) fraction of the program edges that are present in the change-set. Note that the size of the program itself may differ across each stabilization trigger. In Fig. 30, we show the average of these two parameters taken individually across all the stabilization triggers that occurred during the compilation. We observe that the memory requirements for maintaining the change-set are negligible in terms of the program nodes and edges, for real-world optimizations – maximum and geomean values for these fractions are 0.36% and 0.09%, respectively, for program nodes, and 1.26% and 0.31%, respectively, for program edges.

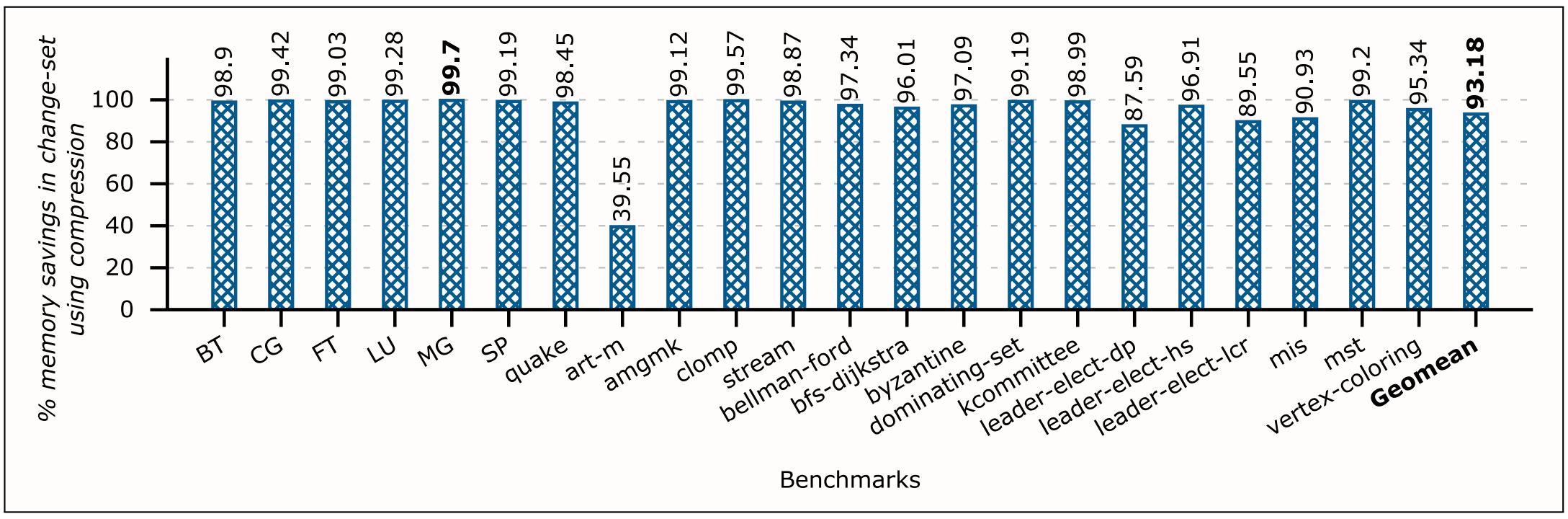

Fig. 31. Percentage reduction in the size of the uncompressed change-set with the help of compression optimization, when applying the client optimization passes from `BarrElim`. Higher is better.

**Impact of compression on change-set.** In order to study the impact of compression optimization employed by *Homeostasis* in order to keep the change-set compact, we ran a set of experiments where we disabled the compression optimization altogether. For the ease of comparison, we consider the size of a change-set to be equal to the summation of (i) its number of program nodes, and (ii) twice the number of its program edges. In Fig. 24, column 10 shows the sum of the size of the uncompressed change-set across all the stabilization triggers, when compilation was performed with compression-optimization disabled.

Recall that *Homeostasis* allows the compression routine to be invoked at any point of compilation, any number of times. For the purpose of our evaluation, we invoked the compression routine at the end of each stabilization trigger (a natural choice for performing compression, as the change-index gets updated there). The total number of stabilization triggers, and hence, the total number of invocations of the compression routine, for each benchmark program when applying the `BarrElim` set of optimization passes is shown in Column 2 of Fig. 24. In Fig. 31, we show the percentage reduction in the sum of the total size of the change-set across all the stabilization triggers, when we ran the experiments with compression-optimization enabled, as compared to the runs where the optimization was disabled. We observe that the overall reduction in memory requirements with the help of compression-optimization are significant – maximum and geomean values are 99.7% and 93.18%, respectively. The overall impact of the compression-optimization on the change-set depends on the exact transformations being performed by the client optimization pass on the program under compilation. For the case of `art-m`, we observe that the compression-optimization yielded the least benefits (39.55%). This observation is attributed to the fact that the absolute size of the uncompressed change-set itself is quite small to begin with (only 134, which is the least among all the programs).

**Impact of compression on compilation time.** We also studied the impact of compression optimization on the compilation time, and found no observable difference resulting from the optimization.

*Summary.* In conclusion, we observe that (i) the compression-optimization is quite efficient at keeping the change-set compact, and as a result, (ii) the memory requirements of the change-set are negligible in terms of the program size under compilation, thereby making the UPD-modes feasible for even large programs which may undergo aggressive optimizations.

### 7.5 Empirical Correctness

In order to empirically validate the correctness of our design/implementation of HIDFA and its six instantiations (see Section 5), we have verified the generated flow-facts under each mode of stabilization, in the context of various optimization passes provided by IMOP [Nougrahiya and Nandivada 2019]. We found that the final flow-facts across all the stabilization-modes match.

Further, to empirically validate correct stabilization in the context of the `BarrElim` optimization passes, we have also verified that for each benchmark the generated optimized code (i) does not differ across the modes of self-stabilization, and (ii) produces the same execution output as that of the unoptimized code.

**Overall evaluation summary.** Our evaluation shows that (i) *Homeostasis* makes it easy to write optimizations and program analysis passes, (ii) the lazy stabilization choices lead to faster compilation times with lesser memory requirements, (iii) the proposed compression optimization employed by *Homeostasis* is effective and leads to negligible space requirements for keeping track of the program-changes, and (iv) our implementation leads to correct analysis and optimized code.

### 7.6 Threats to Validity

In this section, we discuss various external and internal threats to the validity of our evaluation results, and how we mitigate the same.

**Threats to external validity. (1)** The evaluation results obtained for any one client-optimization pass may not depict the nature of results for other optimization passes. To mitigate the impact of this threat to validity, we have performed the evaluation with a large set of optimization passes – involving *nine* diverse optimization passes and numerous program-abstractions – which we collectively term as `BarrElim` (see Section 6.3). **(2)** The evaluation results on selected benchmark programs may not be generalizable to other kinds of input programs. To mitigate this issue, we have performed our evaluation on a set of 24 real-world programs from four popular benchmark suites (see Section 7.1). Some of these are among the largest open-source standard benchmark programs available in OpenMP C. For better coverage, the selected set also contains some medium-/small-sized standard programs. **(3)** The performance results obtained on one compiler framework may not be applicable to other compiler frameworks. In Section 7.3, we notice a strong correlation between the performance of a stabilization-mode (in terms of compilation time and memory requirement) and the number of times transfer-functions are applied when using that mode. Usage of a different underlying compiler framework is not expected to have any considerable impact on the number of transfer-function applications: this number depends on the incremental algorithm, the current program, and the change-set. Consequently, we expect similar relative-performance behaviour across various compilers.

**Threats to internal validity. (4)** For any program-abstraction, the set of relevant change-points in an optimization pass may be an over-approximation of the program-points where an expert compiler writer would manually trigger stabilization (with no redundant stabilization operations). Thus, the `RPINV` stabilization-mode may not exactly represent an expert-written stabilization (see Section 5.4). Such a threat-to-validity is unavoidable, as it would require manual analysis of large unfamiliar code of the selected set of client optimization passes as well as the numerous program-abstractions that these passes may potentially impact – these could span many tens of thousands of lines of code, making it nearly infeasible for independent experts to carry out such an exercise, owing to their busy schedules. To investigate the impact of this threat-to-validity on our evaluation, we manually compared the set of relevant change-points obtained via profiling, and the set of change-points where we (the implementers of `BarrElim` passes) would have inserted stabilization triggers for manual stabilization. We found that these two sets matched. We have made the complete codebase for our client optimization passes and related program-abstractions as open-source [Nougrahiya and Nandivada 2021] for the purpose of cross-checking and reproducing the experiments. **(5)** The measurement of memory consumption can be unreliable due to the non-deterministic behaviour of the garbage collector (GC) of Java VM. In order to mitigate the impact of GC, we report the numbers by taking a geometric mean over 30 runs of each experiment, taking inspiration from the findings of Georges et al. [2007].

## 8 RELATED WORK

We divide the related work into two different subsections: one on the efficient recompilation of programs under development, and another on incremental analyses.

### 8.1 Efficient Recompilation

When a program undergoes edits across its multiple versions, or during the process of its development in IDEs, the full recompilation of the modified program including re-computation of all the analyses and application of optimizations from scratch can be cost-prohibitive. Various approaches have been given to reuse different program-abstractions (including object code, and IR obtained before/after applying optimizations) to minimize the cost of recompilation. Smith et al. [1990] have developed mechanisms to perform incremental update of dependence information during interactive parallelization of Fortran programs. During the recompilation of programs, Pollock and Soffa [1992] incrementally incorporate the changes into globally optimized code in an attempt to reduce redundant analysis that is performed for the purpose of optimizations. For enabling incremental symbolic executions, Person et al. [2011] have provided methods to detect and characterize the effects of program changes, using static analyses. Incrementalization of static analyses that are expressed using logic programming languages, such as Prolog and Datalog, is facilitated by Eichberg et al. [2007], Szabó et al. [2018], Szabó et al. [2021], and Garcia-Contreras et al. [2018], among others. Various approaches, such as those by Szabó et al. [2018, 2016], exist to enable incremental update of static analyses in response to program edits in the context of Integrated Development Environments (IDE)s. Eichberg and Bockisch [2005] provide a constraint-solving based approach for resolving dependencies (explicitly specified) among the analyses, in Eclipse. Kloppenburg [2009] discusses the notion of incremental update for static analyses, in the context of IDEs. To the best of our knowledge, there are no generic object-oriented compiler designs or implementations that address the challenges related to stabilization requirements, and guarantee self-stabilization *during the process* of compilation, especially in the context of parallel programs.

### 8.2 Incremental Analyses

There is a vast literature on the topic of incremental update of program-abstractions of various kinds of program analyses ([Sathyanathan et al. 2017], [Nichols et al. 2019], [Yur et al. 1997], [Liu et al. 2019], [Chen et al. 2015], [Lu and Xue 2019], [Pollock and Soffa 1989], [Marlowe and Ryder 1989; Ryder et al. 1988], [Carroll and Ryder 1987, 1988], [Sreedhar et al. 1996], [Liu and Huang 2022] [Ryder 1983], [Arzt and Bodden 2014]). Though these incremental approaches provide algorithms that take a set of program changes as their argument, and stabilize the stale program-abstraction accordingly, they only address $\mathbf{Q}_3^{\mathrm{T}}$ out of the six key questions raised in Fig. 2. Further, these approaches do not provide any mechanism for efficiently maintaining the program changes either. *Homeostasis* closes this gap by providing automated techniques to address the remaining five key questions. Thus, different incremental analyses can be used along with *Homeostasis*. Under *Homeostasis*, such algorithms can be directly implemented in the `handleUpdate` method. The pass writers need not specify where to invoke the incremental stabilization and with what arguments – these tasks are conveniently automated by *Homeostasis*, for all the existing and future optimizations.

## 9 CONCLUSION

In this paper, we have presented a novel, efficient, and reliable compiler-design framework called *Homeostasis*, for enabling generic self-stabilization of the relevant program-abstractions in response to every possible transformation of the program, in the context of object-oriented compilers. *Homeostasis* decouples the program analysis and optimization passes in a compiler: using *Homeostasis*, neither the optimization writers need to write any code to stabilize the (existing or future) program-abstractions, nor do the writers of a program analysis need to know about the set of optimization passes in the compiler in order to ensure correct stabilization of the corresponding program-abstraction. We added *Homeostasis* to the IMOP compiler framework for OpenMP C programs. To illustrate the benefits of *Homeostasis*, we implemented a generic inter-thread data-flow analysis pass HIDFA, and a set of six standard IDFAs as its instantiations. We also implemented a set of optimization passes, collectively termed as `BarrElim`, which includes a set of four standard optimizations, and is used to remove redundant barriers in OpenMP programs; we did not have to add any stabilization-specific code for any of the optimization passes of `BarrElim`. Our evaluation of these passes on a set of real-world benchmarks has given us encouraging results concerning the performance and feasibility of using *Homeostasis*. While *Homeostasis* has been discussed using a Java-based compiler, we believe that the design of *Homeostasis* is generic enough to be applicable to other object-oriented compilers (including JIT compilers), for serial as well as parallel programs.

## ACKNOWLEDGMENTS

We express our gratitude to the ACM TOPLAS editor Jan Vitek, associate editor Michael G. Burke, and the anonymous reviewers for their insightful suggestions, which greatly contributed to enhancing the quality of this paper. Special appreciation goes to the members of PACE Lab (IIT Madras), for their valuable discussions related to this work. We are especially grateful to Rupesh Nasre, Manas Thakur, and Anju M A, for their thorough reviews of various drafts of this paper. Furthermore, we would also like to thank the anonymous reviewers of previous submissions of this paper, for their constructive feedback, which has helped refine this research. This work is partially supported by SERB CRG grant (sanction number CRG/2022/006971), and NSM research grant (sanction number MeitY/R&D/HPC/2(1)/2014).

## REFERENCES

Raghesh Aloor and V. Krishna Nandivada. 2015. Unique Worker Model for OpenMP. In *Proceedings of the 29th ACM on International Conference on Supercomputing (ICS '15)*. ACM, New York, NY, USA, 47–56.

Lars Ole Andersen. 1994. *Program Analysis and Specialization for the C Programming Language*. Ph.D. Dissertation. DIKU, University of Copenhagen.

Steven Arzt and Eric Bodden. 2014. Reviser: Efficiently Updating IDE-/IFDS-Based Data-Flow Analyses in Response to Incremental Program Changes. In *Proceedings of the 36th International Conference on Software Engineering (ICSE 2014)*. Association for Computing Machinery, New York, NY, USA, 288–298. https://doi.org/10.1145/2568225.2568243

Vishal Aslot, Max Domeika, Rudolf Eigenmann, Greg Gaertner, Wesley B. Jones, and Bodo Parady. 2001. SPEComp: A New Benchmark Suite for Measuring Parallel Computer Performance. In *Workshop on OpenMP Applications and Tools*, Rudolf Eigenmann and Michael J. Voss (Eds.). Springer Berlin Heidelberg, Berlin, Heidelberg, 1–10.

Rajkishore Barik, Jisheng Zhao, and Vivek Sarkar. 2013. Interprocedural Strength Reduction of Critical Sections in Explicitly-parallel Programs. In *Proceedings of the 22nd International Conference on Parallel Architectures and Compilation Techniques, Edinburgh, United Kingdom, September 7-11, 2013*. IEEE Computer Society, 29–40.

William Blume, Rudolf Eigenmann, Keith Faigin, John Grout, Jay Hoeflinger, David Padua, Paul Petersen, William Pottenger, Lawrence Rauchwerger, Peng Tu, and Stephen Weatherford. 1995. Polaris: Improving the effectiveness of parallelizing compilers. In *Languages and Compilers for Parallel Computing*, Keshav Pingali, Utpal Banerjee, David Gelernter, Alex Nicolau, and David Padua (Eds.). Springer Berlin Heidelberg, Berlin, Heidelberg, 141–154.

Neil V. Brewster and Tarek S. Abdelrahman. 2001. A Compiler Infrastructure for High-Performance Java. In *High-Performance Computing and Networking*, Bob Hertzberger, Alfons Hoekstra, and Roy Williams (Eds.). Springer Berlin Heidelberg, Berlin, Heidelberg, 675–684.

Alan Carle and Lori Pollock. 1989. Modular Specification of Incremental Program Transformation Systems. In *Proceedings of the 11th International Conference on Software Engineering (ICSE '89)*. Association for Computing Machinery, New York, NY, USA, 178–187. https://doi.org/10.1145/74587.74612

Martin D. Carroll and Barbara G Ryder. 1987. An Incremental Algorithm for Software Analysis. In *Proceedings of the Second ACM SIGSOFT/SIGPLAN Software Engineering Symposium on Practical Software Development Environments (SDE 2)*. Association for Computing Machinery, New York, NY, USA, 171–179. https://doi.org/10.1145/24208.24228

Martin D. Carroll and Barbara G Ryder. 1988. Incremental Data Flow Analysis via Dominator and Attribute Update. In *Proceedings of the 15th ACM SIGPLAN-SIGACT Symposium on Principles of Programming Languages (POPL '88)*. Association for Computing Machinery, New York, NY, USA, 274–284. https://doi.org/10.1145/73560.73584

Steven Carroll and Constantine Polychronopoulos. 2003. A Framework for Incremental Extensible Compiler Construction. In *Proceedings of the 17th Annual International Conference on Supercomputing (ICS'03)*. Association for Computing Machinery, New York, NY, USA, 53–62. https://doi.org/10.1145/782814.782824

Yuting Chen, Qiuwei Shi, and Weikai Miao. 2015. Incremental Points-to Analysis for Java via Edit Propagation. In *Structured Object-Oriented Formal Language and Method*, Shaoying Liu and Zhenhua Duan (Eds.). Springer International Publishing, Cham, 164–178.

Michael Eichberg and Christoph Bockisch. 2005. *Magellan*. Retrieved August 11, 2022 from http://www.st.informatik.tu-darmstadt.de/Magellan

Michael Eichberg, Matthias Kahl, Diptikalyan Saha, Mira Mezini, and Klaus Ostermann. 2007. Automatic Incrementalization of Prolog Based Static Analyses. In *Proceedings of the 9th International Conference on Practical Aspects of Declarative Languages (PADL'07)*. Springer-Verlag, Berlin, Heidelberg, 109–123. https://doi.org/10.1007/978-3-540-69611-7_7

Erich Gamma, Richard Helm, Ralph Johnson, and John Vlissides. 1995. *Design Patterns: Elements of Reusable Object-Oriented Software*. Addison-Wesley Longman Publishing Co., Inc., USA.

Isabel Garcia-Contreras, José F. Morales, and Manuel V. Hermenegildo. 2018. An Approach to Incremental and Modular Context-sensitive Analysis of Logic Programs. *CoRR* abs/1804.01839 (2018). http://arxiv.org/abs/1804.01839

Andy Georges, Dries Buytaert, and Lieven Eeckhout. 2007. Statistically Rigorous Java Performance Evaluation. *SIGPLAN Not.* 42, 10 (Oct. 2007), 57–76. https://doi.org/10.1145/1297105.1297033

Google. 2001. *Chrome V8*. Retrieved August 11, 2022 from https://github.com/v8/v8

Manish Gupta and Edith Schonberg. 1996. Static Analysis to Reduce Synchronization Costs in Data-Parallel Programs. In *Proceedings of the 23rd ACM SIGPLAN-SIGACT Symposium on Principles of Programming Languages (POPL '96)*. Association for Computing Machinery, New York, NY, USA, 322–332. https://doi.org/10.1145/237721.237799

Suyash Gupta and V. Krishna Nandivada. 2015. IMSuite: A Benchmark Suite for Simulating Distributed Algorithms. *J. Parallel and Distrib. Comput.* 75 (2015), 1–19. https://doi.org/10.1016/j.jpdc.2014.10.010

Suyash Gupta, Rahul Shrivastava, and V. Krishna Nandivada. 2017. Optimizing Recursive Task Parallel Programs. In *Proceedings of the International Conference on Supercomputing, ICS 2017, Chicago, IL, USA, June 14-16, 2017*, William D. Gropp, Pete Beckman, Zhiyuan Li, and Francisco J. Cazorla (Eds.). ACM, 11:1–11:11.

IBM. 2017. *Eclipse OpenJ9*. Retrieved August 11, 2022 from https://github.com/eclipse/openj9

Sven Kloppenburg. 2009. *Incrementalization of Analyses for Next Generation IDEs*. Ph.D. Dissertation. Technische Universität, Darmstadt. http://tuprints.ulb.tu-darmstadt.de/1960/

Gnanambikai Krishnakumar, Kommuru Alekhya Reddy, and Chester Rebeiro. 2019. ALEXIA: A Processor with Lightweight Extensions for Memory Safety. *ACM Trans. Embed. Comput. Syst.* 18, 6, Article Article 122 (Nov. 2019), 27 pages. https://doi.org/10.1145/3362064

Chris Lattner and Vikram Adve. 2004. LLVM: A Compilation Framework for Lifelong Program Analysis & Transformation. In *Proceedings of the International Symposium on Code Generation and Optimization (CGO '04)*. IEEE Computer Society, Washington, DC, USA.

Bozhen Liu and Jeff Huang. 2022. SHARP: Fast Incremental Context-Sensitive Pointer Analysis for Java. *Proc. ACM Program. Lang.* 6, OOPSLA1, Article 88 (Apr 2022), 28 pages. https://doi.org/10.1145/3527332

Bozhen Liu, Jeff Huang, and Lawrence Rauchwerger. 2019. Rethinking Incremental and Parallel Pointer Analysis. *ACM Trans. Program. Lang. Syst.* 41, 1, Article 6 (March 2019), 31 pages. https://doi.org/10.1145/3293606

LLVM-Developer-Community. 2019a. *LLVM GitHub Repository*. Retrieved August 11, 2022 from https://github.com/llvm/llvm-project/commit/8299fd9dee7df7c5f92ab2572aad04ce2fbbf83e

LLVM-Developer-Community. 2019b. *LLVM GitHub Repository*. Retrieved August 11, 2022 from https://github.com/llvm/llvm-project/commit/d2904ccf88e8ed487647feb90cfbf331bd888509

LLVM-Developer-Community. 2019c. *LLVM GitHub Repository*. Retrieved August 11, 2022 from https://github.com/llvm/llvm-project/commit/a95d95d3922e1a24d8b9affdd570c1d8fca00129

LLVM-Developer-Community. 2020a. *LLVM GitHub Repository*. Retrieved August 11, 2022 from https://github.com/llvm/llvm-project/commit/fa8c2ae76f7e4f498d29e2716233bd29025e8827

LLVM-Developer-Community. 2020b. *LLVM GitHub Repository*. Retrieved August 11, 2022 from https://github.com/llvm/llvm-project/commit/de92dc2850c17259090ccf644b2f2375ab8e1664

LLVM-Developer-Community. 2020c. *LLVM GitHub Repository*. Retrieved August 11, 2022 from https://github.com/llvm/llvm-project/commit/e1133179587dd895962a2fe4d6eb0cb1e63b5ee2

LLVM-Developer-Community. 2020d. *LLVM GitHub Repository*. Retrieved August 11, 2022 from https://github.com/llvm/llvm-project/commit/e2fc6a31d347dc96c2dec6acb72045150f525630

LLVM-Developer-Community. 2020e. *LLVM GitHub Repository*. Retrieved August 11, 2022 from https://github.com/llvm/llvm-project/commit/1ccfb52a6174816e450074f65e5f0929a9f046a5

LLVM-Developer-Community. 2020f. *LLVM GitHub Repository*. Retrieved August 11, 2022 from https://github.com/llvm/llvm-project/commit/e6cf796bab7e02d2b8ac7fd495f14f5e21494270

LLVM-Developer-Community. 2020g. *LLVM GitHub Repository*. Retrieved August 11, 2022 from https://github.com/llvm/llvm-project/commit/edccc35e8fa2c546e0ef1c8efde56e6b12e3c175

LLVM-Developer-Community. 2020h. *LLVM GitHub Repository*. Retrieved August 11, 2022 from https://github.com/llvm/llvm-project/commit/d6b05fccb709eb38b5b4b21901cb63825faee83e

LLVM-Developer-Community. 2020i. *LLVM GitHub Repository*. Retrieved August 11, 2022 from https://github.com/llvm/llvm-project/commit/0d90d2457c3b94760df4848941c0e7b93d07b1a2

LLVM-Developer-Community. 2021a. *LLVM GitHub Repository*. Retrieved August 11, 2022 from https://github.com/llvm/llvm-project/commit/ddc4b56eef9fec990915470069a29e70bbde3711

LLVM-Developer-Community. 2021b. *LLVM GitHub Repository*. Retrieved August 11, 2022 from https://github.com/llvm/llvm-project/commit/7c8b8063b66c7b936d41a0c4069c506669e13115

LLVM-Developer-Community. 2021c. *LLVM GitHub Repository*. Retrieved August 11, 2022 from https://github.com/llvm/llvm-project/commit/2461cdb41724298591133c811df82b0064adfa6b

Jingbo Lu and Jingling Xue. 2019. Precision-Preserving yet Fast Object-Sensitive Pointer Analysis with Partial Context Sensitivity. *Proc. ACM Program. Lang.* 3, OOPSLA, Article 148 (Oct. 2019), 29 pages. https://doi.org/10.1145/3360574

Thomas J. Marlowe and Barbara G. Ryder. 1989. An Efficient Hybrid Algorithm for Incremental Data Flow Analysis. In *Proceedings of the 17th ACM SIGPLAN-SIGACT Symposium on Principles of Programming Languages (POPL'90)*. Association for Computing Machinery, New York, NY, USA, 184–196. https://doi.org/10.1145/96709.96728

Steven S. Muchnick. 1998. *Advanced Compiler Design and Implementation*. Morgan Kaufmann Publishers Inc., San Francisco, CA, USA.

V. Krishna Nandivada, Jun Shirako, Jisheng Zhao, and Vivek Sarkar. 2013. A Transformation Framework for Optimizing Task-Parallel Programs. *ACM Trans. Program. Lang. Syst.* 35, 1, Article Article 3 (April 2013), 48 pages. https://doi.org/10.1145/2450136.2450138

Lawton Nichols, Mehmet Emre, and Ben Hardekopf. 2019. Fixpoint Reuse for Incremental JavaScript Analysis. In *Proceedings of the 8th ACM SIGPLAN International Workshop on State Of the Art in Program Analysis (SOAP 2019)*. Association for Computing Machinery, New York, NY, USA, 2–7. https://doi.org/10.1145/3315568.3329964

Emma Nilsson-Nyman, Görel Hedin, Eva Magnusson, and Torbjörn Ekman. 2009. Declarative Intraprocedural Flow Analysis of Java Source Code. *Electronic Notes in Theoretical Computer Science* 238, 5 (2009), 155–171. Proceedings of the 8th Workshop on Language Descriptions, Tools and Applications (LDTA 2008).

Aman Nougrahiya and V. Krishna Nandivada. 2019. *IMOP: IIT Madras OpenMP compiler framework*. Retrieved August 11, 2022 from https://github.com/amannougrahiya/imop-compiler

Aman Nougrahiya and V. Krishna Nandivada. 2021. *Implementation of Homeostasis in the IMOP compiler framework*. Retrieved August 11, 2022 from https://github.com/anonymousoopsla21/homeostasis

Aman Nougrahiya and V. Krishna Nandivada. 2023. *List of Mainstream and Experimental Compiler Frameworks*. https://www.cse.iitm.ac.in/~amannoug/compiler-listing.pdf

Oracle. 1999. *HotSpot*. Retrieved August 11, 2022 from https://github.com/openjdk-mirror/jdk7u-hotspot

Suzette Person, Guowei Yang, Neha Rungta, and Sarfraz Khurshid. 2011. Directed Incremental Symbolic Execution. In *Proceedings of the 32nd ACM SIGPLAN Conference on Programming Language Design and Implementation (PLDI '11)*. Association for Computing Machinery, New York, NY, USA, 504–515. https://doi.org/10.1145/1993498.1993558

Lori L. Pollock and Mary Lou Soffa. 1989. An Incremental Version of Iterative Data Flow Analysis. *IEEE Transactions on Software Engineering* 15, 12 (1989), 1537–1549.

Lori L. Pollock and Mary Lou Soffa. 1992. Incremental Global Reoptimization of Programs. *ACM Trans. Program. Lang. Syst.* 14, 2 (April 1992), 173–200. https://doi.org/10.1145/128861.128865

Daniel Quinlan, Chunhua Liao, Thomas Panas, Robb Matzke, Markus Schordan, Rich Vuduc, and Qing Yi. 2013. *ROSE User Manual: A Tool for Building Source-to-Source Translators.* Technical Report. Lawrence Livermore National Laboratory.

Thomas Reps, Tim Teitelbaum, and Alan Demers. 1983. Incremental Context-Dependent Analysis for Language-Based Editors. *ACM Trans. Program. Lang. Syst.* 5, 3 (July 1983), 449–477. https://doi.org/10.1145/2166.357218

James Rumbaugh, Michael Blaha, William Premerlani, Frederick Eddy, and William Lorensen. 1991. *Object-Oriented Modeling and Design.* Prentice-Hall, Inc., USA.

Barbara G. Ryder. 1983. Incremental Data Flow Analysis. In *Proceedings of the 10th ACM SIGACT-SIGPLAN Symposium on Principles of Programming Languages (POPL'83).* Association for Computing Machinery, New York, NY, USA, 167–176. https://doi.org/10.1145/567067.567084

Barbara G. Ryder, T.J. Marlowe, and M.C. Paull. 1988. Conditions for Incremental Iteration: Examples and Counterexamples. *Science of Computer Programming* 11, 1 (1988), 1–15.

Patrick W. Sathyanathan, Wenlei He, and Ten H. Tzen. 2017. Incremental Whole Program Optimization and Compilation. In *Proceedings of the 2017 International Symposium on Code Generation and Optimization (CGO '17).* IEEE Press, 221–232.

Seager, M. 2008. The ASC Sequoia Programming Model. (8 2008). https://doi.org/10.2172/945684

Kevin Smith, Bill Appelbe, and Kurt Stirewalt. 1990. Incremental Dependence Analysis for Interactive Parallelization. In *Proceedings of the 4th International Conference on Supercomputing (ICS '90).* Association for Computing Machinery, New York, NY, USA, 330–341. https://doi.org/10.1145/77726.255173

Vugranam C. Sreedhar, Guang R. Gao, and Yong-Fong Lee. 1996. A New Framework for Exhaustive and Incremental Data Flow Analysis Using DJ Graphs. In *Proceedings of the ACM SIGPLAN 1996 Conference on Programming Language Design and Implementation (PLDI '96).* Association for Computing Machinery, New York, NY, USA, 278–290. https://doi.org/10.1145/231379.231434

Richard M. Stallman and GCC-Developer-Community. 2009. *Using The GNU Compiler Collection: A GNU Manual For GCC Version 4.3.3.* CreateSpace, Paramount, CA.

Tamás Szabó, Gábor Bergmann, Sebastian Erdweg, and Markus Voelter. 2018. Incrementalizing Lattice-Based Program Analyses in Datalog. *Proc. ACM Program. Lang.* 2, OOPSLA, Article Article 139 (Oct. 2018), 29 pages. https://doi.org/10.1145/3276509

Tamás Szabó, Sebastian Erdweg, and Gábor Bergmann. 2021. Incremental Whole-Program Analysis in Datalog with Lattices. In *Proceedings of the 42nd ACM SIGPLAN International Conference on Programming Language Design and Implementation (PLDI 2021).* Association for Computing Machinery, New York, NY, USA, 1–15. https://doi.org/10.1145/3453483.3454026

Tamás Szabó, Sebastian Erdweg, and Markus Voelter. 2016. IncA: A DSL for the Definition of Incremental Program Analyses. In *Proceedings of the 31st IEEE/ACM International Conference on Automated Software Engineering (ASE 2016).* Association for Computing Machinery, New York, NY, USA, 320–331. https://doi.org/10.1145/2970276.2970298

Chau-Wen Tseng. 1995. Compiler Optimizations for Eliminating Barrier Synchronization. In *Proceedings of the Fifth ACM SIGPLAN Symposium on Principles and Practice of Parallel Programming (PPOPP '95).* Association for Computing Machinery, New York, NY, USA, 144–155. https://doi.org/10.1145/209936.209952

Raja Vallée-Rai, Phong Co, Etienne Gagnon, Laurie Hendren, Patrick Lam, and Vijay Sundaresan. 2010. Soot: A Java Bytecode Optimization Framework. In *CASCON First Decade High Impact Papers (CASCON '10).* IBM Corp., USA, 214–224. https://doi.org/10.1145/1925805.1925818

Rob F Van der Wijngaart and Parkson Wong. 2002. *NAS Parallel Benchmarks Version 3.0.* Technical Report. NAS technical report, NAS-02-007.

Jyothi Krishna Viswakaran Sreelatha and Shankar Balachandran. 2016. Compiler Enhanced Scheduling for OpenMP for Heterogeneous Multiprocessors. In *Workshop on Energy Efficiency with Heterogeneous Computing (EEHCO '16).* ACM, Prague, Czech Republic.

Jyothi Krishna Viswakaran Sreelatha, Shankar Balachandran, and Rupesh Nasre. 2018. CHOAMP: Cost Based Hardware Optimization for Asymmetric Multicore Processors. *IEEE Trans. Multi-Scale Computing Systems* 4, 2 (2018), 163–176.

Jyothi Krishna Viswakaran Sreelatha and Rupesh Nasre. 2018. Optimizing Graph Algorithms in Asymmetric Multicore Processors. *IEEE Trans. on CAD of Integrated Circuits and Systems* 37, 11 (2018), 2673–2684.

Jyh-Shiarn Yur, Barbara G. Ryder, William A. Landi, and Phil Stocks. 1997. Incremental Analysis of Side Effects for C Software System. In *Proceedings of the 19th International Conference on Software Engineering (ICSE '97).* Association for Computing Machinery, New York, NY, USA, 422–432. https://doi.org/10.1145/253228.253369

| 1 | | 2 | 3 | 4 | 5 | 6 | 7 |
|---|---|---|---|---|---|---|---|
| **Benchmark** | | **#Trig** | **#T-Func** | **STB-Time** (s) | | **Total Time** (s) | |
| | | | | Nanda | K2 | Nanda | K2 |
| 1. | BT (NPB) | 77 | 842396 | 12.3 | 22.56 | 18.92 | 34.45 |
| 2. | CG (NPB) | 226 | 601627 | 4.89 | 8.16 | 6.76 | 11.2 |
| 3. | EP (NPB) | 4 | 3304 | 0.07 | 0.15 | 0.6 | 1.28 |
| 4. | FT (NPB) | 153 | 1108996 | 9.54 | 16.8 | 14.98 | 26.28 |
| 5. | IS (NPB) | 0 | 806 | 0 | 0 | 0.33 | 0.8 |
| 6. | LU (NPB) | 139 | 1998450 | 29.65 | 58 | 35.07 | 67.68 |
| 7. | MG (NPB) | 540 | 5097722 | 52.46 | 129.09 | 70.44 | 158.95 |
| 8. | SP (NPB) | 122 | 2252994 | 24.86 | 46.52 | 30.18 | 55.67 |
| 9. | quake (SPEC) | 91 | 1115729 | 15.11 | 25.98 | 17.81 | 30.56 |
| 10. | art-m (SPEC) | 10 | 161550 | 1.92 | 3.63 | 3.39 | 6.51 |
| 11. | amgmk (Sequoia) | 161 | 714256 | 7.69 | 16.17 | 12.73 | 24.53 |
| 12. | clomp (Sequoia) | 330 | 2048957 | 30.57 | 54.78 | 54.14 | 95.16 |
| 13. | stream (Sequoia) | 141 | 158386 | 1.41 | 2.57 | 2.27 | 4.22 |
| 14. | bellman-ford (IMS) | 55 | 34775 | 0.63 | 1.17 | 1.27 | 2.44 |
| 15. | bfs-dijkstra (IMS) | 37 | 10037 | 0.42 | 0.71 | 0.89 | 1.76 |
| 16. | byzantine (IMS) | 47 | 19061 | 0.51 | 0.91 | 0.98 | 1.96 |
| 17. | dominating-set (IMS) | 180 | 442560 | 3.97 | 6.34 | 5.74 | 9.04 |
| 18. | kcommittee (IMS) | 131 | 134113 | 1.38 | 2.3 | 2.8 | 4.68 |
| 19. | leader-elect-dp (IMS) | 10 | 4260 | 0.29 | 0.5 | 0.61 | 1.3 |
| 20. | leader-elect-hs (IMS) | 40 | 32761 | 0.41 | 0.7 | 1.15 | 2.26 |
| 21. | leader-elect-lcr (IMS) | 12 | 3422 | 0.08 | 0.16 | 0.58 | 1.2 |
| 22. | mis (IMS) | 14 | 11315 | 0.38 | 0.68 | 0.8 | 1.58 |
| 23. | mst (IMS) | 169 | 687945 | 5.82 | 8.75 | 7.54 | 11.57 |
| 24. | vertex-coloring (IMS) | 29 | 28527 | 0.44 | 0.81 | 1.17 | 2.27 |

Fig. 32. Evaluation numbers for performing the BarrElim set of optimizations on various benchmark programs, in the context of EGINV mode of stabilization. These numbers serve as the baseline for Section A. Abbreviations: *#Trig*=number of stabilization triggers, *#T-Func*=number of transfer-function applications, and *STB-Time* and *Total Time* refer to the IDFA-stabilization time, and overall compilation time, respectively.

# A PERFORMANCE COMPARISON WITH EAGER MODES OF STABILIZATION (EGINV AND EGUPD)

We now present an evaluation describing the impact of the lazy modes of self-stabilization on compilation while running the BarrElim set of optimizations, taking the weaker baseline of EGINV mode, where the stabilization is performed eagerly at the end of each elementary transformation. Note that such a scheme of EGINV stabilization is easy to automate, but can be quite inefficient, as we show in this section. We present our evaluation in the context of parameters related to stabilization time, and total compilation time. In Fig. 32, we present various parameters related to compilation of the benchmark programs under study, when compiled with EGINV mode of stabilization; these numbers serve as the baseline for the evaluations discussed in this section.

In Fig. 32, columns 4 and 5 show the stabilization time for IDFA flow-maps in the context of EGINV mode of stabilization, while performing the BarrElim set of optimization passes on the benchmark programs under study, for Nanda and K2, respectively. The zero (0) entries for the STB-times of IS are due to the fact that no optimization opportunities were found by BarrElim passes for IS.

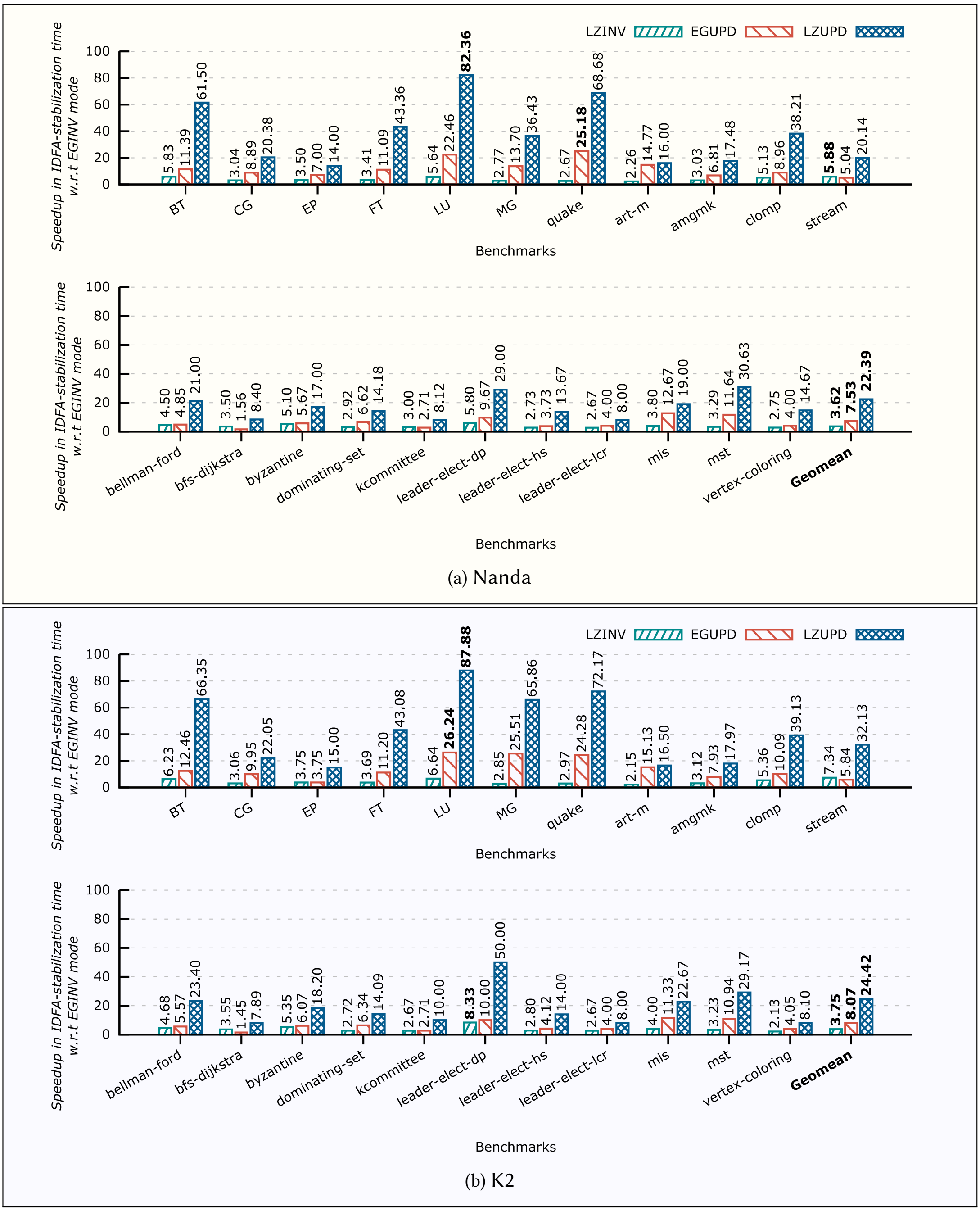

(a) Nanda

(b) K2

Fig. 33. Speedup in IDFA stabilization-time under various modes of stabilization with respect to the EGINV mode, when applying the set of client optimizations, BarrElim. Higher is better.

In Fig. 33, we illustrate the impact of using LZINV, EGUPD, and LZUPD modes of stabilization, by showing their relative speedups with respect to EGINV, in terms of speedups in the IDFA stabilization-time. We exclude SP from this discussion on stabilization time, since there were no stabilization triggers for SP under lazy modes of stabilizations (that is, the stabilization time for lazy modes of stabilization was zero). As expected, the LZUPD mode incurs the least cost for stabilization among all the cases across both the platforms; consequently it results in the maximum speedup with respect to EGINV – with speedups varying between 8.00× and 82.36× (geomean 22.39×) for Nanda; between 7.89× and 87.88× (geomean 24.42×) for K2.

We have noted that the gains in IDFA stabilization time using a particular mode of stabilization depend on multiple stabilization-mode-specific factors, such as (i) number of triggers of stabilization, (ii) number of times transfer-functions are applied on various program nodes during IDFA stabilization, (iii) cost incurred to process each program node per stabilization, and so on. For our baseline mode of stabilization, EGINV, we show the first two factors in columns 2 and 3 of Fig. 32, respectively. In Fig. 34, we show the number of stabilization triggers when using the lazy modes of stabilization normalized with respect to the EG-modes. Similarly, in Fig. 35, for various modes of stabilization we show the number of transfer-function applications across all stabilization triggers normalized with respect to those numbers in case of EGINV mode. We observe that for each benchmark, the speedup obtained across different modes of stabilization closely correlates to the these numbers. Across different benchmarks, the relative impact of these factors may vary, depending upon, for instance, the fraction of time spent in applying transfer-function when processing a program node during IDFA stabilization. We now illustrate our observations by comparing the performance of different modes of stabilization.

**LZUPD vs. EGINV.** The LZUPD mode consistently outperforms the RPINV mode across all benchmarks, for both the platforms, as shown in Fig. 33. This can be attributed to the significant reductions in the number of transfer-function applications, as well as in the number of stabilization triggers, when using the LZUPD mode as compared to the EGINV mode, as shown in Fig. 35 and Fig. 34, respectively. The maximum speedup in IDFA stabilization-time for LZUPD was observed in LU (82.36× in Nanda, and 29.68× in K2), consequent upon the fact that for LU, compared to EGINV, LZUPD re-processes significantly small fraction of the nodes (3.93%) over a reduced number of stabilization triggers (13.67%). In contrast, for leader-elect-lcr and bfs-dijkstra, LZUPD leads to some of the least (though still quite significant) speedups, as it results in reprocessing high fractions of nodes (15.24% and 19.30%, respectively), along with a high count of stabilization triggers (25% and 35.14%, respectively), as compared to the EGINV mode.

**LZUPD vs. EGUPD.** It is clear from Fig. 33 that though the EGUPD mode consistently performs better than EGINV, compared to LZUPD it performs consistently worse, across all the benchmarks, for both the platforms. This is because, as shown in Fig. 34 and Fig. 35, LZUPD results in significantly fewer stabilization triggers (geomean 76.36% lower) than the EGUPD mode, thereby also resulting in a considerably fewer number of applications of the transfer-functions (geomean 14.6% fewer).

**LZINV vs. EGINV.** Across all the benchmarks, and both the platforms, we notice a performance improvement with LZINV mode, as compared to the EGINV mode, with geomean speedup as 3.62×, for Nanda, and 3.75×, for K2. The maximum speedup was observed in case of stream (5.88×) in Nanda, and leader-elect-dp (8.33×) in K2, consequent upon the fact that among the benchmarks shown in Fig. 33, stream and leader-elect-dp correspond to quite large reductions in the number of stabilization triggers (see Fig. 34) with the lazy modes of stabilization.

**LZINV vs. EGUPD.** From Fig. 33, note that the EGUPD mode outperforms the LZINV mode in the context of IDFA-stabilization time, for most of the cases across both the platforms. This trend can be explained as follows. From Fig. 34, it is clear that the LZINV mode consistently results in fewer stabilization triggers than the EGUPD mode. However, it is also expected that the incremental update of IDFA flow-maps in UPD-modes will result in fewer applications of the transfer-functions per trigger, as compared to the same in case of INV-modes which involve invalidation and full recomputation of the IDFA flow-maps. This is evident in Fig. 34, where we notice that EGUPD mode consistently applies significantly fewer transfer-functions (geomean 90.38% fewer) as compared to the LZINV mode. As a result, EGUPD mode outperforms the LZINV mode, for most of the scenarios.

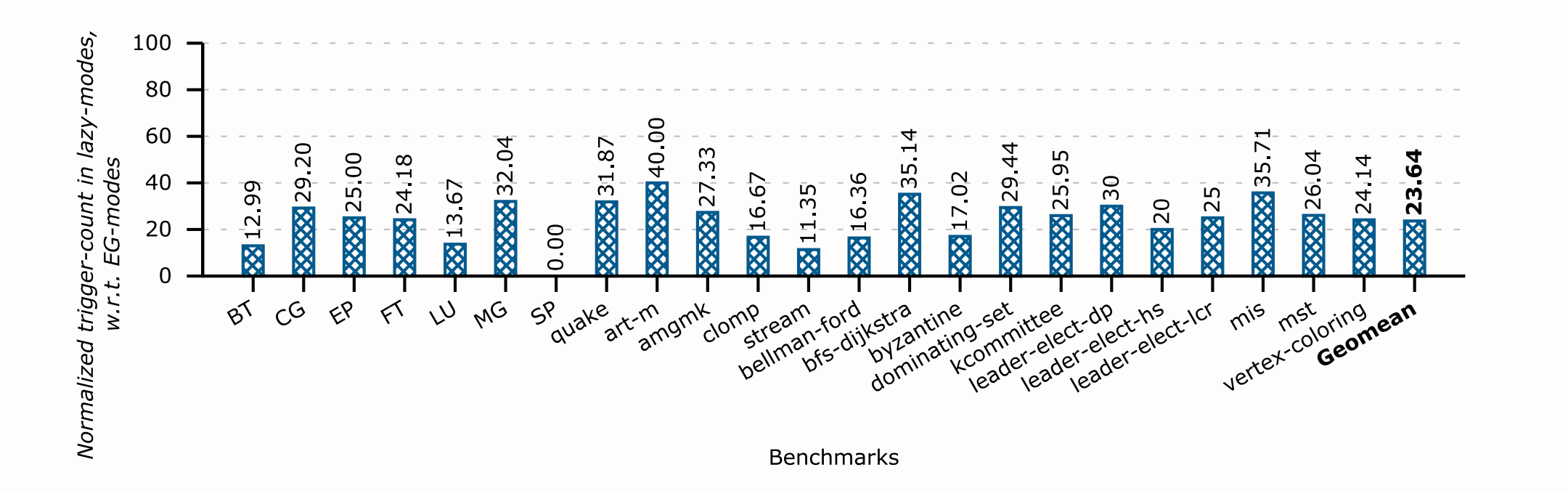

Fig. 34. Number of times IDFA stabilization was triggered under lazy modes of stabilization when performing the BarrElim optimization passes, normalized with respect to the number of triggers in case of EG-modes. Lower is better.

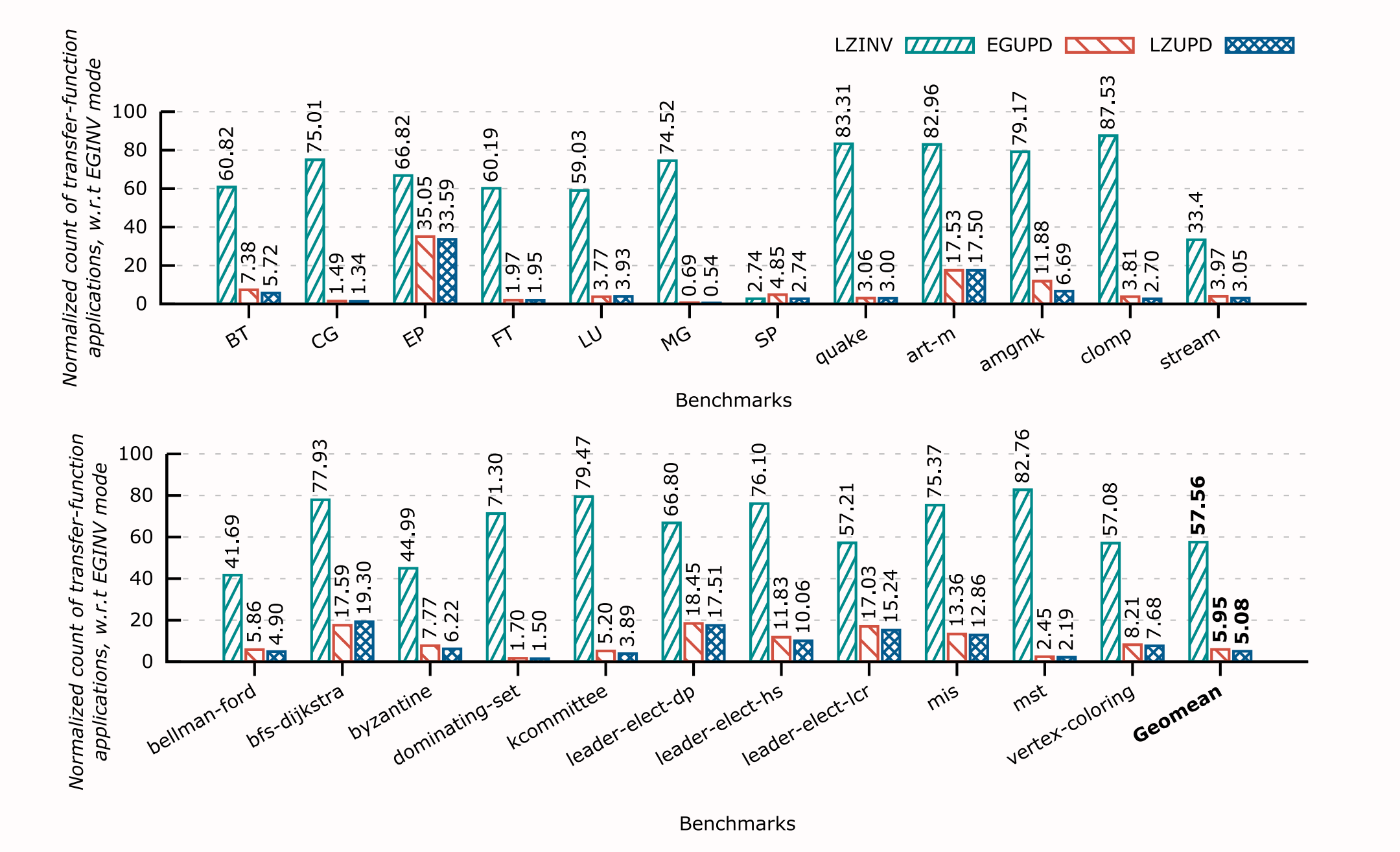

Fig. 35. Total number of applications of transfer-functions across all stabilization triggers when performing the BarrElim set of optimizations, for various modes of stabilization normalized with respect to the number of applications in case of the EGINV mode. Lower is better.

*Summary.* Overall, we found that the LZUPD mode leads to the maximum benefits in stabilization time, across all the four modes of stabilization. Further, we also observed that if the UPD-mode is unavailable for a program-abstraction due to absence of its incremental-update algorithm, the fully-automated stabilization provided by *Homeostasis* in the form of LZINV mode can still provide good performance improvements over the naive automated stabilization attained using the EGINV mode. These observations underline the performance benefits in stabilization time when using

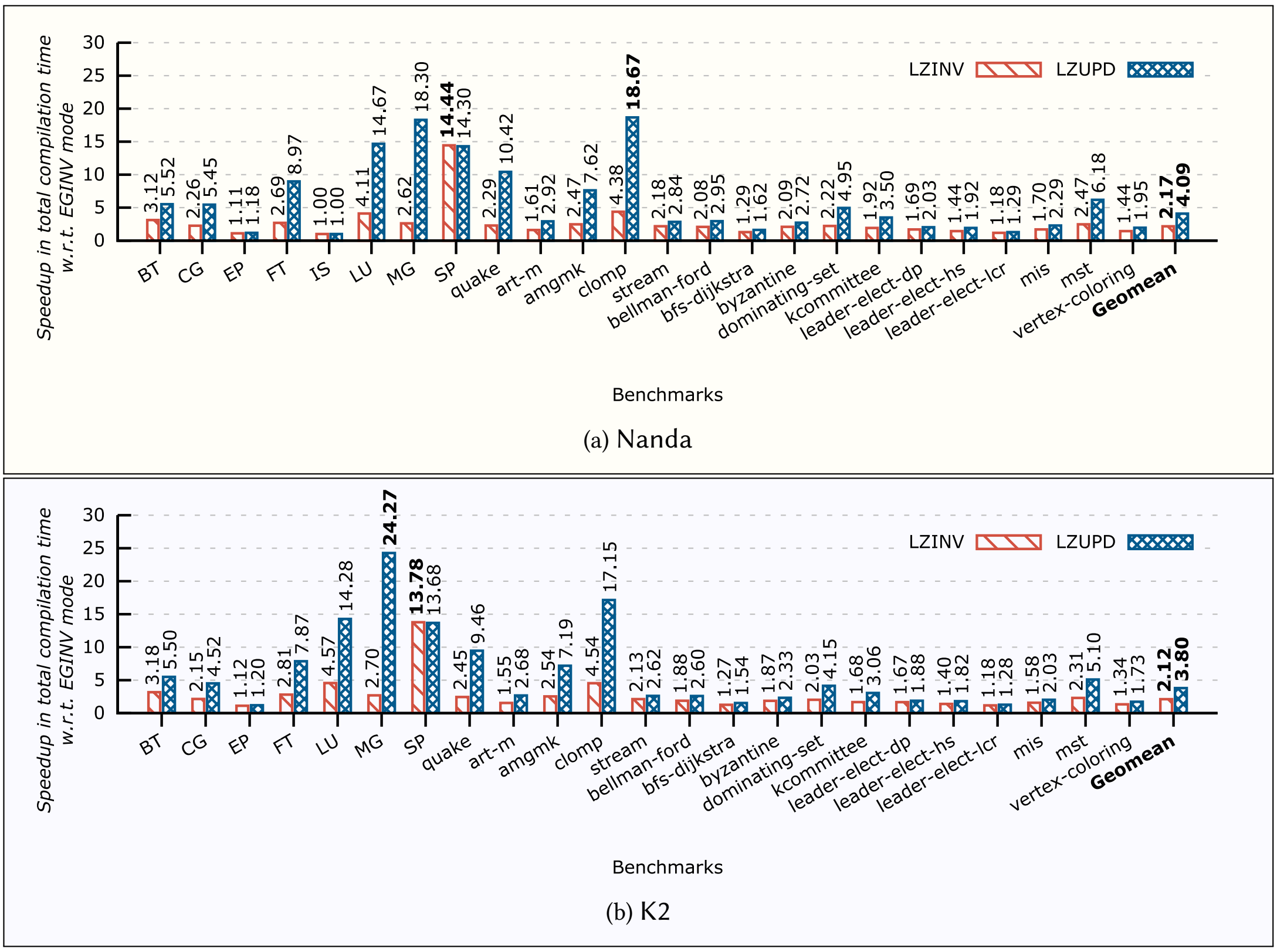

Fig. 36. Speedup in total compilation time under both the lazy modes of stabilization facilitated by *Homeostasis*, with respect to the EGINV mode, when applying the BarrElim set of optimization passes. Higher is better.

*Homeostasis*. This may in turn improve the overall compilation speed, as shown in Fig. 36 (with geomean improvements of 4.09× and 3.80× in Nanda and K2, respectively).

## B GUIDELINES TO RETROFIT *HOMEOSTASIS* IN LLVM

Given the popularity of the LLVM compiler framework, in this section we provide brief guidelines on how LLVM can be retrofitted with *Homeostasis*. This further helps us argue about the generality and applicability of *Homeostasis* to arbitrary IRs. The key components of *Homeostasis* (from Section 3.2) are mapped in LLVM as follows:

- The LLVM IR corresponds to the IR which can be used as the base program-abstraction. The basic blocks (`BasicBlock`), functions (`Function`), and modules (`Module`) of the LLVM IR can be treated as the three specific nodes on which elementary transformations can be specified. Each element (`Instruction`) of the basic-block needs to be made immutable.
- Various fundamental transformations on `BasicBlock` of LLVM IR that can serve as the *elementary transformations* need to be identified. A key requirement here is that one should be able to represent all possible transformations of the basic blocks using these elementary transformations. One such possible set comprises of two fundamental transformations: (i) insertion of an IR instruction at a given index, and (ii) removal of an IR instruction. Note that the updates to an immutable IR instruction can be achieved by creating a new instruction, and replacing the old instruction with the new one. The structure of these elementary transformations should be made conforming to the design shown in Fig. 5. Similar elementary transformations can be identified for the classes `Function` and `Module`.

- For ease of use, various higher-level transformation APIs (such as those that support copying a block, renaming a variable, etc.) can be created using these elementary transformations. Further, the code for existing transformation APIs should be updated such that all translations of the IR are expressed directly or indirectly using the identified set of elementary transformations.
- In *Homeostasis*, all program-abstractions must inherit from a base class, `BasePA`. LLVM already has a `Pass` super-class, which can serve the purpose of `BasePA`. The `Pass` class can be modified to adhere to the design given in Fig. 7. The definition of `stabilize` method from *Homeostasis* can be used as it is in the `Pass` class.
- LLVM already identifies a set of analysis passes, and a set of transformation passes; former corresponds to the *analysis passes* discussed in this text, whereas latter to the *optimization* passes. For each of the identified analysis passes, one needs to ensure that all accesses to the related program-abstractions are done through the appropriate getters that follow the simple structure shown in Fig. 7. As discussed in Section 6.2, one needs to ensure that (1) the existing code to generate the program-abstraction from scratch is invoked from the overridden method `compute`, and (2) the existing code to perform incremental update of the program-abstraction, wherever applicable, is invoked from the overridden method `handleUpdate`.

Note that the core library of LLVM is larger than one million lines of code written in C++, spanning across more than $3k$ files. Consequently, the development effort required to retrofit *Homeostasis* to LLVM is not trivial. However, the proven benefits of attaining self-stabilization in a compiler clearly make this one-time effort worthwhile, more so considering the large size of the code base involving numerous interacting passes of analysis and transformation passes. We leave this as a future work.